\documentclass[a4paper,11pt]{article}
\usepackage{jheppub} 

\usepackage{siunitx}
\usepackage{braket} 
\usepackage{amsmath}
\usepackage{amssymb}
\usepackage{bbold}
\usepackage{bm}

\usepackage{tikz}
\usepackage[compat=1.0.0]{tikz-feynman}

\usepackage{subcaption}

\usepackage{multirow}

\usepackage{xcolor} 

\title{Inclusive semileptonic decays from lattice QCD: analysis of systematic effects}

\author[a]{R.~Kellermann,}
\author[b]{A.~Barone,}
\author[c,d]{A.~Elgaziari,}
\author[a,e]{S.~Hashimoto,}
\author[a]{Z.~Hu,}
\author[c,d,f]{A.~J\"uttner,}
\author[a,e]{T.~Kaneko}
\affiliation[a]{High Energy Accelerator Research Organization (KEK), Ibaraki 305-0801, Japan}
\affiliation[b]{PRISMA+ Cluster of Excellence \& Institut f\"ur Kernphysik, Johannes-Gutenberg-Universit\"at Mainz, D-55099 Mainz, Germany}
\affiliation[c]{School of Physics and Astronomy, University of Southampton, Southampton SO17 1BJ, UK}
\affiliation[d]{STAG Research Center, University of Southampton, Southampton SO17 1BJ, UK}
\affiliation[e]{School of High Energy Accelerator Science, SOKENDAI (The Graduate University for Advanced Studies), Ibaraki 305-0801, Japan}
\affiliation[f]{CERN, Theoretical Physics Department, Geneva, Switzerland}

\emailAdd{kelry@post.kek.jp, abarone@uni-mainz.de, A.Elgaziari@soton.ac.uk, shoji.hashimoto@kek.jp, huzhi@post.kek.jp, andreas.juttner@cern.ch, takashi.kaneko@kek.jp}

\abstract{Lattice QCD calculations of inclusive semileptonic decay rates involve new types of systematic effects, such as truncation errors in the estimation of energy integrals, or finite-volume effects for multi-body final states. We investigate them for the lattice data of $D_s \to X_s \ell\nu$ decays, obtained using M\"obius domain-wall fermions. Separating the ground-state and excited-state contributions results in better control over these systematic effects.  With the Chebyshev polynomial approximation, the truncation error is under control, while the finite-volume effects are estimated using a model to describe two-body final states.}

\begin{document}

\begin{flushright}
 KEK-CP-0407\hspace{1em}CERN-TH-2025-059
\end{flushright}

\maketitle

\flushbottom

\section{Introduction}
\label{sec:Introduction}

Tensions between experimental observations and the Standard Model (SM) expectations may indicate New Physics effects, provided that theoretical uncertainties are understood quantitatively. 
Among these discrepancies is the long-standing tension between the exclusive 
and the inclusive 
measurements of the Cabibbo-Kobayashi-Maskawa (CKM) matrix elements $|V_{cb}|$ and $|V_{ub}|$
(see, for example, \cite{FlavourLatticeAveragingGroupFLAG:2021npn}).
While lattice calculations have proven very successful in computing exclusive form factors with controlled errors (for reviews see {\it e.g.} \cite{FlavourLatticeAveragingGroupFLAG:2021npn, Kaneko:2023kxx}), theory predictions for inclusive decays involve perturbative QCD and Operator Product Expansion (OPE), see {\it e.g.} \cite{Fael:2024rys, Finauri:2023kte}, for which further tests with non-perturbative inputs would be useful to verify the estimated results and their uncertainties.

In \cite{Hashimoto:2017wqo} a proposal was made to study inclusive semileptonic decays on the lattice. It relies on the analytic continuation of the forward Compton amplitude calculated on the Euclidean lattice, which corresponds to the amplitude in an unphysical kinematical regime. Ref.~\cite{Hansen:2017mnd} advocated that decay rates to multi-hadron states can be obtained through the forward Compton amplitude, provided that a method to extract the corresponding spectral function exists, which is unfortunately known to be a numerically ill-posed problem. However, the extraction of the spectral function can be bypassed by identifying the phase-space factor as a smearing of the spectral function, with which the computational requirement is less severe, paving the way for a computation of the inclusive decay rate within lattice QCD \cite{Gambino:2020crt}. The method has been used for the lattice calculation of the semileptonic decays of~heavy mesons \cite{Gambino:2022dvu,Barone:2023tbl} and hadronic tau decays \cite{Evangelista:2023fmt,ExtendedTwistedMass:2024myu}. It can also be applied for inclusive lepton-nucleon scattering \cite{Fukaya:2020wpp}.

In this paper, following previous studies \cite{Gambino:2020crt,Barone:2022gkn,Kellermann:2022mms, Barone:2023iat,Kellermann:2023yec,Barone:2023tbl,Kellermann:2024jqg}, we investigate the systematic effects specific to this new method for calculating the inclusive decay rate. As an example, we take the inclusive semileptonic decay rate of the $D_s$ meson into hadronic final states containing a strange quark and its antiquark, {\it i.e.} $D_s \to X_{s} \ell\nu_\ell$, which is related to the determination of $|V_{cs}|$. We employ the Chebyshev-polynomial method \cite{Barata:1990rn,Bailas:2020qmv,Gambino:2020crt} to realize the energy integral over possible hadronic final states, which is a key to bypass the extraction of the spectral density, and calculate the decay rate directly. Although not discussed in this work, we refer to \cite{Barone:2023tbl} for a comparison between the Chebyshev approximation and the Hansen-Lupo-Tantalo (HLT) reconstruction \cite{Hansen:2019idp,Gambino:2022dvu}. In this work, we further examine the error from truncated higher order terms, which can be estimated in a systematic manner in the Chebyshev approach.

The forward Compton amplitude is a matrix element of two flavor-changing weak currents sandwiched between two $D_s$-meson states. Depending on the current (vector or axial-vector), its polarization (parallel or perpendicular to the momentum inserted) and the momentum carried away by the lepton pairs, the amplitude is saturated by different hadronic final states. The major contribution comes from the $S$-wave meson $\eta_s$\footnote{We ignore the disconnected strange-quark-loop diagram in this work. The $s\bar{s}$ pseudoscalar state is called $\eta_s$ instead of $\eta$ or $\eta'$.} and $\phi$ meson, but other states such as radial and orbital excitations and multi-particle states also contribute. In the inclusive analysis, these states are all taken into account automatically, but associated systematic errors may depend on the composition of the states. For example, whether the total contribution is dominated by a single narrow peak or distributed over many two-body states, would affect how well the Chebyshev approximation works. We therefore discuss the composition of the final states in some detail.

Finite-volume effects are a major concern in the inclusive analysis on the lattice, since the final states include multi-body hadronic states. In the infinite-volume limit, multi-body states exhibit a continuum spectrum depending on the relative momentum among the hadrons in the final state. In finite volumes, on the other hand, only the states that satisfy the boundary conditions, typically periodic boundary conditions, are allowed, and the spectrum becomes discrete. For two-body states, which give the dominant contribution for the $D_s$ semileptonic decay into states other than a single hadron, the spectrum is determined from the phase shift of the two particles through L\"uscher's formula \cite{Luscher:1985dn,Luscher:1986pf}. In this work, neglecting the scattering phase shift, we construct a model to estimate the corresponding systematic effect and verify the model using the actual lattice data for the excited states. We find that finite-volume effects result only in small corrections to the final result.

Our simulations employ M\"obius domain-wall fermions \cite{Shamir:1993zy,Furman:1994ky,Brower:2012vk} for all quarks, including the $c$ and $s$ quarks near their physical values, on one of the lattice ensembles generated for the study of $B$ meson semileptonic decays \cite{Colquhoun:2022atw,Aoki:2023qpa}.

Although the aforementioned discrepancy of the CKM matrix elements is observed in the bottom sector, its lattice calculation comes with potentially large discretization effects for bottom quarks. We therefore focus on the charm sector to validate the proposed method. The formalism developed and discussed in this work can be applied straightforwardly to the bottom sector \cite{Barone:2023tbl}.

The remainder of this paper is structured as follows. In Section~\ref{sec:InclusiveDecays} we review the theoretical framework necessary for this work as well as details on the lattice implementation, which were introduced in \cite{Gambino:2020crt} and built on in \cite{Barone:2023tbl}. Following this, Section~\ref{sec:Numerical} presents details of the simulation. We analyze the structure of the Compton amplitude, such as the limit where only the ground-state contributes, in Section~\ref{sec:Amplitudes}, before addressing the analysis strategy in Section~\ref{sec:LatticeDataAnalysis}. As part of this discussion, we will address the systematic error associated with the analysis strategy starting from Section~\ref{sec:SysErrApproximation}. Section~\ref{sec:SystematicErrors} then follows up with a discussion of our methodology to estimate finite-volume effects and how it is applied in the analysis. Finally, we present our results in Section~\ref{sec:Results}, before presenting a summary and future prospects in Section~\ref{sec:ConclusionOutlook}.

\section{Inclusive decays on the lattice}
\label{sec:InclusiveDecays}

We begin with a brief review of the formalism, introducing all tools and necessary notation required to compute inclusive decay rates from Euclidean correlation functions obtained from lattice simulations. In Section \ref{sec:InclusiveDecay}, we present the formulation in the continuum,
which is followed in Section~\ref{sec:EuclideanInclusive} by a discussion on how hadron correlators are used to obtain an estimate for the energy integral over hadronic final states in the total decay rate.

\subsection{Inclusive decay rate}
\label{sec:InclusiveDecay}

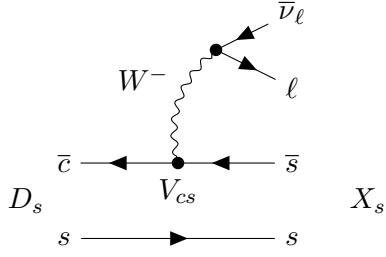
\begin{figure}[tb]
  \centering
  \begin{tikzpicture}
    \begin{feynman}
      \vertex (i1) {\(\overline{c}\)};
      \vertex[right=1.5cm of i1,dot,label=270:{\(V_{cs}\)}] (a1) {};
      \vertex[right=1.5cm of a1] (f1) {\(\overline{s}\)};

      \vertex[below=1cm of i1] (i2) {\(s\)};
      \vertex[below=1cm of f1] (f2) {\(s\)};

      \vertex[above=1cm of f1] (f4) {\(\ell\)};
      \vertex[above=1cm of f4] (f3) {\(\overline{\nu}_\ell\)};
      \vertex[dot] at ($(f4)!0.5!(f3) - (1cm,0)$) (a2) {};

      \node at ($(i1)!0.5!(i2) - (0.5cm,0)$) {\(D_s\)};
      \node at ($(f1)!0.5!(f2) - (-1cm,0)$) {\(X_s\)};

      \diagram*{
        (i1) -- [anti fermion] (a1)[dot] -- [anti fermion] (f1),
        (i2) -- [fermion] (f2),
        (a1) -- [boson,bend left, edge label=\(W^{-}\)] (a2),
        (f3) -- [fermion] (a2) -- [fermion] (f4),
      };
    \end{feynman}
  \end{tikzpicture}
  \caption{Feynman digram for the decay $D_s \rightarrow X_s \ell \nu_\ell$.}
  \label{fig:InclusiveDecayFeynman}
\end{figure}

In contrast to \cite{Gambino:2020crt, Barone:2023tbl}, where the focus was on the $B_s \rightarrow X_c \ell\nu_\ell$ channel, in this work we study the $D_s \rightarrow X_s \ell\nu_\ell$ process as depicted in Fig.~\ref{fig:InclusiveDecayFeynman}, where $X_s$ represents all possible final states containing an $s$- and a $\bar{s}$-quark. The formalism for both cases is essentially the same, and the application to the bottom sector is straightforward.

For the $\bar{c} \rightarrow \bar{s}$ process, the weak Hamiltonian is given by
\begin{align}
  H_W = \frac{4G_F}{\sqrt{2}} V_{cs} [\overline{c}_L\gamma^\mu s_L] [\overline{\nu}_{\ell L} \gamma_\mu \ell_L] \, ,
\end{align}
where $G_F$ is the Fermi constant and $V_{cs}$ is the CKM matrix element governing the charged-current flavor-changing $\bar{c}\to \bar{s}$ quark transition. The electroweak quark current is given by $J_\mu = \overline{c}_L\gamma^\mu s_L = \overline{c}\gamma_{\mu}(1-\gamma_5)s/2$, which can also be written in the form $J_\mu = V_\mu - A_\mu$ where $V_\mu$ and $A_\mu$ are given by $\overline{c} \gamma_\mu s/2$ and $\overline{c} \gamma_\mu \gamma_5 s/2$, respectively.

Compared to its exclusive counterpart, the decay rate of inclusive processes has one more kinematical variable due to the freedom in the mass of the outgoing hadrons. The differential decay rate can be written as
\begin{align}
  \frac{d\Gamma}{dq^2 dq_0 dE_\ell} = \frac{G_F^2 |V_{cs}|^2}{8\pi^3} L_{\mu\nu} W^{\mu\nu} \, ,
  \label{equ:DiffRateInclusive}
\end{align}
where we neglect QED corrections. Here, $L_{\mu\nu}$ is the leptonic tensor
\begin{align}
  L^{\mu\nu} = p_\ell^\mu p_{\nu_\ell}^\nu + p_\ell^\nu p_{\nu_\ell}^\mu - g^{\mu\nu} p_\ell \cdot p_{\nu_\ell} - i\epsilon^{\mu\alpha\nu\beta} p_{\ell, \alpha} p_{\nu_\ell, \beta} \, ,
  \label{equ:LeptonTensor}
\end{align}
where $p_\ell \text{ and } p_{\nu_\ell}$ denote the four-momenta of the lepton and neutrino, respectively. The momentum transfer between initial and final mesons is $q = p_\ell + p_{\nu_\ell}$. The hadronic tensor $W^{\mu\nu}$ is given by
\begin{align}
  \begin{split}
    W^{\mu\nu}(p_{D_s}, q) = &\frac{1}{2\pi} \frac{1}{2E_{D_s}} \int\! d^4x\, e^{iqx} \braket{D_s(\bm{p}_{D_s})|\mathcal{T}\{J^{\mu\dagger}(x)J^{\nu}(0)\}|D_s(\bm{p}_{D_s})} \\
    = &\frac{1}{2E_{D_s}} \sum_{X_s} (2\pi)^3 \delta^{(4)}(p_{D_s} - q - p_{X_s}) \\
    &\times \braket{D_s(\bm{p}_{D_s})|J^{\mu\dagger}(x)|X_s(\bm{p}_{X_s})} \braket{X_s(\bm{p}_{X_s})|J^{\nu}(0)|D_s(\bm{p}_{D_s})} \, ,
    \label{equ:HadronicTensor}
  \end{split}
\end{align}
where we inserted a complete set of states $\mathbb{1} = \sum_{X_s} \ket{X_{s}(\bm{p}_{X_s})} \bra{X_{s}(\bm{p}_{X_s})}$ in the second line. The operators are ordered such that the amplitude corresponds to the desired semileptonic decay, {\it i.e. $x_0 > 0$}; the opposite ordering represents intermediate states having a $c\bar{c}s\bar{s}$ quantum number. It implicitly includes an integration over all possible momenta $\bm{p}_{X_s}$ under a Lorentz-invariant phase-space integral, and $q = p_{D_s} - p_{X_s}$ is the momentum transfer between the initial and final hadronic states. Throughout this paper, we assume the $D_s$ meson at rest, {\it i.e.} $\bm{p}_{D_s} = (0,0,0)$.
The hadronic tensor can be decomposed as a sum of five scalar structure functions 
$W_i(q^2)$ 
given by
\begin{align}
  W^{\mu\nu} = -g^{\mu\nu} W_1 + v^\mu v^\nu W_2 -i\epsilon^{\mu\nu\alpha\beta} v_{\alpha} q_{\beta} W_3 + q^{\mu}q^{\nu} W_4 + (v^{\mu}q^{\nu} + v^{\nu}q^{\mu}) W_5 \, ,
  \label{equ:HadronicDecomposition}
\end{align}
where $v = p_{D_s}/m_{D_s} = (1,0,0,0)$ is the four-velocity of the initial $D_s$ meson at rest and $q = (q_0, \bm{q}) = (m_{D_s}-\omega, -\bm{p}_{X_s})$. Hereafter, $\omega$ will denote the energy of the final-state hadron, {\it i.e.} $\omega = E_{X_s}$. We use the relations to decompose each component:
\begin{align}
  W^{00} &= -W_1 + W_2 + q_0^2 W_4 + 2q_0 W_5 \label{equ:W00} \, , \\
  W^{ij} &= \delta_{ij} W_1 + q_i q_j W_4 - i \epsilon_{ij0k} q^k W_3 \label{equ:Wij} \, , \\
  W^{0i} &= q_i (q_0 W_4 + W_5) \label{equ:W0i} \, ,
\end{align}
where we denote the spatial indices by $i$, $j$, and $k$. We can contract the spatial indices with the three-momentum components $q_i$ to invert these relations to obtain expressions for the structure functions $W_i$ in terms of the hadronic tensor $W^{\mu\nu}$ and $\bm{q}$.

To obtain the total decay rate, we first integrate \eqref{equ:DiffRateInclusive} over the lepton energy $E_\ell$. We ignore the lepton mass, {\it i.e.} $m_l \simeq 0$. We also rewrite the integrals over $q^2$ and $q_0$ in terms of $\bm{q}^2 = q_0^2 - q^2$ and $\omega = m_{D_s} - q_0$, {\it i.e.} the three-momentum and energy of the final hadronic state $X_s$; the Jacobian of this transformation being $1$. The total decay rate is then written in terms of integrals over $\omega$ and $\bm{q}^2$:
\begin{align}
  \Gamma = \frac{G_F^2 |V_{cs}|^2}{24\pi^3} \int_0^{\bm{q}_{\text{max}}^2} d\bm{q}^2 \sqrt{\bm{q}^2} \bar{X}(\bm{q}^2) \, .
  \label{equ:TotalRateInclusive}
\end{align}
Here, the energy integral is given by
\begin{align}
  \bar{X}(\bm{q}^2) = \sum_{l=0}^{2} \bar{X}^{(l)}(\bm{q}^2) \, , \quad \bar{X}^{(l)}(\bm{q}^2) \equiv \int_{\omega_{\text{min}}}^{\omega_{\text{max}}} d\omega \, X^{(l)}(\bm{q}^2, \omega) \, ,
  \label{equ:XDecomposition}
\end{align}
where we use 
\begin{align}
  \begin{split}
    X^{(0)}(\bm{q}^2, \omega) &= \bm{q}^2 W_{00} + \sum_{i} (q_i^2 - \bm{q}^2) W_{ii} + \sum_{i \neq j} q^{i} W_{ij} q^{j} \, , \\
    X^{(1)}(\bm{q}^2, \omega) &= -q_0 \sum_{i} q^{i} \left(W_{0i} + W_{i0}\right) \, ,\\
    X^{(2)}(\bm{q}^2, \omega) &= q_0^2 \sum_{i} W_{ii} \, .\\
  \end{split}
  \label{equ:XComponents}
\end{align}
The integral limits are $\bm{q}_{\text{max}}^2 = (m_{D_s}^2 - m_{\eta_s}^2)^2/(4M_{D_S}^2)$, $\omega_{\text{min}} = \sqrt{m_{\eta_s}^2 + \bm{q}^2}$ and $\omega_{\text{max}} = m_{D_s} - \sqrt{\bm{q}^2}$, by imposing four-momentum conservation as well as fixing the lightest final state of this inclusive decay to the $\eta_s$-meson.

These equations play the central role in the analysis of inclusive decays, as they allow us to express the total decay rate through an integral over the energy $\omega$ of the hadronic final state, as well as the corresponding three-momentum $\bm{q}^2$; all information of the hadronic dynamics is encoded in $X^{(l)}(\bm{q}^2, \omega)$. They are linear combinations of the hadronic tensor and some kinematical factors. 
The $V-A$ nature of the charged current in $X^{(l)}$ is realised by writing the hadronic tensor and the corresponding $X^{(l)}(\bm{q}^2,\omega)$ as
\begin{align}
  X^{(l)} = X^{(l), VV} + X^{(l), AA} - X^{(l), VA} - X^{(l), AV} \, ,
  \label{equ:Decomposition}
\end{align}
where $V$ and $A$ denote the insertion of the vector and axial-vector currents, respectively. The decomposition for $\bar{X}^{(l)}$ follows in a similar manner.

\subsection{Euclidean formulation of the inclusive decay}
\label{sec:EuclideanInclusive}

Here we discuss the implementation of the energy integral \eqref{equ:XDecomposition} in the lattice analysis. This section follows \cite{Hashimoto:2017wqo, Gambino:2020crt, Gambino:2022dvu, Barone:2023tbl}. We start from the hadronic tensor \eqref{equ:HadronicTensor}
\begin{align}
  W^{\mu\nu}(q) = \frac{1}{4\pi} \frac{1}{m_{D_s}} \int d^4x\, e^{iqx} \braket{D_s|J^{\mu\dagger}(x) J^{\nu}(0)|D_s} \, .
  \label{equ:HadronicTensor2}
\end{align}
On the lattice, we compute the time dependence of the Euclidean four-point function
\begin{align}
  C^{SJJS}_{\mu\nu}(\bm{q}, t_{\text{snk}}, t_2, t_1, t_{\text{src}}) \stackrel{t_2 \geq t_1}{=} \sum_{\bm{x}_{\text{snk}}, \bm{x}_{\text{src}}} \braket{\mathcal{O}^{S}_{D_s}(x_{\text{snk}}) \tilde{J}_\mu^\dagger(\bm{q},t_2) \tilde{J}_\nu(\bm{q},t_1) \mathcal{O}^{S\dagger}_{D_s}(x_{\text{src}})} \, ,
  \label{equ:FourPointDefinition}
\end{align} 
where $\mathcal{O}_{D_s}^S$ defines an interpolating operator of the quantum numbers of the $D_s$ meson, and $\tilde{J}_\nu(\bm{q}, t) \equiv \sum_{\bm{x}} \exp(-i\bm{q}\cdot\bm{x}) J_\nu(\bm{x},t)$ is a discrete Fourier transform, projecting the currents onto a specific three-momentum. This setup creates a $D_s$ meson that carries zero momentum at the source time slice $t_{\text{src}}$, which is annihilated at the sink $t_{\text{snk}}$. The corresponding quark flow diagram is shown in Fig. \ref{fig:QuarkFlow4Pt}. The propagator of the $c$ quark from position $x_1$ to $x_{\text{src}}$, $G_c(x_{\text{src}},x_1)$, is represented by the black line. The blue line, $\Sigma_{scs}(x_1,x_{\text{src}})$, defines a sequential propagator, propagating the $s$ quark from $x_{\text{src}}$ to $x_{\text{snk}}$, the $c$ quark from $x_{\text{snk}}$ to $x_2$ and finally the $s$ quark from $x_2$ to $x_1$. The source and sink positions $x_{\text{src}}$ and $x_{\text{snk}}$ are summed over the spatial volume.

\begin{figure}[tb]
  \centering
  \begin{tikzpicture}
    \begin{feynman}
      \vertex[dot, label=180:{\(x_{\text{src}}\)}] (Source) {};
      \vertex[right=8cm of Source, dot,label=0:{\(x_{\text{snk}}\)}] (Sink) {};
      \vertex[above right= 1.6cm and 2.5cm of Source, dot, label=135:{\(J^{\nu}(t_1)\)}] (C1) {};
      \vertex[above left= 1.6cm and 2.5cm of Sink, dot, label=45:{\(J^{\mu\dagger}(t_2)\)}] (C2) {};
      \vertex[above right= 1.6cm and 2.5cm of Source, dot, label=315:{\(x_1\)}] (C3) {};
      \vertex[above left= 1.6cm and 2.5cm of Sink, dot, label=225:{\(x_2\)}] (C4) {};
      \vertex[above=1cm of C1] (aux1) {};
      \vertex[above=1cm of C2] (aux2) {};
      \vertex[below=4cm of C1] (aux3) {};
      \vertex[below=4cm of C2] (aux4) {};

      \node (aux5) at ($(C1)!0.5!(Source)+ (0.35cm,-0.1cm)$) {\(G_c(x_{\text{src}},x_1)\)};
      \node (aux6) at ($(Sink)!0.5!(Source) - (0cm, 1.5cm)$) {\(s\)};
      \node (aux7) at ($(C1)!0.5!(C2) + (0cm,0.25cm)$ ) {\(s\)};

      \diagram*{
        (Source) -- [fermion, bend right,blue, thick,edge label={\(\Sigma_{scs}(x_1,x_{\text{src}})\)}] (Sink),
        (Source) -- [anti fermion, bend left, thick, edge label=\(c\)] (C1),
        (C2) -- [anti fermion, bend left, blue, thick , edge label=\(\textcolor{black}{c}\)] (Sink),
        (C2) -- [charged scalar, thick , blue] (C1),
        (aux1) -- [lightgray, very thick, scalar] (aux3),
        (aux2) -- [lightgray, very thick, scalar] (aux4),

     };

    \end{feynman}
  \end{tikzpicture}
  \caption{Schematic representation of the diagram for the four-point correlator. The contraction depicted here is based on two propagators. First, we have $G_c(x_{\text{src}},x_1)$ depicted by the black line, which propagates the $c$ quark from $x_1$ to $x_{\text{src}}$. Secondly, we have $\sum_{scs} (x_1, x_{\text{src}})$ depicted by the blue line, which is a sequential propagator used to propagate the $s$ quark from $x_{\text{src}}$ to $x_{\text{snk}}$, the $c$ quark from $x_{\text{snk}}$ to $x_2$ and the $s$ quark from $x_2$ to $x_1$.}
  \label{fig:QuarkFlow4Pt}
\end{figure}

In order to relate \eqref{equ:FourPointDefinition} to the matrix element \eqref{equ:HadronicTensor2}, we need the condition $t_{\text{snk}} - t_2 \gg 1$, $t_1 - t_{\text{src}} \gg 1$ and $t_2 > t_1$ to ensure that the excited states of the $D_s$ meson have decayed sufficiently. To enlarge the time window we employ operator smearing, {\it i.e.} enhancing the overlap of the operator $\mathcal{O}_{D_s}^S$ with the ground-state $D_s$ meson. This is specified by the superscripts $S$ and $L$, denoting smeared and unsmeared operators, respectively. More details will be given in Section \ref{sec:Numerical}.

Within the valid time window for the ground-state saturation, the four-point function takes the form
\begin{align}
    \begin{split}
  &C_{\mu\nu}^{SJJS}(\bm{q},t_{\text{snk}}, t_2, t_1, t_{\text{src}})  \\ &= \frac{1}{4m_{D_s}^2} \braket{0|\mathcal{O}^{S}_{D_s}(t_{\text{snk}})|D_s} \braket{D_s|\tilde{J}_\mu^\dagger(\bm{q},t_2) \tilde{J}_\nu(\bm{q},t_1)|D_s} \braket{D_s|\mathcal{O}^{S\dagger}_{D_s}(t_{\text{src}})|0} \, .
  \end{split}
  \label{equ:FourPointWindow}
\end{align}
To identify the forward-scattering matrix element \eqref{equ:HadronicTensor2} in \eqref{equ:FourPointWindow} we have to cancel the factors $\braket{0|\mathcal{O}^{S}_{D_s}|D_s}$ and $\braket{D_s|\mathcal{O}^{S\dagger}_{D_s}|0}$. They are obtained from zero-momentum two-point functions
\begin{align}
  C^{SS}(t_2,t_1) &= \sum_{\bm{x}_1, \bm{x}_2} \braket{\mathcal{O}^S_{D_s}(\bm{x}_2,t_2) \mathcal{O}_{D_s}^{S\dagger} (\bm{x}_1,t_1)} \\
  &\stackrel{t_2 - t_1 \gg 0}{=} \frac{1}{2m_{D_s}} \braket{0|\mathcal{O}^S_{D_s}|D_s} \braket{D_s|\mathcal{O}_{D_s}^{S\dagger}|0} e^{-(t_2 - t_1) m_{D_s}} \, .
\end{align}
through the ratio
\begin{align}
  R_{\mu\nu}(\bm{q}, t_2, t_1) \equiv\frac{C_{\mu\nu}^{SJJS}(\bm{q},t_{\text{snk}}, t_2, t_1, t_{\text{src}})}{C^{SS}(t_{\text{snk}},t_2) C^{SS}(t_1, t_{\text{src}})} \rightarrow \frac{\frac{1}{2m_{D_s}} \braket{D_s|\tilde{J}_\mu^\dagger(\bm{q},t_2) \tilde{J}_\nu(\bm{q},t_1)|D_s}}{\frac{1}{2m_{D_s}} \left|\braket{0|\mathcal{O}_{D_s}^S|D_s}\right|^2} \, ,
  \label{equ:RatioFourAndTwoFunction}
\end{align}
where the additional factor $\frac{1}{2m_{D_s}} \left|\braket{0|\mathcal{O}_{D_s}^S|D_s}\right|^2$ appearing in the denominator on the r.h.s. can be evaluated from fits to the time dependence of the $C^{SL}$, $C^{LS}$, $C^{LL}$ and $C^{SS}$ two-point functions. It is also possible to define different ratios,  {\it e.g.}, employing a combination of $C^{SL}$ and $C^{LS}$, as was done in \cite{Barone:2023tbl}. We then use the invariance under a shift in time maintaining $t \equiv t_2 -t_1$ to obtain
\begin{align}
  C_{\mu\nu}(\bm{q},t) \equiv \frac{1}{2m_{D_s}} \left|\braket{0|\mathcal{O}_{D_s}^S|D_s}\right|^2 R_{\mu\nu}(\bm{q}, t_2-t_1, 0)  = \frac{1}{2m_{D_s}} \braket{D_s|\tilde{J}_\mu^\dagger(\bm{q},0) e^{-\hat{H}t} \tilde{J}_\nu(\bm{q},0)|D_s} \, .
  \label{equ:LatticeCorrelatorRatio}
\end{align}
The relation to the hadronic tensor in \eqref{equ:HadronicTensor2} is then given through the Laplace transform
\begin{align}
  \begin{split}
    C_{\mu\nu}(\bm{q},t) &= \int_0^\infty d\omega\,\frac{1}{2m_{D_s}} \braket{D_s|\tilde{J}_\mu^\dagger(\bm{q},0) \delta(\hat{H} -\omega) \tilde{J}_\nu(\bm{q},0)|D_s} e^{-\omega t} \\
    &= \int_0^\infty d\omega\, W_{\mu\nu} (\bm{q}, \omega) e^{-\omega t} \, .
  \end{split}
  \label{equ:SpectralRepresentation}
\end{align}
Here, we have
\begin{align}
  W_{\mu\nu}(\bm{q},\omega) = \frac{1}{2m_{D_s}} \sum_{X_s} \delta(\omega - E_{X_s}) \braket{D_s|\tilde{J}_\mu^\dagger(\bm{q},0)|X_s} \braket{X_s|\tilde{J}_\nu(\bm{q},0)|D_s} \, ,
  \label{equ:HadronicTensor3}
\end{align}
as a spectral representation of $C_{\mu\nu}(\bm{q},t)$. 

The extraction of spectral densities from hadronic correlators obtained from lattice simulations is an infamous ill-posed problem, {\it i.e.} the reconstruction of $C_{\mu\nu}$ is trivial if $W_{\mu\nu}$ is known, while the other way round is not. Fortunately, what we really need here, are the integrals $\bar{X}^{(l)}(\bm{q}^2)$ \eqref{equ:XDecomposition}, in which the hadronic tensor is smeared over energy. The energy integral $\bar{X}^{(l)}(\bm{q}^2)$ can be generically written as
\begin{align}
  \bar{X}^{(l)}(\bm{q}^2) = \int_{\omega_{\text{min}}}^{\omega_{\text{max}}} d\omega\, W^{\mu\nu}(\bm{q}, \omega) k_{\mu\nu}^{(l)}(\bm{q}, \omega) \, ,
\end{align}
where $k_{\mu\nu}^{(l)}(\bm{q},\omega)$ is a known function depending only on the three-momentum $\bm{q}$ and the energy $\omega$ of the hadronic final state. We manipulate the integral by shifting the limits of integration as $\omega_{\text{min}} \rightarrow \omega_0$, with $\omega_0 \leq \omega_{\text{min}}$ and $\omega_{\text{max}} \rightarrow \infty$, introducing a step function $\theta(\omega_{\text{max}} - \omega)$ to cut off all contributions above $\omega_{\text{max}}$:
\begin{align}
  \begin{split}
    \bar{X}^{(l)}(\bm{q}^2) &= \int_{\omega_{0}}^{\infty} d\omega\, W^{\mu\nu}(\bm{q}, \omega) k_{\mu\nu}^{(l)}(\bm{q}, \omega) \theta(\omega_{\text{max}} - \omega) \\
    &= \int_{\omega_{0}}^{\infty} d\omega\, W^{\mu\nu}(\bm{q}, \omega) K_{\mu\nu}^{(l)}(\bm{q}, \omega) \, ,
  \end{split}
  \label{equ:EnergyIntegralXBar}
\end{align}
where in the second step we defined $K_{\mu\nu}^{(l)}(\bm{q},\omega)\equiv k_{\mu\nu}^{(l)}(\bm{q},\omega) \theta(\omega_{\text{max}} - \omega)$. We refer to it as the kernel function. The lower limit $\omega_0$ can be freely chosen in the range $0 \leq \omega_0 \leq \omega_{\text{min}}$, because there is no state below the lowest-lying energy state $\omega_{\text{min}}$. In the case of $D_s \rightarrow X_s\ell\nu_\ell$, this corresponds to $\omega_{\text{min}} = \sqrt{m_{\eta_s}^2 + \bm{q}^2}$. In Section \ref{sec:Results}, this freedom of choosing $\omega_0$ will be further exploited.

In the lattice implementation of the energy integral, we replace the sharp cut of the step function $\theta(\omega_{\text{max}} - \omega)$ by a smooth one, {\it e.g.} a sigmoid function
\begin{align}
  \theta_{\sigma}(x) = \frac{1}{1 + e^{-x/\sigma}} \, ,
  \label{equ:SigmoidFunction}
\end{align}  
where the range of smoothing is controlled by the smearing parameter $\sigma$. The limit $\sigma\rightarrow 0$ is needed to restore the physical decay rate, but smearing is, as we will show below, a useful tool for better controlling and understanding systematic effects.  In Section~\ref{sec:SysErrApproximation}, we discuss a method to estimate the corrections due to the $\sigma\rightarrow 0$ extrapolation. Following \cite{Gambino:2020crt}, we expand the kernel $K_{\sigma, \mu\nu}^{(l)}(\bm{q}, \omega)$ in polynomials of $e^{-a\omega}$ (for simplicity, we set $a = 1$ in the following) up to an order $N$, 
\begin{align}
  K_{\sigma,\mu\nu}^{(l)}(\bm{q}, \omega) \simeq c_{\mu\nu,0}^{(l)} (\bm{q};\sigma) + c_{\mu\nu,1}^{(l)} (\bm{q};\sigma) e^{-\omega} + \cdots + c_{\mu\nu,N}^{(l)} (\bm{q};\sigma) e^{-N\omega} \, .
\end{align}
The subscript $\sigma$ on $\bar{X}$ indicates its dependence on the smearing parameter.
Our target quantity $\bar{X}_\sigma^{(l)}(\bm{q}^2)$ can then be calculated as
\begin{align}
  \begin{split}
    &\bar{X}_{\sigma}^{(l)}(\bm{q}^2) \simeq \int_{\omega_0}^{\infty} d\omega\, W^{\mu\nu}(\bm{q},\omega) e^{-2\omega t_0} K_{\sigma, \mu\nu}^{(l)}(\bm{q}, \omega; t_0) \\
    &\simeq c_{\mu\nu,0}^{(l)}(\bm{q};\sigma; t_0) \int_{\omega_0}^{\infty} d\omega\, W^{\mu\nu}(\bm{q},\omega) e^{-2\omega t_0} + c_{\mu\nu,1}^{(l)}(\bm{q};\sigma; t_0) \int_{\omega_0}^{\infty} d\omega\, W^{\mu\nu}(\bm{q},\omega) e^{-2\omega t_0} e^{-\omega} \\
    &+ \cdots + c_{\mu\nu,N}^{(l)}(\bm{q};\sigma; t_0) \int_{\omega_0}^{\infty} d\omega\, W^{\mu\nu}(\bm{q},\omega) e^{-2\omega t_0} e^{-N\omega} \, .
  \end{split}
  \label{equ:PolynomialApproximationStart}
\end{align}
We have furthermore introduced the factor $e^{-2\omega t_0}$ in the first line of \eqref{equ:PolynomialApproximationStart}, which we compensate for in the kernel function $K_{\sigma, \mu\nu}^{(l)}(\bm{q},\omega; t_0) \equiv e^{2\omega t_0} K_{\sigma, \mu\nu}^{(l)}(\bm{q},\omega)$. The purpose of this term is to avoid the contact term of $t_1 = t_2$ appearing in \eqref{equ:FourPointDefinition}, since this element receives contributions from the opposite time ordering corresponding to unphysical $\bar{c}ss\bar{c}$ final states. Details on the suitable choices of $t_0$ will follow in Section~\ref{sec:Results}. 
The coefficients $c_{\mu\nu,k}^{(l)}(\bm{q};\sigma; t_0)$ reflect the change of the kernel including $e^{2\omega t_0}$.
By comparing \eqref{equ:SpectralRepresentation} and \eqref{equ:PolynomialApproximationStart} we obtain
\begin{align}
  \bar{X}_{\sigma}^{(l)}(\bm{q}^2) \simeq \sum_{k=0}^{N} c_{\mu\nu,k}^{(l)}(\bm{q};\sigma; t_0) C^{\mu\nu}(\bm{q}, k+ 2 t_0) \, .
  \label{equ:ApproximationIntegral}
\end{align}

The expression \eqref{equ:ApproximationIntegral} gives a relation between $C^{\mu\nu}(\bm{q}, t)$, which is a quantity calculated on the lattice, and $\bar{X}_{\sigma}^{(l)}(\bm{q}^2)$ appearing in the smeared total decay rate. It is an approximation with the  truncation at a finite order $N$. With this setup, the order $N$ of the polynomial corresponds to the maximal Euclidean time separation that we take between the inserted currents in the four-point function \eqref{equ:FourPointDefinition}. For any given value of $\sigma$, the only remaining task to compute the decay rate is to perform the phase-space or $\bm{q}^2$ integration in \eqref{equ:TotalRateInclusive}.

\section{Numerical setup}
\label{sec:Numerical}

We employ gauge ensembles generated by the JLQCD collaboration including $2+1$ flavors of dynamical quarks. Our simulations are performed on a $48^3 \times 96$ lattice, corresponding to a lattice spacing of $a \simeq \SI{0.055}{\femto\meter}$ or a lattice cutoff of $a^{-1} \sim \SI{3.610(9)}{\giga\electronvolt}$. These parameters are determined through the Yang-Mills gradient flow \cite{BMW:2012hcm}. To achieve better control over the discretization errors we employ the tree-level improved Symanzik gauge action and stout smearing \cite{Morningstar:2003gk} to the gauge field when coupled to fermions. We use the M\"obius domain-wall action \cite{Brower:2012vk, Tomii:2016xiv} for both heavy and light quarks. For additional information on the practical implementation and formulation of the quark action in five dimensions we refer to \cite{Boyle:2015vda, Colquhoun:2022atw, Brower:2012vk, Aoki:2023qpa}. The choice of light-quark masses used in this work corresponds to a pion mass of $M_\pi \simeq \SI{300}{\mega\electronvolt}$. Our ensemble satisfies the condition $M_\pi L > 4$, where $L$ is the spatial extent of the lattice. This condition ensures a sufficient suppression of finite-volume effects to a regime at the few-percent level for meson masses and form factors.
Due to the finite fifth dimension $L_5$, M\"obius domain-wall fermions have no exact chiral symmetry. A measure of the violation of chiral symmetry is the residual quark mass, which for the lattice used in this work is below $\SI{0.2}{\mega\electronvolt}$, much smaller than the physical masses of the up and down quarks.
We use the renormalization constant $Z_V$ from \cite{Tomii:2016xiv, Nakayama:2016atf} determined for the analysis of short-distance current correlator of light quarks. For the ensemble considered in this work the numerical value is $0.9636(58)$, where statistical and systematic errors are added in quadrature.
For our simulations, we average over 50 statistically independent gauge configurations, perform the measurements for each configuration with 8 evenly distributed choices of the time source. We introduce four different momenta in the four-point function \eqref{equ:FourPointDefinition}. In terms of of $\bm{q} = (2\pi/L) \bm{n}$, we take $\bm{n} = (0,0,0)$, $(0,0,1)$, $(0,1,1)$, and $(1,1,1)$. All correlation functions analyzed in this work have been computed using the Grid \cite{BoyleGrid, Boyle:2015tjk, Yamaguchi:2022feu} and Hadrons \cite{PortelliHadrons} software packages. For most of the fits we have employed python packages lsqfit \cite{LepageLSQ, Lepage:2001ym} and corrfitter \cite{LepageCORR}. The gvar class \cite{LepageGVAR} is used to capture statistical correlations between data points as well as correlations between data points and priors, allowing for a straightforward treatment of Gaussian-distributed random variables.

For the two-point correlation functions of the $D_s$ meson that appear in the denominator of \eqref{equ:RatioFourAndTwoFunction}, we consider different combinations of the smearing for the zero-momentum correlator $C_{D_s}^{LL}(t,t_{\text{src}})$, $C_{D_s}^{LS}(t,t_{\text{src}})$, $C_{D_s}^{SL}(t,t_{\text{src}})$ and $C_{D_s}^{SS}(t,t_{\text{src}})$. The superscripts $L$ and $S$ specify local or smeared operators $\mathcal{O}^{X}_{D_s} = \bar{c}^X \gamma_5 s$, with $X = L$ or $S$. The fit function for the two-point correlator is 
\begin{align}
	C_{P}(t) = \sum_{n} a_{n,P} b^{*}_{n,P} \left( e^{-E_{n,P} t} + e^{-E_{n,P} (T-t)}\right) \, ,
	\label{equ:TwoPointFit}
\end{align}
where the subscript $n,P$ corresponds to the $n$-th state of pseudoscalar $P$, so that $n=0$ corresponds to the ground state, and $T$ is the extent of the lattice in the temporal direction. In the case where the interpolating operators at source and sink are both smeared or local, the amplitudes $a_{n,P}$ and $b^{*}_{n,P}$ are identical. To extract the amplitude $\braket{0|\mathcal{O}_{D_s}^S|D_s}$ in the denominator of \eqref{equ:RatioFourAndTwoFunction} we perform a combined fit of all correlators.

For the computation of the ratio between two- and four-point correlation functions \eqref{equ:RatioFourAndTwoFunction}, we take a source-sink separation of $T = t_{\text{snk}} - t_{\text{src}} = 42$ in lattice units. We have the freedom to keep either $t_1$ or $t_2$ constant while varying its counterpart (see Fig.~\ref{fig:QuarkFlow4Pt}). We compute both cases and average over them. The $t_1$ and $t_2$ are chosen such that a sufficient ground-state saturation is achieved for $D_s$.
For one case, we keep the current $J_\mu^\dagger$ fixed at $t_2 = t_{\text{src}} + 26$, so that the time dependence in \eqref{equ:LatticeCorrelatorRatio} is chosen to be in the range $0 \leq t \leq 26$, where $t = t_2 - t_1$. This choice of time separation requires a sufficient ground-state saturation at $t' = t_{\text{snk}} - t_2 = 16$. The same holds when the current $J_\nu$ at time slice $t_1 = t_{\text{src}} + 16$ is kept fixed. 

Since currents have a $(V-A)$ structure, we take all possible combinations of $J_\nu(x_1)$ and $J_\mu^\dagger(x_2)$, namely $V_\mu^\dagger V_\nu$, $V_\mu^\dagger A_\nu$, $A_\mu^\dagger V_\nu$ and $A_\mu^\dagger A_\nu$. But, in the limit of massless leptons, {\it i.e.} $m_\ell = 0$, the combinations of $V_\mu^\dagger A_\nu$ and $A_\mu^\dagger V_\nu$ do not contribute to the total decay rate. These current combinations are related to the structure function $W_3$ through the relation $W_{ij}^{AV} + W_{ij}^{VA} = i\epsilon_{ij0k}q^k W_3$, which does not contribute in the case of massless leptons due to parity.

\section{Structure of the Compton Amplitude}
\label{sec:Amplitudes}

The hadronic final states that saturate the forward Compton matrix element \eqref{equ:FourPointDefinition} depend on the current, their polarization and the injected momentum. We discuss the decomposition of $\bar{X}(\bm{q}^2)$ into different final states including the possible ground state for each channel. In Section~\ref{sec:GSLimit}, we first focus on the $S$-wave final states and their form factors, {\it i.e.} the $s\bar{s}$ pseudoscalar $\eta_s$ and vector $\phi$ mesons. In Section~\ref{sec:GroundStateEstimate}, we show the effective-mass plots for each channel. We further scrutinize the interferences from excited states by extending the decomposition of the Compton amplitude to include contributions from the $P$-wave final states. Finally, in Section~\ref{sec:InclusiveGS}, we apply the previous discussions to extract the ground-state contribution from the inclusive data set, to determine the form factors of the exclusive decay channel. This serves as an important cross-check of the inclusive analysis. 

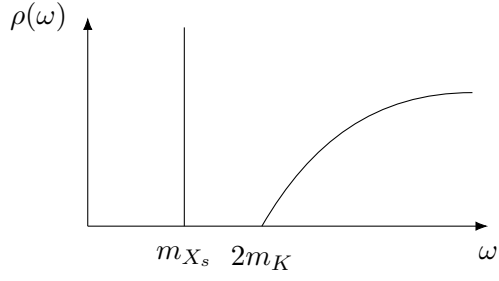
\begin{figure}[tb!]
	\centering
	\begin{tikzpicture}
	\node (A) {};
	\node[above=2.5cm of A,label=180:{\(\rho(\omega)\)}] (B) {};
	\node[right=5cm of A,label=270:{\(\omega\)}] (C) {};
	
	\node[right=1cm of A,label=270:{\(m_{X_s}\)}] (D) {};
	\node[above=2.5cm of D] (E) {};
	
	\node[right=0.75cm of D,label=270:{\(2m_K\)}] (F) {};
	\node[above right=1.5cm and 2.5cm of F] (G) {};
	
	\draw[-{Latex}] (A.center) -- (B.center);
	\draw[-{Latex}] (A.center) -- (C.center);
	\draw (D.center) -- (E);
	\draw[out=60,in=180] (F.center) to (G.center);
	\end{tikzpicture}
	\caption{Spectral density $\rho(\omega)$ including the ground state $\omega = m_X$.}
	\label{fig:SpecD}
\end{figure}

The ground-state contribution can also be used in the analysis to achieve better control over systematic effects. We recall our discussions in Section~\ref{sec:EuclideanInclusive}, especially Eq.~\eqref{equ:SigmoidFunction}, where the sharp cut of the kernel function is replaced by a sigmoid function when applying the polynomial expansion of the kernel. This results in critical situations, especially for higher recoil momenta $\bm{q}$, where the inclusive rate is nearly dominated by the ground state due to the increasingly restricted phase space. In this kinematical setup, a very small smearing width is required to properly treat its contribution to $\bar{X}(\bm{q}^2)$. This situation can be circumvented in the analysis by assuming a decomposition of the spectral density following the sketch in Fig.~\ref{fig:SpecD} as
\begin{align}
    \rho(\omega) = \rho_0 \delta(\omega - m_X) + \rho_{\text{Ex}}(\omega) \, ,
    \label{equ:SpectralDeconstruction}
\end{align}
where the first term on the r.h.s. is the ground-state contribution and the second term is the spectral function containing all excited-state contributions. Then, assuming that the ground-state contribution can be extracted precisely from the lattice data, it is possible to treat its contribution to $\bar{X}(\bm{q}^2)$ as exact under the energy integral, and perform the inclusive analysis only on the remaining excited-state contributions, which should only be subdominant, and hence reduce the systematic effects compared to an analysis using the full data set.

The fit results discussed in this section will be used for this purpose of extracting the ground-state contributions and we will show a comparison between both approaches in Section~\ref{sec:SysErrApproximation}.

\subsection{Ground-state contributions ($S$-wave states)}
\label{sec:GSLimit}

First, we decompose $X(\bm{q}^2, \omega)$ into longitudinal and transverse components to identify contributions of different final states. It can also be used for comparison with OPE results \cite{Gambino:2022dvu}. 

Introducing a basis in three-dimensional space, $e_\parallel$, $e_1$ and $e_2$, such that
\begin{align}
  e_\parallel = \frac{\bm{q}}{\sqrt{\bm{q}^2}} \, , \quad e_i e_\parallel = 0 \, , \quad e_i e_j = \delta_{ij} \, , \quad
   i = \{1, 2\} \, ,
   \label{equ:3DSpaceLongTrans}
\end{align}
we construct the longitudinal ($\parallel$) and transverse ($\perp$) projectors
\begin{align}
  \Pi_\parallel^{ij} = \frac{q_iq_j}{\bm{q}^2} \, , \quad \Pi_\perp^{ij} = \sum_{a=1}^{2} e_a^i e_a^j = \delta^{ij} - \frac{q_iq_j}{\bm{q}^2} \, , \quad \delta^{ij} = \Pi_\parallel^{ij} + \Pi_\perp^{ij} \, .
  \label{equ:ProjectorsLongTran}
\end{align}
Using these projectors, $X(\bm{q}^2, \omega)$ in Eqs.~\eqref{equ:XDecomposition}--\eqref{equ:XComponents} is given by
\begin{align}
  \begin{split}
    X(\bm{q}^2, \omega) &= \bm{q}^2 W_{00} - q_0 \sum_i q_i (W_{i0} + W_{0i}) + (q_0^2 - \bm{q}^2) \sum_{i,j} \delta^{ij} W_{ij} + \bm{q}^2 \sum_{i,j} \frac{q_i W_{ij} q_j}{\bm{q}^2} \\
    &= \bm{q}^2 W_{00} - q_0 \sum_i q_i (W_{i0} + W_{0i}) + (q_0^2 - \bm{q}^2) \sum_{i,j} \Pi_\parallel^{ij} W_{ij} + \bm{q}^2 \sum_{i,j}  \Pi_\parallel^{ij} W_{ij} \\
    &+ (q_0^2 - \bm{q}^2) \sum_{i,j} \Pi_\perp^{ij} W_{ij} \, .
  \end{split}
\end{align}
We decompose it as $X(\bm{q}^2,\omega) = X_\parallel(\bm{q}^2,\omega)+X_\perp(\bm{q}^2,\omega)$,  defining the longitudinal and transverse contributions, $X_\parallel(\bm{q}^2, \omega)$ and $X_\perp(\bm{q}^2, \omega)$, respectively, as
\begin{align}
  X_\parallel(\bm{q}^2, \omega) &= \bm{q}^2 W_{00} - q_0 \sum_i q_i (W_{i0} + W_{0i}) + \frac{q_0^2}{\bm{q}^2} \sum_{i,j} q_i W_{ij} q_j \, , \\
  \begin{split}
    X_\perp(\bm{q}^2, \omega) &= (q_0^2 - \bm{q}^2) \sum_{i,j} \left[ \delta^{ij} - \frac{q_iq_j}{\bm{q}^2}\right] W_{ij}  \\
    &= \left(1 - \frac{q_0^2}{\bm{q}^2}\right) \left[\sum_{i} (q_i^2 - \bm{q}^2) W_{ii} + \sum_{i \neq j} q_i W_{ij} q_j \right] \, .
  \end{split}
\end{align}
These can be expressed in terms of the structure functions as
\begin{align}
  X_\parallel(\bm{q}^2, \omega) &= q^2 W_1 + \bm{q}^2 W_2 \label{equ:ParallelDecomposition} \, , \\
  X_\perp(\bm{q}^2, \omega) &= 2q^2 W_1 \label{equ:PerpDecomposition} \, .
\end{align}

Here, we consider the case where only the lowest-lying energy state contributes to the inclusive rate, and relate the total decay rate to the corresponding form factors. The hadronic tensor can be written as
\begin{align}
  W^{\mu\nu} \rightarrow \delta(\omega- E_{\text{GS}}) \frac{1}{4m_{D_s}E_{H_{\text{GS}}}} \braket{D_s|J_\mu^\dagger|H_{\text{GS}}(\bm{p}_{H_{\text{GS}}})}\braket{H_{\text{GS}}(\bm{p}_{H_{\text{GS}}})|J_\nu|D_s} \, ,
  \label{equ:HadronicTensorGs}
\end{align}
where $J_\mu = V_\mu$ or $A_\mu$, and $H_{\text{GS}}$ is the lightest hadronic final state for the corresponding channel. 

We use a parameterization motivated by the heavy-quark effective theory (HQET) for simplicity, although the final states considered in this work have a quantum number $s\bar{s}$. They can be rewritten using the standard form factors if necessary. They are
\begin{align}
  \frac{\braket{\eta_s (v')|V^{\mu}|D_s(v)}}{\sqrt{m_{D_s} m_{\eta_s}}} &= h_+(w) (v + v')^\mu + h_-(w) (v-v')^\mu \label{equ:PSFF} \, , \\
  \frac{\braket{\phi (v')|V^{\mu}|D_s(v)}}{\sqrt{m_{D_s} m_{\phi}}} &= h_V (w) \epsilon^{\mu\nu\lambda\sigma} v_\nu v^{'}_{\lambda} \varepsilon^{*}_\sigma \label{equ:VecFF} \, , \\
  \frac{\braket{\phi (v')|A^{\mu}|D_s(v)}}{\sqrt{m_{D_s} m_{\phi}}} &= i h_{A_1}(w) (1+w) \varepsilon^{*\mu} -i \left[ h_{A_2}(w) v^\mu + h_{A_3}(w) v^{'\mu} \right] (\varepsilon^{*} \cdot v) \label{equ:AxVecFF} \, ,
\end{align}
where four-velocities $v = p/m_{D_s}$ and $v' = p'/m_{\eta_s,\phi}$ are introduced, as well as $w=v\cdot v'$.
The HQET form factors are $h_i(w)$ with $i = +$, $-$, $V$, $A_1$, $A_2$, $A_3$. The relevant ground states are $\eta_s$ and $\phi$ depending on the channel, and $\varepsilon^{*}$ denotes the polarization vector of the $\phi$ meson.

Taking the momentum $\bm{q}$ in the $z$-direction, {\it i.e.} $\bm{q}=(0,0,q_z)$, we define the longitudinal and transverse components as $p_\parallel = p_z$ and $p_\perp=p_x=p_y$. Non-vanishing components of the hadronic tensor \eqref{equ:XComponents} are $W_\perp = (W_{11} + W_{22})/2$, $W_\parallel = W_{33}$ and $W_{0\parallel, \parallel0} = W_{03, 30}$, with which we can write
\begin{align}
  X^{(0)}(\bm{q}^2, \omega) &= \bm{q}^2 \left(W_{00} - 2W_\perp\right) \, , \\ 
  X^{(1)}(\bm{q}^2, \omega) &= - q_0 q_\parallel (W_{\parallel0} + W_{0\parallel}) \, , \\
  X^{(2)}(\bm{q}^2, \omega) &= q_0^2 (W_\parallel - 2W_\perp) \, .
\end{align}
The longitudinal and transverse contributions \eqref{equ:ParallelDecomposition} and \eqref{equ:PerpDecomposition} read
\begin{align}
  X_\parallel(\bm{q}^2, \omega) &= \bm{q}^2 W_{00} - q_0 q_\parallel (W_{\parallel0} + W_{0\parallel}) + q_0^2 W_\parallel \label{equ:ParallelDecomposition2} \, , \\
  X_\perp(\bm{q}^2, \omega) &= 2 (q_0^2 - \bm{q}^2) W_\perp \label{equ:PerpDecomposition2} \, .
\end{align}

Combining with the decomposition \eqref{equ:Decomposition} and inserting the definitions \eqref{equ:PSFF}--\eqref{equ:AxVecFF} into \eqref{equ:HadronicTensorGs}, we obtain expressions for the longitudinal and transverse components for the vector ($VV$) or axial-vector ($AA$) insertions of the hadronic tensor
\begin{align}
  \bar{X}^{VV}_{\parallel}(\bm{q}^2) &= \frac{\bm{q}^2}{4m_{\eta_s} m_{D_s}} \left[h_+(w) (m_{D_s} + m_{\eta_s}) - h_-(w) (m_{D_s} - m_{\eta_s})\right]^2 \, , \label{equ:XVVPar} \\
  \bar{X}^{VV}_{\perp}(\bm{q}^2) &= \frac{\bm{q}^2}{2 m_{\phi} E_{\phi}} \left[(m_{D_s} - m_{\phi})^2 - 2m_{D_s}(E_{\phi} - m_{\phi})\right] h_V(w)^2 \, ,\label{equ:XVVPerp}\\
  \begin{split}
  \bar{X}^{AA}_{\parallel}(\bm{q}^2) &= \frac{1}{4m_{\phi} E_{\phi}} \bigg[E_\phi(m_{D_s} - E_{\phi})(1+w)h_{A_1}(w)  \\
  &+ \bm{q}^2\left(h_{A_1}(w)(1+w) - h_{A_2}(w) - \frac{m_{D_s}}{m_{\phi}}h_{A_3}(w)\right)\bigg]^2 \, , 
  \end{split}
  \label{equ:XAAPar} \\
  \bar{X}^{AA}_{\perp}(\bm{q}^2) &= \left[(m_{D_s} - m_{\phi})^2 - 2m_{D_s}(E_{\phi} - m_{\phi})\right] \frac{(1+w)^2}{2w} h_{A_1}(w)^2\label{equ:XAAPerp} \, .
\end{align}
For $\bar{X}^{VV}_{\parallel}$ this expression can be simplified to
\begin{align}
  \bar{X}^{VV}_{\parallel}(\bm{q}^2) = \frac{m_{D_s}}{E_{\eta_s}} \bm{q}^2 |f_{+}(\bm{q}^2)|^2 \, ,
  \label{equ:GSXVVPar}
\end{align}
when we replace the HQET definition of the form factors with the more conventional definition $f_{+}(\bm{q}^2)$ defined through
\begin{align}
  \braket{\eta_s(p')|V^{\mu}|D_s(p)} = f_+(q^2) (p + p')^{\mu} + f_-(q^2) (p-p')^{\mu} \, .
  \label{eq:f+f-}
\end{align}
Below, we will compute the ground-state contribution to $\bar{X}^{VV}_{\parallel}(\bm{q}^2)$ directly from 3-point functions but also from 4-point functions. The consistency of results will serve as a cross-checl of our method.

\subsection{Four-point functions including $P$-wave states}
\label{sec:GroundStateEstimate}


Some excited states, the $P$-wave states in particular, can also be identified, although they are unstable for physical light quarks. Their form factors are mostly unknown. Nevertheless, we summarize the corresponding expressions in order to understand the possible contributions in each channel.
The relevant processes are the $D_s\to f_0$ and $D_s\to f_1$ decays, where the final-state mesons have $J^P = 0^+, 1^+$, respectively. The matrix elements are then parameterized as \cite{Leibovich:1997em}
\begin{align}
    \frac{\braket{f_0(v')|A^{\mu}|D_s(v)}}{\sqrt{m_{f_0} m_{D_s}}} &= g_+(w) (v^\mu + v'^\mu) + g_-(w) (v^\mu - v'^\mu) \, , \\
    \frac{\braket{f_1(v',\epsilon)|V^{\mu}|D_s(v)}}{\sqrt{m_{f_1} m_{D_s}}} &= g_{V_1}(w)\epsilon^{*\mu} + \left(g_{V_2}(w) v^{\mu} + g_{V_3}(w) v'^{\mu}\right) \left(\epsilon^{*} \cdot v\right) \, , \\
    \frac{\braket{f_1(v',\epsilon)|A^{\mu}|D_s(v)}}{\sqrt{m_{f_1} m_{D_s}}} &= i g_A(w) \varepsilon^{\mu\alpha\beta\gamma} \epsilon^*_\alpha v_\beta v'_\gamma \, .
\end{align}

The relevant correlators are then written in terms of the low-lying $S$-wave and $P$-wave contributions as follows:
\newcommand{\prefactor}[1]{\frac{e^{-E_{#1}t}}{4E_{#1} m_{#1}}}
{\allowdisplaybreaks
\begin{align}
        C^{VV}_{00}(\bm{q},t)
        =& \prefactor{\eta_s} \left[ h_+(w) \left( E_{\eta_s} + m_{\eta_s} \right) - h_-(w) \left( E_{\eta_s} - m_{\eta_s} \right) \right]^2 \nonumber\\
        +& \prefactor{{f_1}} \bm{q}^2 \left[ g_{V_1}(w) + g_{V_2}(w) + \frac{E_{f_1}}{m_{f_1}} g_{V_3}(w) \right]^2 + \ldots \, , \label{equ:C_V0V0_heavy_meson_FFs_after}\\     
        C^{VV}_{\parallel\parallel}(\bm{q},t)
        =& \prefactor{{\eta_s}} \bm{q}^2 \left[ h_+(w) - h_-(w) \right]^2 \nonumber\\
        +& \prefactor{{f_1}} \left[ g_{V_1}(w) E_{f_1} + g_{V_3}(w) \frac{\bm{q}^2}{m_{f_1}} \right]^2 + \ldots \, , \\
        C^{VV}_{0\parallel}(\bm{q},t) =& C^{VV}_{\parallel 0}(\bm{q},t) \nonumber\\
        =& \prefactor{{\eta_s}} \sqrt{\bm{q}^2} \bigg[ \left[ h_+(w) \left( E_{\eta_s} + m_{\eta_s} \right) - h_-(w) \left( E_{\eta_s} - m_{\eta_s} \right) \right] \left[ h_+(w) - h_-(w) \right] \bigg] \nonumber\\
        +& \prefactor{{f_1}} \sqrt{\bm{q}^2} \Bigg[ \left[ g_{V_1}(w) + g_{V_2}(w) + \frac{E_{f_1}}{m_{f_1}} g_{V_3}(w) \right] \left[ g_{V_1}(w) E_{f_1} + g_{V_3}(w) \frac{\bm{q}^2}{m_{f_1}} \right] \Bigg]  \nonumber\\
        +& \ldots \, , \label{equ:C_VParV0_heavy_meson_FFs_after} \\
        C^{VV}_{\perp \perp}(\bm{q},t) =& \prefactor{\phi} \bm{q}^2 \left( h_V(w) \right)^2 + \prefactor{{f_1}} \left( g_{V_1}(w) m_{f_1} \right)^2 + \ldots \, , \label{equ:C_VPerpVPerp_heavy_meson_FFs_after}
\end{align}}
{\allowdisplaybreaks
\begin{align}
        C^{AA}_{00}(\bm{q},t)
        =& \prefactor{\phi} \bm{q}^2 \left[ \left(\frac{E_{\phi} + m_{\phi}}{m_{\phi}}\right)h_{A_1}(w) - h_{A_2}(w) - \frac{E_{\phi}}{m_{\phi}} h_{A_3}(w) \right]^2 \nonumber\\
        +& \prefactor{{f_0}} \left[ g_+(w) \left( E_{f_0} + m_{f_0} \right) - g_-(w) \left( E_{f_0} - m_{f_0} \right) \right]^2 + \ldots \, , \label{equ:C_A0A0} \\
        C^{AA}_{\parallel\parallel}(\bm{q},t)
        =& \prefactor{{\phi}} \left[ h_{A_1}(w) \frac{\left(E_{\phi} + m_{\phi}\right)E_{\phi}}{m_{\phi}} - h_{A_3}(w) \frac{\bm{q}^2}{m_{\phi}} \right]^2 \nonumber\\
        +& \prefactor{{f_0}} \bm{q}^2 \left[ g_+(w) - g_-(w) \right]^2 + \ldots \, , \\
        C^{AA}_{0\parallel}(\bm{q},t) =& C^{AA}_{\parallel 0}(\bm{q},t) \nonumber\\
        =& \prefactor{{\phi}} \sqrt{\bm{q}^2} \Bigg[ \left[ \left(\frac{E_{\phi} + m_{\phi}}{m_{\phi}}\right)h_{A_1}(w) - h_{A_2}(w) - \frac{E_{\phi}}{m_{\phi}} h_{A_3}(w) \right] \nonumber \\
        &\left[ h_{A_1}(w) \frac{\left(E_{\phi} + m_{\phi}\right)E_{\phi}}{m_{\phi}} - h_{A_3}(w) \frac{\bm{q}^2}{m_{\phi}} \right] \Bigg] \nonumber\\
        +& \prefactor{{f_0}} \sqrt{\bm{q}^2} \Bigg[ \left[ g_+(w) \left( E_{f_0} + m_{f_0} \right) - g_-(w) \left( E_{f_0} - m_{f_0} \right) \right] \left[ g_+(w) - g_-(w) \right] \Bigg] \nonumber \\
        +& \ldots \, , \label{eq:C_A0Apara_heavy_meson_FFs_after} \\
        C^{AA}_{\perp \perp }(\bm{q},t) 
        =& \prefactor{{\phi}} \left[ h_{A_1}(w) \left(E_{\phi} + m_{\phi}\right) \right]^2 + \prefactor{f_1} \bm{q}^2 \left( g_A(w) \right)^2 + \ldots \, . \label{equ:C_APerpAPerp_heavy_meson_FFs_after}
\end{align}}
We see that the correlation functions are in general a mixture of $S$-wave and $P$-wave state contributions. 

At vanishing recoil momentum $\bm{q}=\bm{0}$, the correlators $C^{VV}_{00}(\bm{q},t)$ and $C^{AA}_{\parallel\parallel}(\bm{q},t)$ are solely described by the $S$-wave contributions. This also applies to $C^{AA}_{\perp\perp}(\bm{q},t)$, as there is no distinction between $\parallel$ and $\perp$ directions at zero-recoil. On the other hand, the parity-partner correlators $C^{VV}_{\parallel\parallel}(\bm{q},t)$ and $C^{AA}_{00}(\bm{q},t)$ only receive $P$-wave contributions in the zero-recoil limit, while the $S$-wave contributions are suppressed by a factor of $\bm{q}^2$. 

\begin{figure}[tbp]
  \centering
    \includegraphics[width=0.6\textwidth]{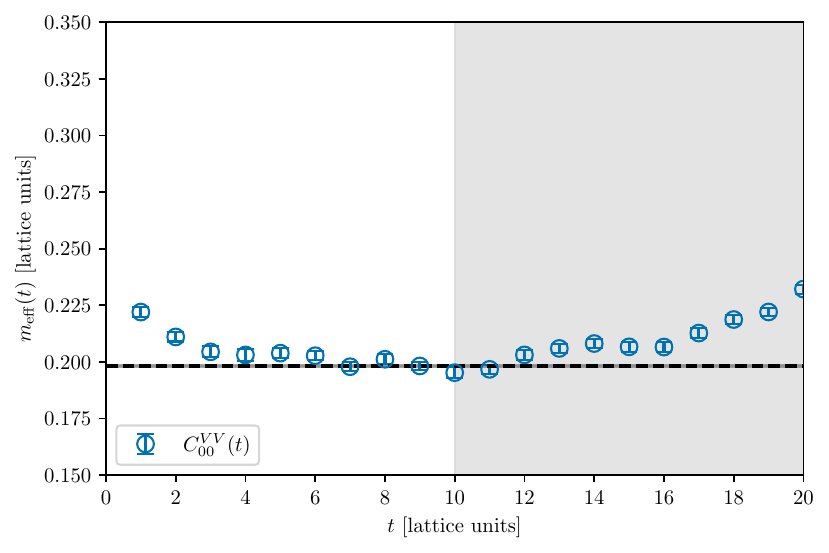}
    \includegraphics[width=0.6\textwidth]{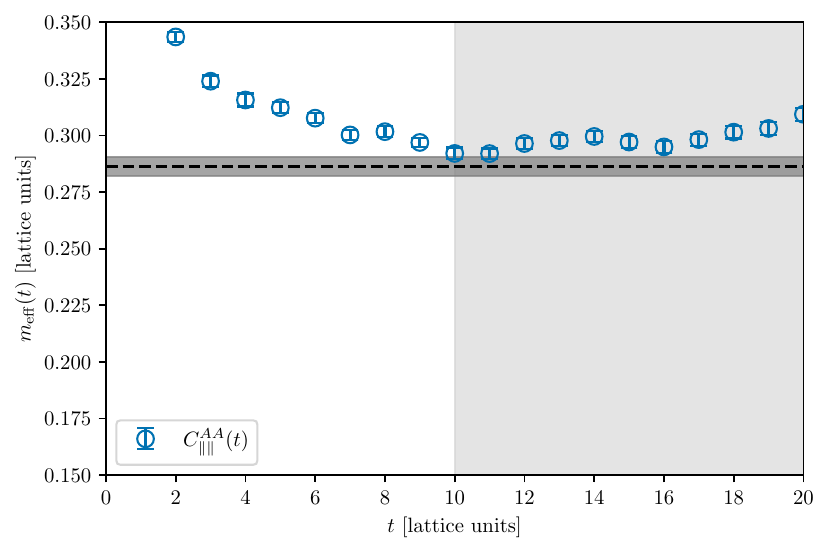}
  \caption{Effective mass for $C_{00}^{VV}(\bm{q},t)$ (top panel) and $C_{\parallel\parallel}^{AA}(\bm{q},t)$ (bottom panel) at zero recoil momentum. The horizontal bands represent a separate determination from two-point functions. (Their values are $m_{\eta_s} = \num{0.19821(30)}$ and $m_{\phi} = \num{0.2864(42)}$, obtained from uncorrelated fits using a single exponential \cite{Fahy:2016qji}.) The plots are shaded for $t_{\text{max}}>10$, as the correlators are affected by the excited state contamination from the initial state $D_s$ meson.}
  \label{fig:SWaveGroundState}
\end{figure}

In Fig.~\ref{fig:SWaveGroundState} we show the effective masses for the correlators $C^{VV}_{00}(\bm{q},t)$ and $C^{AA}_{\parallel\parallel}(\bm{q},t)$ at zero-recoil momentum. They are not contaminated by the $P$-wave states and dominated by the $S$-wave pseudoscalar $\eta_s$ and vector $\phi$ meson, respectively.
We observe that the correlators reach the expected plateau region at $t=$ 7--10, or at 10, for $C^{VV}_{00}(\bm{q},t)$ and $C^{AA}_{\parallel\parallel}(\bm{q},t)$, respectively.
Outside of this region, they rapidly deviate upward. This is due to the fixed source-sink separation $t_{\text{snk}}-t_{\text{src}}=42$ while fixing $t_1-t_{\text{src}}=16$. Too large $t=t_2-t_1$ (the shaded area in the plots) implies that $t_{\text{snk}}-t_2$ becomes too short to ensure ground-state dominance for the $D_s$ meson. To keep $t_{\text{snk}}-t_2\ge 16$, we have to restrict ourselves to $t$ smaller than 10. (See the diagram in Fig.~\ref{fig:QuarkFlow4Pt}.)  

We also find significant excited-state contributions at smaller $t$'s for both $C^{VV}_{00}(\bm{q},t)$ and $C^{AA}_{\parallel\parallel}(\bm{q},t)$. They are expected to be due to the radial excitation or multi-hadron states, which are of interest in the study of the inclusive decay rate.

\begin{figure}[tbp]
  \centering
  \includegraphics[width=0.6\textwidth]{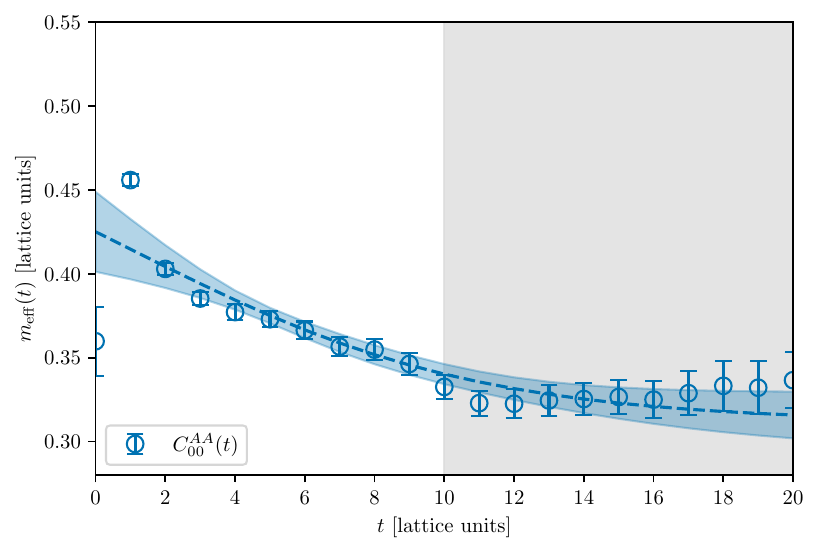}
  \includegraphics[width=0.6\textwidth]{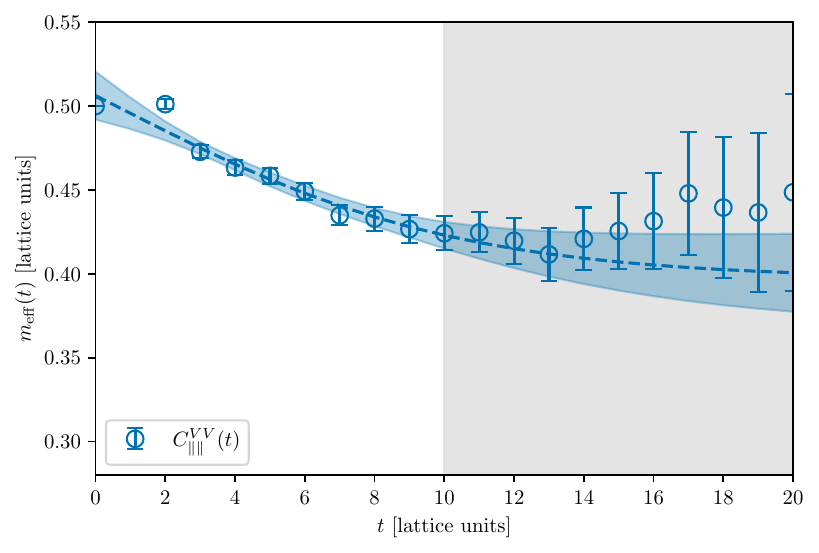}
  \caption{Effective mass for $C_{00}^{AA}(\bm{q},t)$ (top panel) and $C_{\parallel\parallel}^{VV}(\bm{q},t)$ (bottom panel) that are dominated by the $P$-wave states at zero-momentum. Two exponential fits are also shown by the bands. The fit ranges are $t_{\text{min}} = 5$ and $t_{\text{max}} = 17$ or $t_{\text{min}} = 3$ and $t_{\text{max}} = 13$, respectively. The plots are shaded for $t_{\text{max}}>10$, as the correlators are affected by the excited state contamination from the initial state $D_s$ meson. 
  }
  \label{fig:PWaveGroundState}
\end{figure}

The effective mass for the parity-partner correlators $C_{00}^{AA}(\bm{q},t)$ and $C_{\parallel\parallel}^{VV}(\bm{q},t)$ are shown in Fig.~\ref{fig:PWaveGroundState}. Here, the same comment applies about the valid time range to avoid the excited-state contamination for the initial-state $D_s$ meson, {\it i.e.} we should focus on $t<10$. In this case, we are unable to identify a clear plateau to isolate the $P$-wave ground state. Nevertheless, we fit the data with two exponentials and found the lowest energy states 1.13(7)~GeV and 1.43(10)~GeV for the $0^+$ and $1^+$ channels, respectively. The Review of Particle Properties \cite{ParticleDataGroup:2024cfk} lists several $0^+$ states: $f_0(500)$, $f_0(980)$, $f_0(1370)$, etc. They are unstable and are not straightforward to identify experimentally. Moreover, our lattice calculation does not include the disconnected diagrams, so a direct comparison with the experimental data would not be appropriate. For the $1^+$ states, the listed candidates are $f_1(1285)$, $f_1(1420)$. Given the incomplete treatment in the lattice calculation, the rough agreement we found is encouraging.

\begin{figure}[tbp]
  \centering
    \centering
    \includegraphics[width=0.6\textwidth]{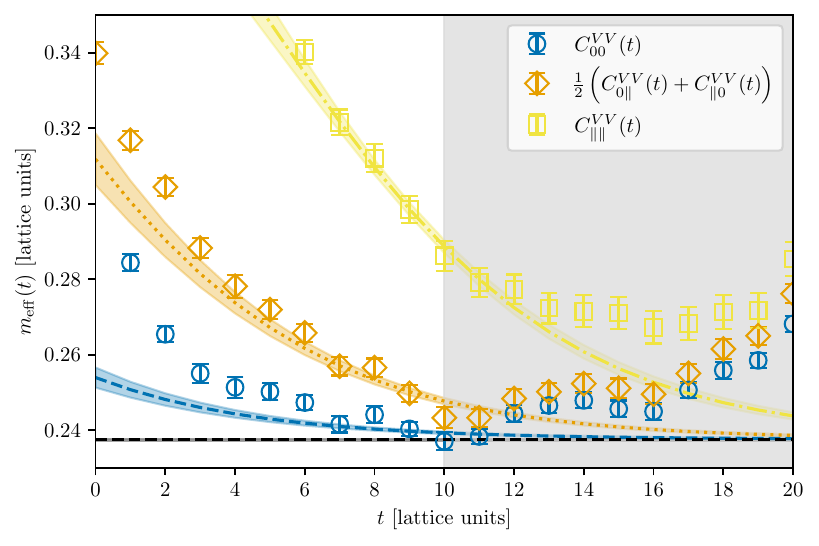}
    \centering
    \includegraphics[width=0.6\textwidth]{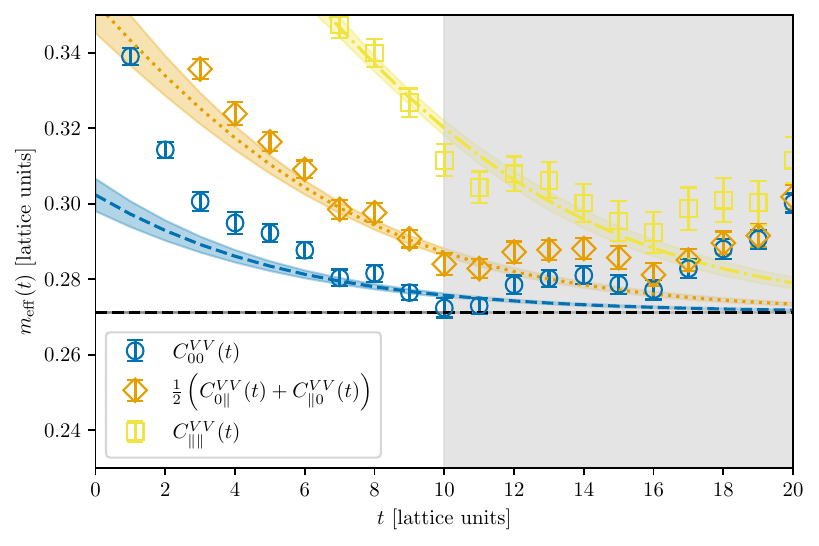}
  
    \centering
    \includegraphics[width=0.6\textwidth]{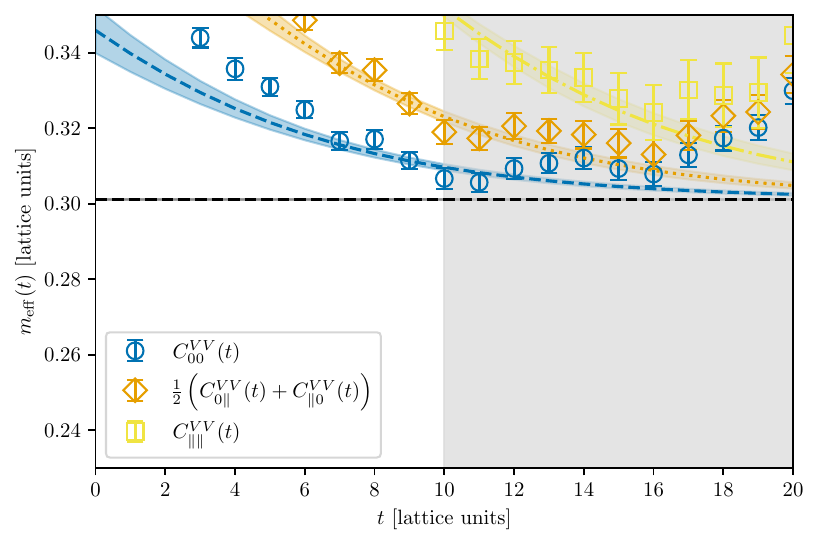}
  \caption{Effective mass for $C^{VV}_{00}(\bm{q},t)$ (circles), $(C_{0\parallel}^{VV}(\bm{q},t)+C_{\parallel 0}^{VV}(\bm{q},t))/2$ (diamond) and $C^{VV}_{\parallel}(\bm{q},t)$ (squares) contributing to $\bar{X}^{VV}_{\parallel}(\bm{q}^2)$ for all non-zero values of $\bm{q}$. A double-exponential fit using all correlators is also shown. The fit range is chosen as $t_{\text{min}} = 7$ and $t_{\text{max}} = 13$. The horizontal lines represent the expected $S$-wave ground state. The recoil momentum is $\bm{q} = (0,0,1)$ (top), $\bm{q} = (0,1,1)$ (middle), $\bm{q} = (1,1,1)$ (bottom).} 
  \label{fig:CorrelatorsVectorPar}
\end{figure}
\begin{figure}[tbp]
  \centering
    \centering
    \includegraphics[width=0.6\textwidth]{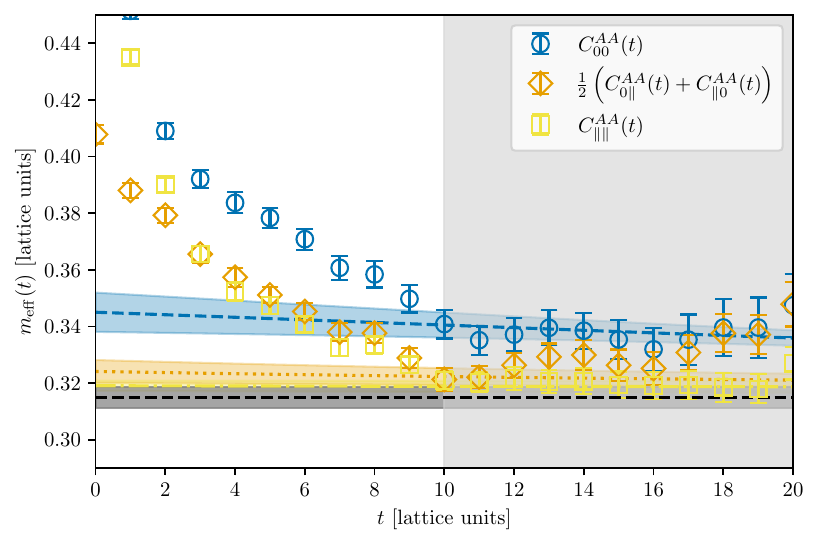}
    \centering
    \includegraphics[width=0.6\textwidth]{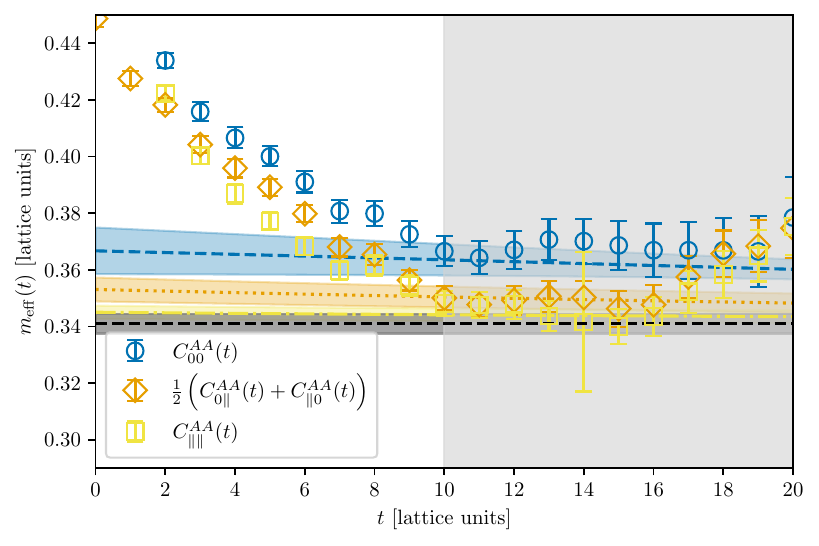}
  
    \centering
    \includegraphics[width=0.6\textwidth]{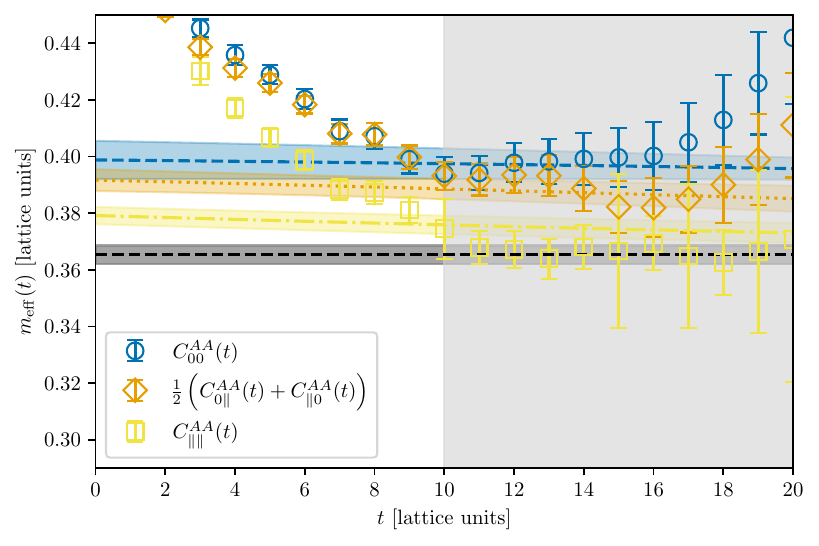}
  \caption{Same as Fig.~\ref{fig:CorrelatorsVectorPar}, but for $C^{AA}_{00}(\bm{q},t)$ (circles), $(C_{0\parallel}^{AA}(\bm{q},t)+C_{\parallel 0}^{AA}(\bm{q},t))/2$ (diamond) and $C^{AA}_{\parallel}(\bm{q},t)$ (squares) contributing to $\bar{X}^{AA}_{\parallel}(\bm{q}^2)$. The fit range is $t_{\text{min}} = 10$ and $t_{\text{max}} = 17$.}
  \label{fig:CorrelatorsAxialVectorPar}
\end{figure}
\begin{figure}[!tbp]
  \centering
    \centering
    \includegraphics[width=0.6\textwidth]{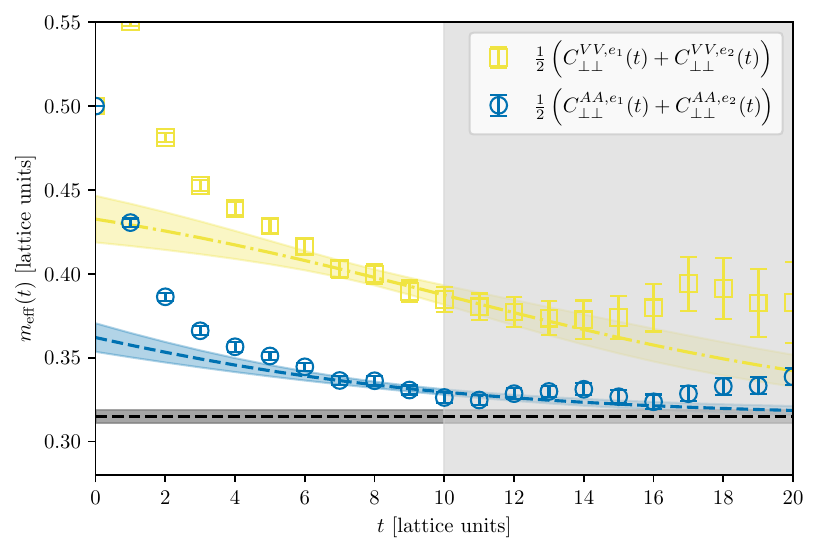}

    \centering
    \includegraphics[width=0.6\textwidth]{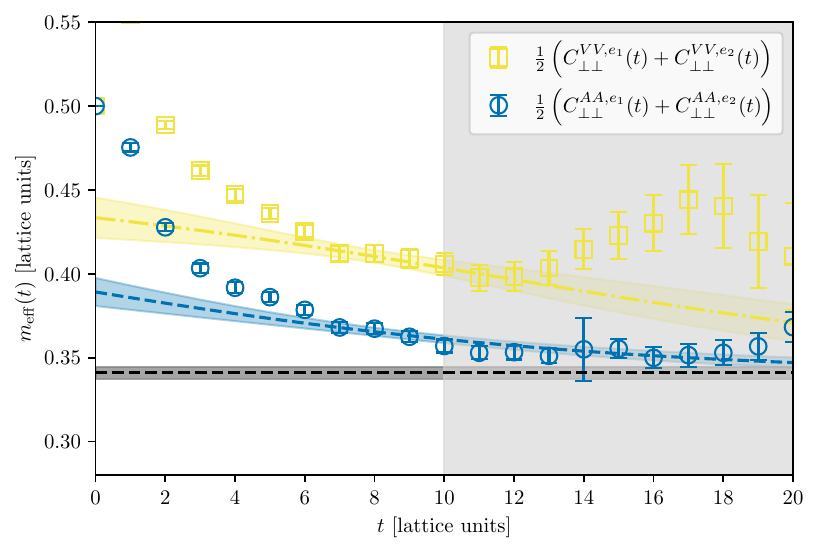}
    \centering
    \includegraphics[width=0.6\textwidth]{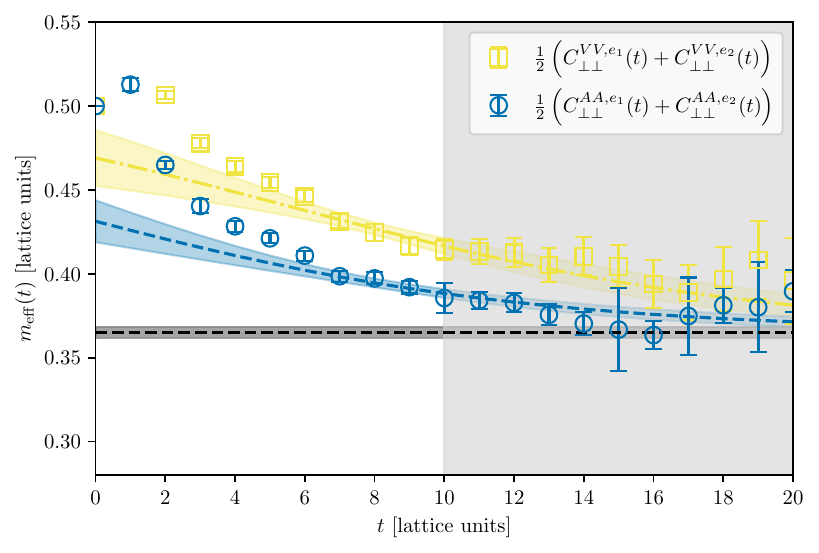}
  \caption{Effective mass for the average over both transverse directions $e_1$ and $e_2$ constituting $C^{VV}_{\perp\perp}(\bm{q},t)$ (squares) and $C^{AA}_{\perp\perp}(\bm{q},t)$ (circles). The recoil momentum is $\bm{q} = (0,0,1)$ (top), $\bm{q} = (0,1,1)$ (middle), $\bm{q} = (1,1,1)$ (bottom). The effective masses for a simultaneous double-exponential fit with fit range $t_{\text{min}} = 7$ and $t_{\text{max}} = 13$ are also shown. The horizontal lines represent the expected $S$-wave ground state.}
  \label{fig:CorrelatorsVectorPerp}
\end{figure}

Figs.~\ref{fig:CorrelatorsVectorPar}--\ref{fig:CorrelatorsVectorPerp} show the effective mass for the correlators involving $V_0$ and $V_\parallel$ (Fig.~\ref{fig:CorrelatorsVectorPar}), $A_0$ and $A_\parallel$ (Fig.~\ref{fig:CorrelatorsAxialVectorPar}), as well as $V_\perp$ and $A_\perp$ (Fig.~\ref{fig:CorrelatorsVectorPerp}), with finite recoil momenta. 
For each of them, we expect contributions from both the $S$-wave and $P$-wave states for non-zero recoil momenta, as Eqs.~\eqref{equ:C_V0V0_heavy_meson_FFs_after}--\eqref{equ:C_APerpAPerp_heavy_meson_FFs_after} suggest. For example, Fig.~\ref{fig:CorrelatorsVectorPar} shows the effective masses of the correlators $C_{00}^{VV}(\bm{q},t)$, $(C_{0\parallel}^{VV}(\bm{q},t)+C_{\parallel 0}^{VV}(\bm{q},t))/2$, $C_{\parallel\parallel}^{VV}(\bm{q},t)$, which share contributions from the same set of intermediate states. We therefore fit them simultaneously, assuming common intermediate-state energies of the $\eta_s$ and $f_1$ mesons. The $\eta_s$ mass is constrained from the corresponding two-point functions and input as a prior for the fit of the four-point functions. They also share the amplitudes, {\it e.g.} the $S$-wave contribution to the three correlators are made of two amplitudes $[h_+(w)(E_{\eta_s}+m_{\eta_s})-h_-(w)(E_{\eta_s}-m_{\eta_s})]$ and $[h_+(w)-h_-(w)]$, corresponding to the matrix elements of $V_0$ and $V_\parallel$ respectively. We implemented this structure for both the $S$-wave and the $P$-wave states to reduce the number of free parameters. The energies of the $P$-wave states are determined by the fit of the four-point corelators. In the plots, we include the result of this combined fit, as well as the $S$-wave ground-state energy. The upward trend beyond $t>10$ is due to the $D_s$ excited-state contamination as discussed before.

We observe some discrepancies with the expected $S$-wave ground-state energies. They are noticable for $C_{\parallel\parallel}^{VV}$ (squares in Fig.~\ref{fig:CorrelatorsVectorPar}), $C_{00}^{AA}$ (circles in Fig.~\ref{fig:CorrelatorsAxialVectorPar}) and $C_{\perp\perp}^{VV}$ (squares in Fig.~\ref{fig:CorrelatorsVectorPerp}). We interpret this as a consequence of the suppressed mixture of the $S$-wave state to the dominant $P$-wave states. In fact, for these correlators, the $S$-wave contributions are suppressed by a factor of $\bm{q}^2$, and larger time separations would be required to achieve ground-state saturation. They are different for $C_{00}^{VV}$ (circles in Fig.~\ref{fig:CorrelatorsVectorPar}), $C_{\parallel\parallel}^{AA}$ (squares in Fig.~\ref{fig:CorrelatorsAxialVectorPar}) and $C_{\perp\perp}^{AA}$ (circles in Fig.~\ref{fig:CorrelatorsVectorPerp}), which are dominated by the $S$-wave contributions.
The observed discrepancies decrease as $\bm{q}$ increases.

For the axial-vector correlators (Fig.~\ref{fig:CorrelatorsAxialVectorPar}) we fit the data beyond $t=10$,  because the lowest-lying state saturation is slow for the zero-recoil correlators (see the effective mass of $C_{\parallel\parallel}^{AA}=C_{\perp\perp}^{AA}$ in  Fig.~\ref{fig:SWaveGroundState}). The fits in Fig.~\ref{fig:CorrelatorsAxialVectorPar} assume two exponentials, each of which represents the $S$-wave and $P$-wave states and do not take account of their radial excitations. (The $S$-wave state, {\it i.e.} the $\phi$ mass, is an input.) It turned out that the $S$-wave and $P$-wave masses are not much separated in this case. There is a risk that the fit is contaminated by the excited states of the initial $D_s$ meson as discussed earlier. The result of the fit is only used for estimating the systematic error in the analysis of the inclusive decays. 


For the transverse components, we perform a simultaneous fit of $C^{VV}_{\perp\perp}$ and $C^{AA}_{\perp\perp}$. Eqs.~\eqref{equ:C_VPerpVPerp_heavy_meson_FFs_after} and \eqref{equ:C_APerpAPerp_heavy_meson_FFs_after} show that they share the same lowest-lying $S$- and $P$-wave states but their dominant contributions are switched. The simultaneous fit allows us to obtain a better constraint on the energies.

\subsection{Decay rate to the ground state}
\label{sec:InclusiveGS}
As a case study, we consider the $VV$ channels, for which the expected ground-state contribution is from $D_s \rightarrow \eta_s \ell \nu_\ell$. The same argument holds for other channels. In terms of the four-point function, the ground-state limit (denoted by $\text{GS}$ in the superscript) translates to restricting the correlators to
\begin{align}
	C_{\mu\nu}^{\text{GS}}(\bm{q},t) = \frac{1}{4m_{D_s} E_{\eta_s}} \braket{D_s|V_\mu^\dagger|\eta_s}\braket{\eta_s|V_\nu|D_s} e^{-E_{\eta_s} t} \, .
	\label{equ:GSOnly}
\end{align}
The standard procedure to determine the relevant matrix elements involves the ratio of two- and three-point functions 
\begin{align}
	R_{D_s \eta_s, \mu} (\bm{q},t) = \sqrt{4m_{D_s} E_{\eta_s}} \sqrt{\frac{C_{D_s\eta_s, \mu}^{SS} (\bm{q}, t_{\text{snk}}, t, t_{\text{src}}) C_{\eta_s D_s, \mu}^{SS} (\bm{q}, t_{\text{snk}}, t, t_{\text{src}}) }{C_{\eta_s}^{SS} (t_{\text{snk}}, t_{\text{src}}) C_{D_s}^{SS} (\bm{q},t_{\text{snk}}, t_{\text{src}})}},
\end{align}
which, for $t \gg t_{\text{src}}$ and $t \ll t_{\text{snk}}$, converges to $|\braket{\eta_s|V_\mu|D_s}|$. $f_+(\bm{q}^2)$ is then extracted using a constant fit, see \cite{Barone:2023tbl}.

\begin{figure}[tb]
	\centering
	\includegraphics[width=0.7\textwidth]{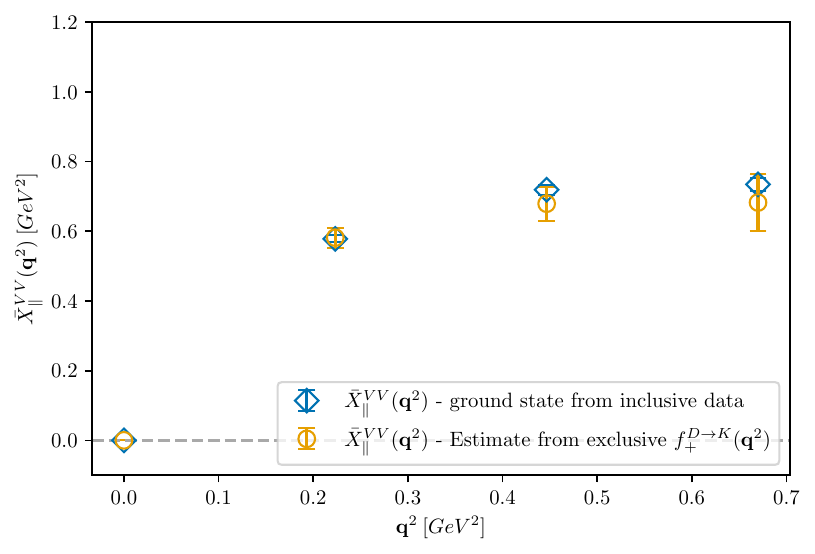}
	\caption{Contribution of the ground state to $\bar{X}_{\parallel}^{VV}(\bm{q}^2)$. The expected contribution from $f_{+}(q^2)$ for the exclusive $D\to K$ decay are calculated using \eqref{equ:GSXVVPar} for the $\bm{q}^2$ values used in the simulations (orange circles). We compare the results to the ground-state contribution extracted from the inclusive data (blue diamonds) as determined from the analysis in Sec.~\ref{sec:GroundStateEstimate}. 
    }
	\label{fig:ContributionsParPerpwGS}
\end{figure}

Here, we use the form factors extracted for a similar process, the $D \rightarrow K \ell \nu_\ell$ decay, available from \cite{Kaneko:2017xgg}, which is obtained on the same lattice ensemble used in this work. We compare our results for $\bar{X}_{\parallel}^{VV}(\bm{q}^2)$ obtained from the ground-state limit of the four-point functions with those constructed from the $D\to K$ decay using \eqref{equ:GSXVVPar}. The results are shown in Fig.~\ref{fig:ContributionsParPerpwGS}. We find a good agreement. Note that the dependence of the $D$ meson semileptonic decay form factors on the spectator quark mass is very small \cite{Kaneko:2017xgg}.

\section{Data analysis for the inclusive decays}
\label{sec:LatticeDataAnalysis}

The discussion in Section~\ref{sec:EuclideanInclusive} reduces the problem of calculating the inclusive decay rate to finding a suitable polynomial approximation for the kernel function $K_{\sigma, \mu\nu}^{(l)}(\bm{q}, \omega; t_0)$. In the literature, two methods have been  proposed to determine the expansion coefficients $c_{\mu\nu, k}^{(l)}(\bm{q};\sigma; t_0)$, namely the Hansen-Lupo-Tantalo (HLT) approach \cite{Hansen:2019idp} based on the Backus-Gilbert method \cite{Backus:1968}, and the Chebyshev-polynomial approach \cite{Gambino:2020crt,Barata:1990rn,Bailas:2020qmv}. This work employs the latter. For a comparison between the two approaches in the context of inclusive decays we refer to \cite{Barone:2023tbl}.


In principle, one can compute $\bar{X}_\sigma^{(l)}(\bm{q}^2)$ as defined in \eqref{equ:ApproximationIntegral}, directly from the lattice data of $C_{\mu\nu}(\bm{q},t)$. Since the kernel $K_{\sigma, \mu\nu}^{(l)}(\bm{q}, \omega; t_0)$ is analytically known, the coefficients $c_{\mu\nu, k}^{(l)}(\bm{q};\sigma; t_0)$ in the Chebyshev approximation can be easily calculated and the reconstruction of $\bar{X}_\sigma^{(l)}(\bm{q}^2)$ from the lattice data of $C_{\mu\nu}(\bm{q},t)$ is straightforward. The limiting factor is the finite order $N$ of the expansion, as it corresponds to the number of available time slices for $C_{\mu\nu}(\bm{q},t)$. Unfortunately, the signal-to-noise ratio deteriorates exponentially with increasing time separation in Euclidean time $t$, which means that the extraction of a meaningful signal becomes out of reach for large $N$. This necessitates some control of the trade-off between statistical noise and systematic error due to the truncation in the expansion. 

Before we proceed, following \cite{Barone:2023tbl}, let us introduce the following notation
\begin{align}
  \begin{split}
   \bar{X}_{\sigma}^{(l)}(\bm{q}^2) &= \int_{\omega_0}^\infty d\omega\, W^{\mu\nu}(\bm{q},\omega) e^{-2\omega t_0}  K_{\sigma, \mu\nu}^{(l)}(\bm{q}, \omega; t_0) \\
   &= \frac{1}{2m_{D_s}} \int_{\omega_0}^\infty  d\omega\, K_{\sigma, \mu\nu}^{(l)}(\bm{q}, \omega; t_0) \braket{D_s|\tilde{J}^{\mu\dagger}(\bm{q},0)e^{-\omega t_0}\delta(\hat{H}-\omega)e^{-\omega t_0}\tilde{J}^{\nu}(\bm{q},0)|D_s} \\ 
   &= \braket{\psi^{\mu}(\bm{q})|K_{\sigma, \mu\nu}^{(l)}(\bm{q}, \hat{H}; t_0)|\psi^{\nu}(\bm{q})}\, ,
  \end{split}
  \label{equ:NewNotationXBar}
\end{align}
using \eqref{equ:HadronicTensor3} and introducing $\ket{\psi^{\nu}(\bm{q})} = e^{-\hat{H} t_0} \tilde{J}^{\nu}(\bm{q}, 0)\ket{D_s} / \sqrt{2m_{D_s}}$. In \eqref{equ:NewNotationXBar} the kernel has been promoted to an operator, $K_{\sigma, \mu\nu}^{(l)}(\bm{q}, \hat{H}; t_0)$. In the following we omit the dependence on $t_0$ for simplicity.

\subsection{Chebyshev approximation of the kernel: formulae}
\label{sec:ChebyshevData}

The Chebyshev polynomials $T_k(x)$ with $x = \exp(-\omega)$ are defined for $-1 \leq x \leq 1$, and provide a nearly optimal approximation of functions, {\it i.e.} the maximal deviation from the target function is minimal in the given range of $x$. We refer to Appendix~\ref{sec:AppChebyshev} for a discussion on the properties of the Chebyshev polynomials useful for this work.
We define shifted Chebyshev polynomials $\tilde{T}_k(x)$ in the interval $\omega_0 \leq \omega \leq \infty$, which are related to the standard definition through $\tilde{T}_k(x) = T_k(h(x))$, where $h(x) = Ax + B$ is a mapping function $h : [e^{-\omega_0},0) \rightarrow [-1,1]$, with $A = -2e^{\omega_0}$ and $B=1$.
With this, we can expand the kernel defined in the previous section as
\begin{align}
  K_{\sigma, \mu\nu}^{(l)}(\bm{q},\omega) = \frac{1}{2} \tilde{c}_{\mu\nu,0}^{(l)}(\bm{q};\sigma) \tilde{T}_0(e^{-\omega}) + \sum_{k=1}^{N} \tilde{c}_{\mu\nu,k}^{(l)}(\bm{q};\sigma) \tilde{T}_k(e^{-\omega}) \, ,
\end{align}
up to the order $N$. By definition, we have $\tilde{T}_0(x) = 1$ and the $k$-th term is given by
\begin{align}
  \tilde{T}_k(x) = \sum_{j=0}^{k} \tilde{t}_j^{(k)} x^j \, ,
\end{align}
where the definition of the coefficients $\tilde{t}_j^{(k)}$ can be found in \eqref{equ:CoefficientExpandedSum}.
By employing the orthogonality properties of the Chebyshev polynomials, the coefficients $\tilde{c}_{\mu\nu,k}^{(l)}(\bm{q};\sigma)$ are given by projections as shown in \eqref{equ:AppChebyshevCoeffs}:
\begin{align}
  \tilde{c}_{\mu\nu,k}^{(l)}(\bm{q};\sigma) = \int_{\omega_0}^{\infty} d\omega\, K_{\sigma,\mu\nu}^{(l)}(\bm{q},\omega) \tilde{T}_{k}(e^{-\omega})\Omega_h(e^{-\omega}) \, .
\end{align}
The definition of the weight function $\Omega_h\left(\exp(-\omega)\right)$ depends on the choice of the map $h$ and is given in App.~\ref{sec:AppChebyshev}. 

The expectation value of the kernel operator can be written as
\begin{align}
  \braket{\psi^\mu(\bm{q})|K_{\sigma,\mu\nu}^{(l)}(\bm{q},\hat{H})|\psi^\nu(\bm{q})} = & 
  \frac{1}{2} \tilde{c}_{\mu\nu,0}^{(l)}(\bm{q};\sigma) \braket{\psi^\mu(\bm{q})|\tilde{T}_0(e^{-\hat{H}})|\psi^\nu(\bm{q})} \nonumber\\
  & + \sum_{k=1}^{N} \tilde{c}_{\mu\nu,k}^{(l)}(\bm{q};\sigma) \braket{\psi^\mu(\bm{q})|\tilde{T}_k(e^{-\hat{H}})|\psi^\nu(\bm{q})} \, .
  \label{equ:ExpectatioNKernelOp}
\end{align}
The shifted Chebyshev polynomials are bounded $|\tilde{T}_k| \leq 1$ by definition, as can be seen through the condition \eqref{equ:GeneralCoefficients}. This property is an important ingredient in the data analysis; more details can be found in Section~\ref{sec:Results}. For convenience, we normalize the terms $\braket{\psi^\mu(\bm{q})|\tilde{T}_k(e^{-\hat{H}})|\psi^\nu(\bm{q})}$ by $\braket{\psi^\mu(\bm{q})|\psi^\nu(\bm{q})} = C^{\mu\nu}(\bm{q},2t_0)$ and introduce the short-hand notation
\begin{align}
\braket{K_\sigma^{(l)}(\bm{q})}_{\mu\nu} \equiv \frac{\braket{\psi^\mu(\bm{q})|K_{\sigma,\mu\nu}^{(l)}(\bm{q},\omega)|\psi^\nu(\bm{q})}}{\braket{\psi^\mu(\bm{q})|\psi^\nu(\bm{q})}} \, , \quad \braket{\tilde{T}_k(\bm{q})}_{\mu\nu} \equiv \frac{\braket{\psi^\mu(\bm{q})|\tilde{T}_k(e^{-\hat{H}})|\psi^\nu(\bm{q})}}{\braket{\psi^\mu(\bm{q})|\psi^\nu(\bm{q})}} \, ,
\end{align}
so that \eqref{equ:ExpectatioNKernelOp} can be rewritten as
\begin{align}
  \braket{K_\sigma^{(l)}(\bm{q})}_{\mu\nu} = \frac{1}{2} \tilde{c}_{\mu\nu,0}^{(l)} (\bm{q};\sigma)\braket{\tilde{T}_0(\bm{q})}_{\mu\nu} + \sum_{k=1}^{N} \tilde{c}_{\mu\nu,k}^{(l)}(\bm{q};\sigma) \braket{\tilde{T}_k(\bm{q})}_{\mu\nu} \label{equ:ChebyshevApproximationKernel} \, .
\end{align}
No summation over $\mu, \nu$ is assumed. The terms $\braket{\tilde{T}_k(\bm{q})}_{\mu\nu}$ are referred to as \textit{Chebyshev matrix elements}, which satisfy the condition $|\braket{\tilde{T}_k(\bm{q})}_{\mu\nu}| \leq 1$. The expression for $\bar{X}_\sigma^{(l)}(\bm{q}^2)$ can be given in terms of the Chebyshev expansion
\begin{align}
  \bar{X}_\sigma^{(l)}(\bm{q}^2) = \sum_{\{\mu,\nu\}} \braket{\psi^\mu(\bm{q})|\psi^\nu(\bm{q})} \braket{K_\sigma^{(l)}(\bm{q})}_{\mu\nu} \, .
\end{align}
The explicit relation for each value of $l$ then reads
\begin{align}
  \bar{X}_\sigma^{(0)}(\bm{q}^2) &= C_{00}(\bm{q},2t_0) \braket{K_\sigma^{(0)}(\bm{q})}_{00} + \sum_{i} C_{ii}(\bm{q},2t_0) \braket{K_\sigma^{(0)}(\bm{q})}_{ii} + \sum_{i \neq j} C_{ij}(\bm{q},2t_0) \braket{K_\sigma^{(0)}(\bm{q})}_{ij} \, , \\
  \bar{X}_\sigma^{(1)}(\bm{q}^2) &= \sum_{i} \left(C_{0i}(\bm{q},2t_0) \braket{K_\sigma^{(1)}(\bm{q})}_{0i} + C_{i0}(\bm{q},2t_0) \braket{K_\sigma^{(1)}(\bm{q})}_{i0}\right)\, , \\
  \bar{X}_\sigma^{(2)}(\bm{q}^2) &= \sum_{i} C_{ii}(\bm{q},2t_0) \braket{K_\sigma^{(2)}(\bm{q})}_{ii}\, .
\end{align}

It is possible to directly relate the lattice data to the Chebyshev matrix elements through the relation
\begin{align}
  \frac{\braket{\psi^\mu(\bm{q})|e^{-\hat{H}t}|\psi^\nu(\bm{q})}}{\braket{\psi^\mu(\bm{q})|\psi^\nu(\bm{q})}} = \frac{C_{\mu\nu}(\bm{q},t+2t_0)}{C_{\mu\nu}(\bm{q},2t_0)} \equiv \bar{C}_{\mu\nu}(\bm{q},t) \, .
\end{align}
By employing the properties of the shifted Chebyshev polynomials highlighted in App.~\ref{sec:AppShiftedChebyshev}, we obtain
\begin{align}
  \begin{split}
    \braket{\tilde{T}_k(\bm{q})}_{\mu\nu} &= \frac{\braket{\psi^\mu(\bm{q})|\tilde{T}_k(e^{-\hat{H}})|\psi^\nu(\bm{q})}}{\braket{\psi^\mu(\bm{q})|\psi^\nu(\bm{q})}} = \sum_{X_s} \frac{\braket{\psi^\mu(\bm{q})|\tilde{T}_k(e^{-\hat{H}})|X_s}\braket{X_s|\psi^\nu(\bm{q})}}{\braket{\psi^\mu(\bm{q})|\psi^\nu(\bm{q})}} \\
    &= \sum_{X_s} \sum_{j=0}^{k} \tilde{t}_j^{(k)} e^{-jE_{X_s}} \frac{\braket{\psi^\mu(\bm{q})|X_s}\braket{X_s|\psi^\nu(\bm{q})}}{\braket{\psi^\mu(\bm{q})|\psi^\nu(\bm{q})}} \\
    &= \sum_{j=0}^{k} \tilde{t}_j^{(k)} \bar{C}_{\mu\nu}(\bm{q},j) \, .
  \end{split}
  \label{equ:LinearSystemChebyshev}
\end{align}
Here, we have inserted $\mathbb{1} = \sum_{X_s} \ket{X_s}\bra{X_s}$ and the $\tilde{t}_{j}^{(k)}$'s are defined in \eqref{equ:CoefficientExpandedSum}.

Combining everything, the full expression of the kernel in terms of the Chebyshev expansion reads
\begin{align}
  \begin{split}
    \braket{K_\sigma^{(l)}(\bm{q})}_{\mu\nu} &= \frac{1}{2} \tilde{c}_{\mu\nu,0}^{(l)} (\bm{q};\sigma)\braket{\tilde{T}_0(\bm{q})}_{\mu\nu} + \sum_{k=1}^{N} \tilde{c}_{\mu\nu,k}^{(l)}(\bm{q};\sigma) \braket{\tilde{T}_k(\bm{q})}_{\mu\nu} \\
    &= \sum_{k=0}^{N} \bar{C}_{\mu\nu} (\bm{q},k) \sum_{j=k}^{N} \tilde{c}_{\mu\nu,j}^{(l)}(\bm{q};\sigma) \left(1 - \frac{1}{2}\delta_{0j} \right) \tilde{t}_{k}^{(j)} \, ,
  \end{split}
  \label{equ:KernelApproxCHebyshevMatrix}
\end{align}
where the expressions for the coefficients $\tilde{c}_{\mu\nu,j}^{(l)}(\bm{q};\sigma)$ and $\tilde{t}_{k}^{(j)}$ are known and can be evaluated analytically. Finally, let us present a short-hand notation
\begin{align}
  \braket{K_\sigma^{(l)}(\bm{q})}_{\mu\nu} = \sum_{k=0}^{N} \bar{c}_{\mu\nu,k}^{(l)}(\bm{q};\sigma) \bar{C}_{\mu\nu}(\bm{q},k) \, ,
  \label{equ:KernelApproxCorrelator}
\end{align}
with
\begin{align}
  \bar{c}_{\mu\nu,k}^{(l)}(\bm{q};\sigma) \equiv \sum_{j=k}^{N} \tilde{c}_{\mu\nu,j}^{(l)}(\bm{q};\sigma) \,\tilde{t}_{k}^{(j)} \left(1 - \frac{1}{2}\delta_{0j} \right) \, .
\end{align}

Let us close this section with some remarks. Although the coefficients $\bar{c}_{\mu\nu,k}^{(l)}(\bm{q};\sigma)$ can be obtained by solving the corresponding analytical expression, the lattice computation of $\bar{C}_{\mu\nu}(k)$ relies on Monte Carlo simulations. This means that $\bar{C}_{\mu\nu}(\bm{q},k)$ is associated with a statistical error, which, in return, means that solving the linear system in \eqref{equ:LinearSystemChebyshev} might result in violations of the bound $|\braket{\tilde{T}_k(\bm{q})}_{\mu\nu}| \leq 1$. One viable option to avoid this problem is to introduce a bound through priors during the fitting of the correlator data. An example is to impose a constraint using a Gaussian prior $\braket{\tilde{\tau}}_{\mu\nu} \sim \mathcal{N}(0,1)$, which is then mapped into a flat prior for the Chebyshev matrix elements in the interval $[-1,1]$ using $f(x) = \text{erf}(x/\sqrt{2})$, so that $\braket{\tilde{T}_k(\bm{q})}_{\mu\nu} = f(\braket{\tilde{\tau}}_{\mu\nu})$.

\subsection{Chebyshev approximation of the kernel: truncation errors}
\label{sec:ChebyshevKernel}

Here, we analyze the potential errors due to the truncation of the Chebyshev approximation, as well as those due to the choice of the lower limit $\omega_0$ and the smearing width $\sigma$.

The kernel function has the generic form (see Secs.~\ref{sec:InclusiveDecay} and~\ref{sec:EuclideanInclusive})
\begin{align}
  K^{(l)}_\sigma(\bm{q}^2,\omega) = e^{2\omega t_0} \sqrt{\bm{q}^2}^{2-l} (m_{D_s} - \omega)^l \theta_\sigma(m_{D_s} - \sqrt{\bm{q}^2} - \omega) \, ,
  \label{equ:KernelErrorDefinition}
\end{align}
where we choose $t_0 = 1/2$. Hereafter, we refer to the kernel functions with the Heaviside step-function $\theta(x)$ or the sigmoid function $\theta_\sigma(x)$ as the \textit{unsmeared} and \textit{smeared} kernel, respectively.
The Chebyshev approximation of the kernel is then given by \eqref{equ:ChebyshevApproximationKernel}, where, in the case of $\omega_0 = 0$, the coefficients $\tilde{c}_j(\bm{q};\sigma)$ can be obtained as
\begin{align}
  \tilde{c}^{(l)}_j(\bm{q};\sigma) = \frac{2}{\pi} \int_{0}^{\pi} d\theta K^{(l)}_\sigma\left(\bm{q}^2, -\ln \left(\frac{1-\cos\theta}{2}\right)\right) \cos j\theta \, .
  \label{equ:ChebyshevCoefficientsError}
\end{align}
The general expression for any $\omega_0$ is given in \eqref{equ:ChebyshevCoeffDefinition}. 

\begin{figure}
	\centering
	\begin{subfigure}{0.49\textwidth}
    	\centering
    	\includegraphics[width=\textwidth]{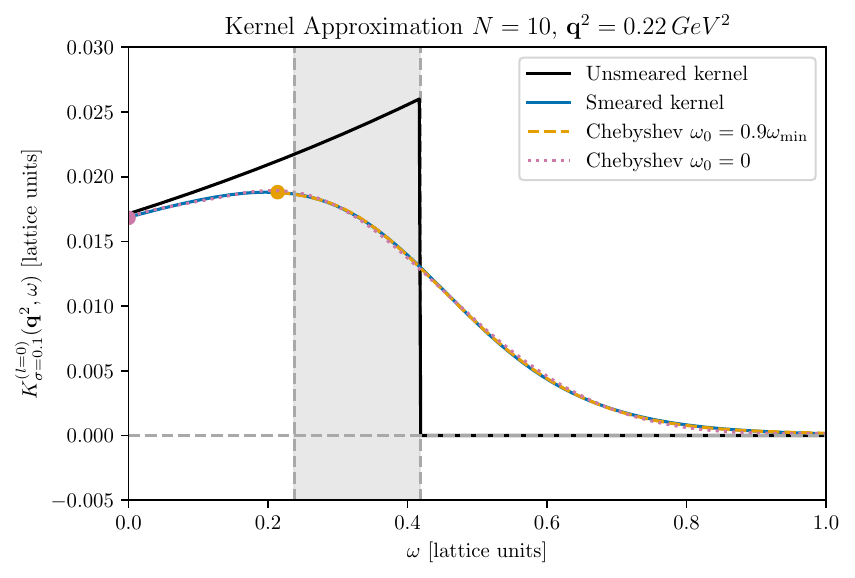}
 	\end{subfigure}
 	\begin{subfigure}{0.49\textwidth}
   		\centering
    	\includegraphics[width=\textwidth]{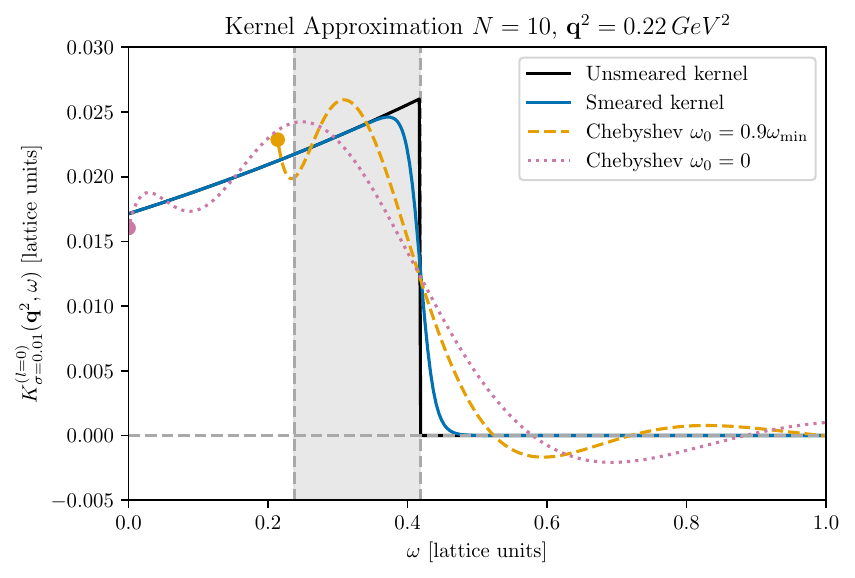}
  	\end{subfigure}

  	\begin{subfigure}{0.49\textwidth}
    	\centering
    	\includegraphics[width=\textwidth]{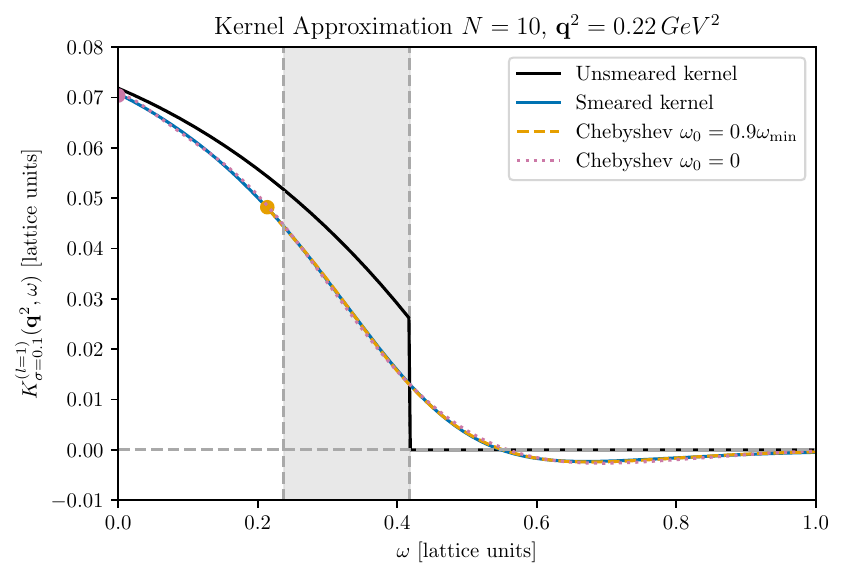}
 	\end{subfigure}
 	\begin{subfigure}{0.49\textwidth}
   		\centering
    	\includegraphics[width=\textwidth]{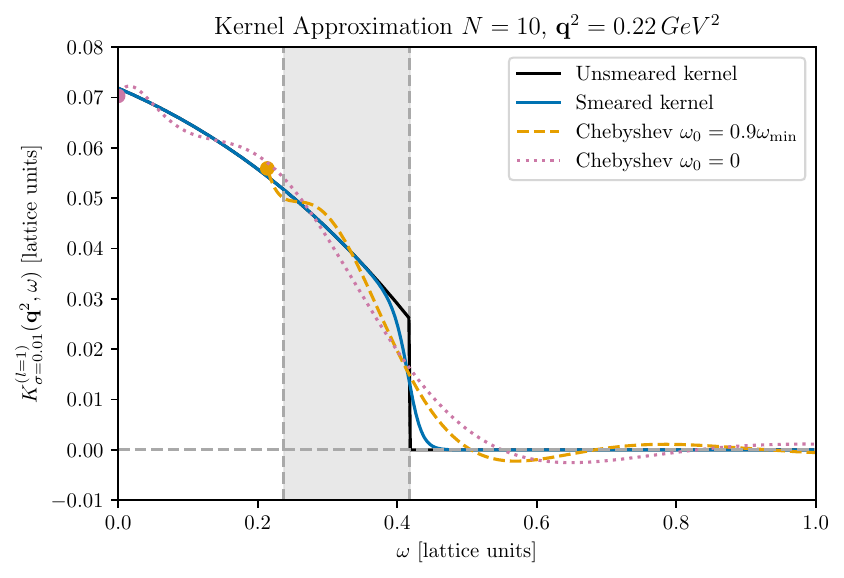}
  	\end{subfigure}

  	\begin{subfigure}{0.49\textwidth}
    	\centering
    	\includegraphics[width=\textwidth]{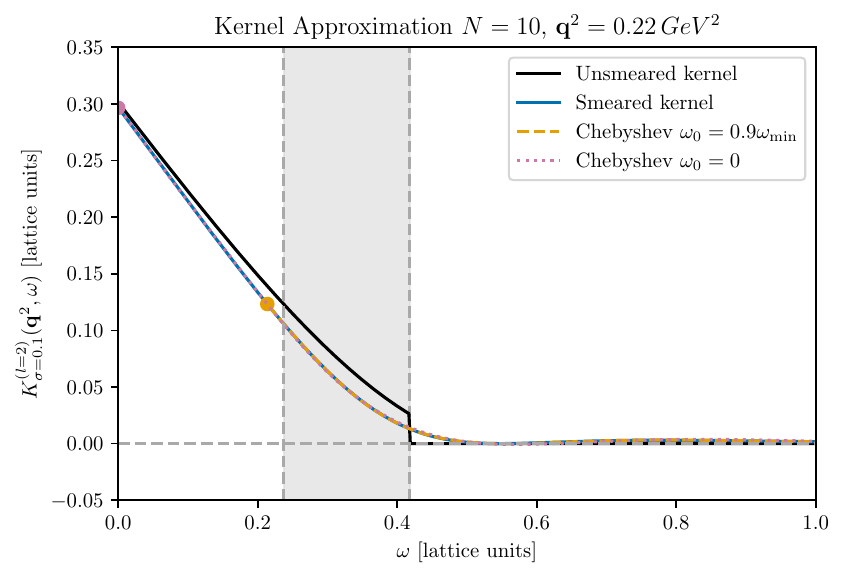}
 	\end{subfigure}
 	\begin{subfigure}{0.49\textwidth}
   		\centering
    	\includegraphics[width=\textwidth]{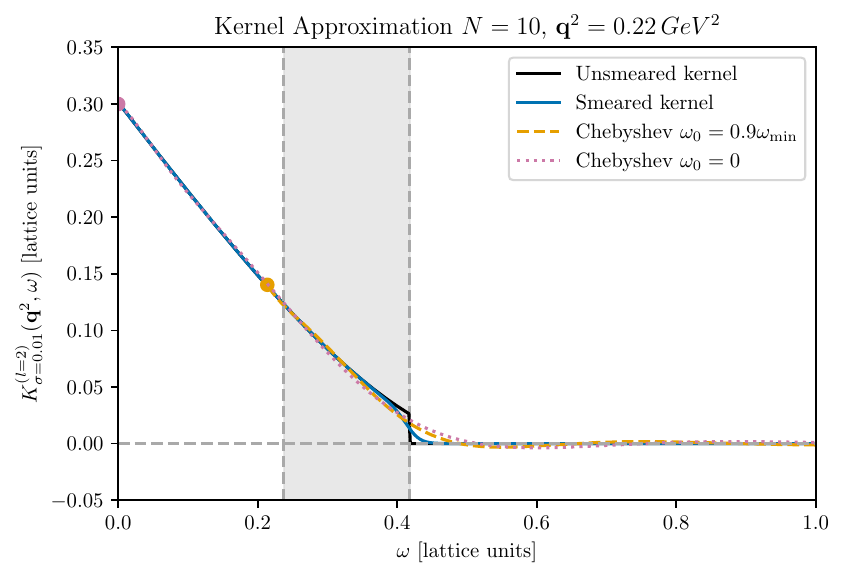}
  	\end{subfigure}
  	\caption{Polynomial approximation of the kernel $K_{\sigma}^{(l)}(\bm{q}^2,\omega)$ at order $N =10$ for $l=0$ (top), $l=1$ (middle), $l=2$ (bottom) with smearing $\sigma = 0.1$ (left) and $\sigma = 0.01$ (right). The momentum is chosen as  $\bm{q}^2 = \SI{0.22}{\giga\electronvolt^2}$. The kinematically allowed region $\omega_{\text{min}} \leq \omega \leq \omega_{\text{max}}$ is represented by the gray shaded area. The solid lines represents the unsmeared (black) and smeared (blue) kernel respectively. The dotted and dashed lines represent the approximation of the smeared kernel depending on the starting point $\omega_0 = 0$ and $\omega_0 = 0.9\omega_{\text{min}}$ shown by the filled circles, respectively.}
  	\label{fig:KernelApproximation}
\end{figure}

The approximation of the kernel function $K_\sigma^{(l)}(\bm{q}^2,\omega)$ for $l =0,1,2$ is shown in Fig.~\ref{fig:KernelApproximation} for $\bm{q}^2=\SI{0.22}{\giga\electronvolt^2}$ representing the momentum $\bm{q} = (0,0,1)$. In order to visualize the differences between the unsmeared and smeared kernel functions, as well as the quality of the approximation, we employ two choices of the smearing width $\sigma = 0.1$ and $0.01$, with other choices discussed later. The polynomial order used in the approximation is $N=10$, which determines the required number of $\bar{C}_{\mu\nu}(\bm{q},t)$ data points (cf. Eq.~\eqref{equ:KernelApproxCorrelator}). We also consider two choices of the lower limit of the approximation $\omega_0$, {\it i.e.} $\omega_0=0$ or $0.9\omega_{\text{min}}$. Here, $\omega_{\text{min}}$ stands for the energy of the lowest-lying state in the corresponding channel. 

The difference of the kernel with and without the smearing is significant especially for $\sigma=0.1$ (plots on the left). The difference is most significant for $l=0$ and decreases with increasing $l$. On the other hand, the approximation with $N=10$ reproduces the smeared kernel fairly precisely for $\sigma=0.1$ with either choice of $\omega_0$, while the approximation curves for $\sigma=0.01$ oscillate wildly around the kernel function. Choosing $\omega_0$ close to $\omega_{\text{min}}$ reduces the deviation from the target kernel function. However, the oscillating behavior remains quite evident, which may introduce an uncontrolled systematic error in the final result. The source of the oscillation is the rapid change of the kernel function, which depends on the smearing width $\sigma$. We conclude that the choice of $\sigma=0.01$ for $N=10$ is too aggressive; we discuss suitable choices of $N$ and how the systematic error can be kept under control in Sec.~\ref{sec:SysErrApproximation}.

For the final systematic error due to the polynomial approximation, one might expect from Fig.~\ref{fig:KernelApproximation} that the contribution from $l=0$ is most sensitive, as its kernel shows the sharpest drop to zero around the threshold. It becomes more relaxed for $l=1$ and $l=2$, as these kernels receive an additional factor of $(m_{D_s} - \omega)^l$ that approaches zero linearly or quadratically, respectively. On the other hand, the contribution to $\bar{X}(\bm{q}^2)$ increases with $l$. Therefore, while $l=0$ receives large systematic corrections, its overall impact on the final result is not dominant.


\subsection{Chebyshev matrix elements from lattice data}
\label{sec:PracticalChebyshev}

To obtain the Chebyshev matrix elements from lattice data, we invert the linear equations as
\begin{align}
	\bar{C}_{\mu\nu}(\bm{q},t) = \sum_{j=0}^{t} \tilde{a}_j^{(t)} \braket{\tilde{T}_j(\bm{q})}_{\mu\nu} \, 
	\label{equ:FitCorrelator}
\end{align}
for a set of correlators $\bar{C}_{\mu\nu}(\bm{q},t)$ of $t$ from 0 to $N$. The coefficients $\tilde{a}_j^{(t)}$ can be determined through the power representation of the Chebyshev polynomials (see App.~\ref{sec:AppChebyshev} for more details). As we mentioned earlier, we introduce a prior to restrict the range of $\braket{\tilde{T}_j(\bm{q})}_{\mu\nu}$ to be uniform within $[-1,+1]$, and obtain the best fit to satisfy \eqref{equ:FitCorrelator}.

We note that, while our prior restricts the Chebyshev matrix elements in a uniform distribution in $[-1,+1]$, we treat the fit results as Gaussian in the error propagation based on the {\it gvar} package \cite{LepageGVAR}. While higher-order Chebyshev matrix elements, which are not well constrained by the data, inherit the uniform prior-distribution, they are suppressed by the rapidly decaying coefficients $\tilde{c}_j(\bm{q};\sigma)$, and Gaussian error propagation can be applied.


\begin{figure}[tbp!]
	\centering
	\begin{subfigure}{0.49\textwidth}
    	\centering
    	\includegraphics[width=\textwidth]{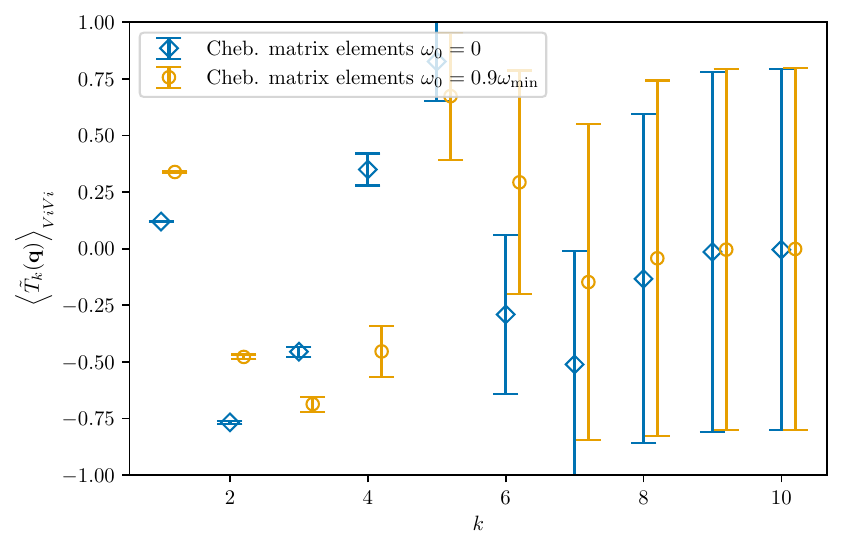}
 	\end{subfigure}
 	\begin{subfigure}{0.49\textwidth}
   		\centering
    	\includegraphics[width=\textwidth]{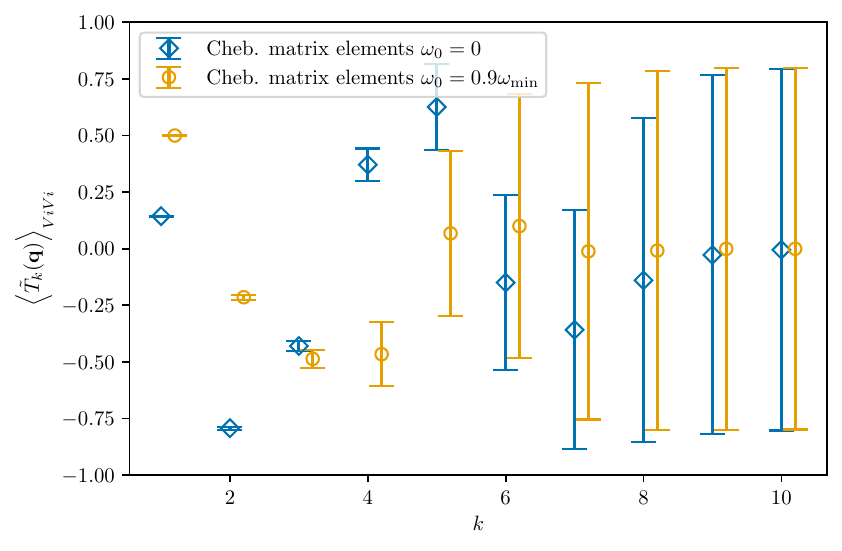}
  	\end{subfigure}

  	\begin{subfigure}{0.49\textwidth}
    	\centering
    	\includegraphics[width=\textwidth]{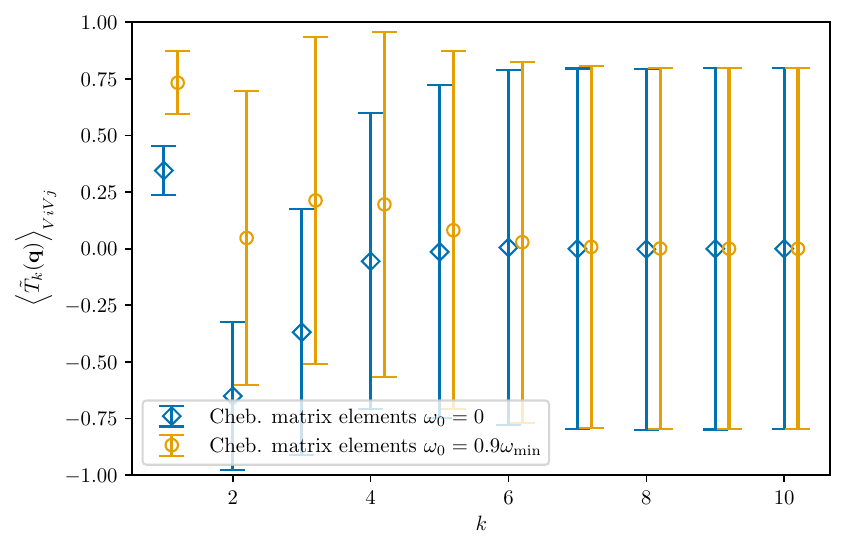}
 	\end{subfigure}
 	\begin{subfigure}{0.49\textwidth}
   		\centering
    	\includegraphics[width=\textwidth]{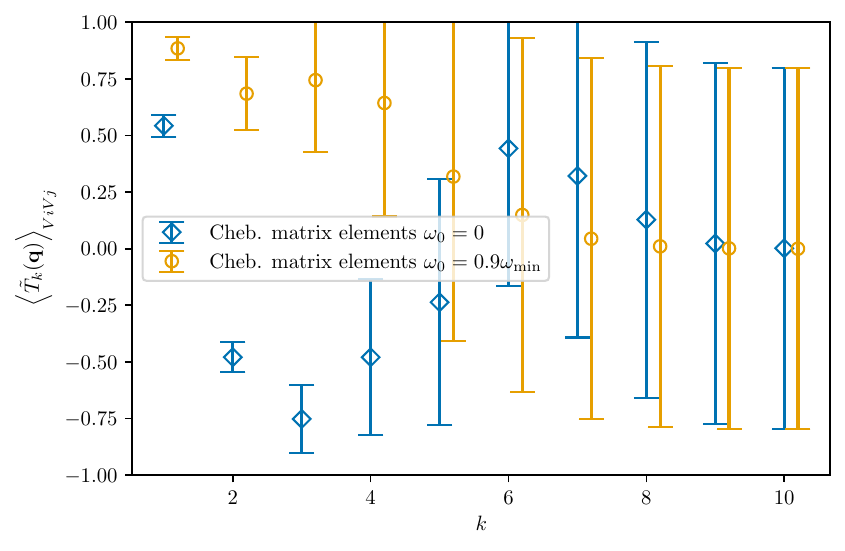}
  	\end{subfigure}
  	\caption{Extracted Chebyshev matrix elements for $\braket{\tilde{T}_k(\bm{q})}_{V_i V_i}$ (first row) and $\braket{\tilde{T}_k(\bm{q})}_{V_i V_j}$ (second row) for $i \neq j$ for $k = 1,2,\cdots, N$ with $N=10$. Results fo $\bm{q}^2=\SI{0}{\giga\electronvolt^2}$ (left column) and $\SI{0.67}{\giga\electronvolt^2}$ (right column). Two choices of the starting point of the approximation $\omega_0 = 0$ (blue diamonds) and $\omega_0 = 0.9\omega_{\text{min}}$ (yellow circles) are shown.}
  	\label{fig:ExtractedChebyshevMatrixVector}
\end{figure}

\begin{figure}[tbp!]
	\centering
	\begin{subfigure}{0.49\textwidth}
    	\centering
    	\includegraphics[width=\textwidth]{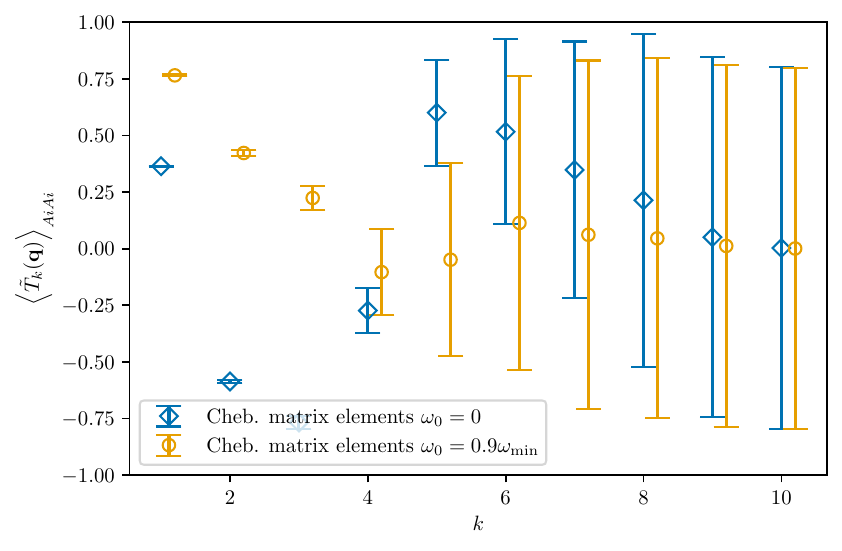}
 	\end{subfigure}
 	\begin{subfigure}{0.49\textwidth}
   		\centering
    	\includegraphics[width=\textwidth]{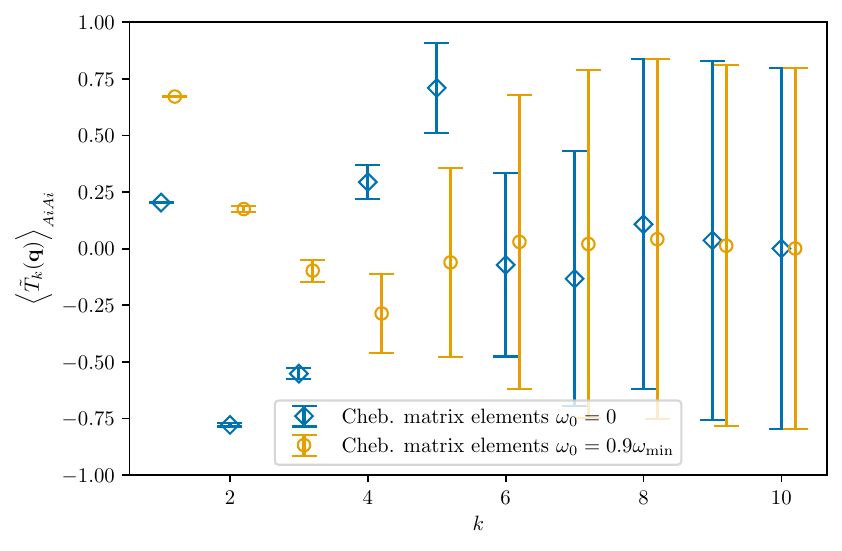}
  	\end{subfigure}

  	\begin{subfigure}{0.49\textwidth}
    	\centering
    	\includegraphics[width=\textwidth]{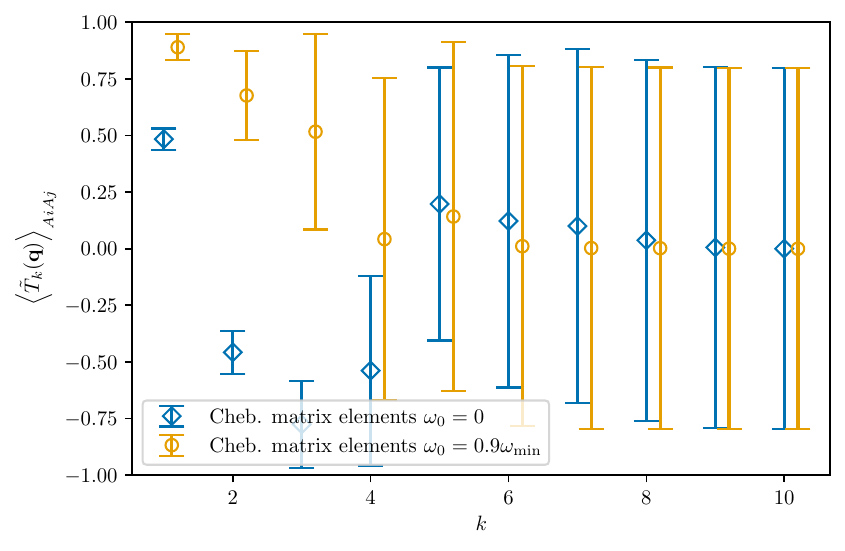}
 	\end{subfigure}
 	\begin{subfigure}{0.49\textwidth}
   		\centering
    	\includegraphics[width=\textwidth]{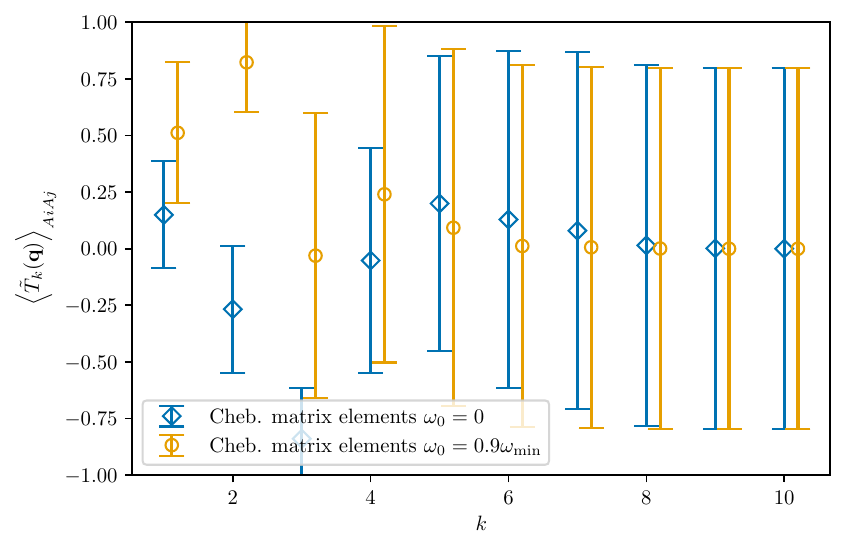}
  	\end{subfigure}
  	\caption{Same as Fig. \ref{fig:ExtractedChebyshevMatrixVector}, but for $\braket{\tilde{T}_k(\bm{q})}_{A_i A_i}$ (first row) and $\braket{\tilde{T}_k(\bm{q})}_{A_i A_j}$ (second row).}
  	\label{fig:ExtractedChebyshevMatrixAxial}
\end{figure}

The examples for the $V_i V_i$ and $V_i V_j$ channels with $i \neq j$ are shown in Fig.~\ref{fig:ExtractedChebyshevMatrixVector}; similar plots are shown in Fig. \ref{fig:ExtractedChebyshevMatrixAxial} for the $A_i A_i$ and $A_i A_j$ channels. We show the Chebyshev matrix elements for two choices of the inserted momentum $\bm{q}^2$, 0 and $\SI{0.67}{\giga\electronvolt^2}$ representing the momentum $\bm{q} = (0,0,0)$ and $\bm{q} = (1,1,1)$, respectively. In these plots, we compare the results for two choices of the lower limit $\omega_0$ of the Chebyshev approximation. We find that the $V_i V_j$ and $A_i A_j$ channels are noisier and the order at which the Chebyshev matrix elements can be extracted meaningfully is lower.
In \cite{Barone:2023tbl}, we monitored the distribution of the Chebyshev matrix elements for bootstrap samples, and observed that the number of the Chebyshev matrix elements extracted significantly depends on the choice of $\omega_0$. Our finding here is consistent with the observation, despite the use of a more simplified statistical analysis. From Figs.~\ref{fig:ExtractedChebyshevMatrixVector} and \ref{fig:ExtractedChebyshevMatrixAxial} we observe that the error of the Chebyshev matrix elements with $\omega_0 = 0.9\omega_{\text{min}}$ saturates one or two orders earlier compared to $\omega_0=0$. This is due to the relation \cite{Barone:2023tbl}
\begin{align*}
	\tilde{a}_j^{(k)}|_{\omega_0=0} = e^{-0.9\omega_{\text{min}}} \tilde{a}_j^{(k)}|_{\omega_0=0.9\omega_{\text{min}}} \, ,
\end{align*}
where the additional exponential factor on the r.h.s. cancels the exponential decay of the ground-state contribution in \eqref{equ:FitCorrelator}. Although the Chebyshev matrix elements become undetermined earlier for $\omega_0=0.9\omega_{\text{min}}$, the reconstructed $\bar{X}(\bm{q}^2)$, as we will see, turns out to have a similar statistical error.


\subsection{Truncation error and its bound}
\label{sec:SysErrApproximation}

We now address the systematic error originating from the Chebyshev approximation and the smearing of the kernel function, which we first discussed in \cite{Kellermann:2022mms}. We need to consider two limits: $\sigma \to 0$ and $N \to \infty$. 

Each order of Chebyshev polynomials can be identified as a frequency component of the target function. The corresponding ``wave length'' determines the length scale that the approximation can accommodate. In general, with Chebyshev polynomials at order $N$, where $N$ translates into the time separation of the two $V-A$ currents in $\bar{C}_{\mu\nu}(\bm{q},t)$, the smallest length scale that can be represented is about $1/N$, so that, in order to keep the systematic error under control, we need to keep the smearing parameter $\sigma$ not too small compared to $1/N$. The discontinuity of the Heaviside function therefore requires infinitely large polynomial orders, which is impractical to implement. With smearing, the length scale is made finite of order $\sigma$. To be consistent with the available orders of the polynomial $N$, we impose the scaling
\begin{align}
  \sigma = \frac{1}{\alpha N} \, ,
  \label{equ:WaveRatioAssumption}
\end{align}
where $\alpha$ is a proportionality factor that we set $\alpha=1$. In this way, the $N \to \infty$  and the $\sigma \to 0$ limits are combined.

In practical applications we cannot reach arbitrarily large $N$ (or small $\sigma$), given that the Chebyshev matrix elements are reconstructed from lattice data, {\it i.e.} the polynomial order $N$ is limited by the finite number of time slices $t$ as well as the statistical error of the correlator. Instead, one can derive a mathematical bound on the errors due to the truncated higher order terms using the property of Chebyshev polynomials, $|\tilde{T}_j(x)| \leq 1$  \cite{Kellermann:2022mms}. 
Namely, above a certain polynomial order $N_{\text{Cut}}$, we assume the values on the edge of the bounds, {\it i.e.} $ \tilde{T}_{k} = \tilde{T}^{\mathbb{Z}_2}_{k}$ for $k > N_{\text{Cut}}$, where $\tilde{T}^{\mathbb{Z}_2}_{k}$ takes values only at $\pm 1$. The kernel can then be estimated as
\begin{align}
  K_{\sigma}^{(l)}(\bm{q}^2, \omega) \simeq \frac{\tilde{c}_0(\bm{q};\sigma)}{2} + \sum_{j=1}^{N_{\text{Cut}}} \tilde{c}_j(\bm{q};\sigma) \tilde{T}_j(e^{-\omega}) + \sum_{k=N_{\text{Cut}}+1}^{N} \tilde{c}_k(\bm{q};\sigma) \tilde{T}^{\mathbb{Z}_2}_{k} \, ,
  \label{equ:ApproxErrorEstimate}
\end{align}
for an arbitrary value of $N$. 
The largest possible error of the last term is then given by
$\sum_{k=N_{\text{Cut}}+1}^{N} |\tilde{c}_k(\bm{q};\sigma)|$,
which represents the mathematical upper bound on the error without affecting the mean value, as the expectation value of $\langle \tilde{T}^{\mathbb{Z}_2}_{k} \rangle=0$. 

\begin{figure}[tb]
	\centering
    	\includegraphics[width=0.6\textwidth]{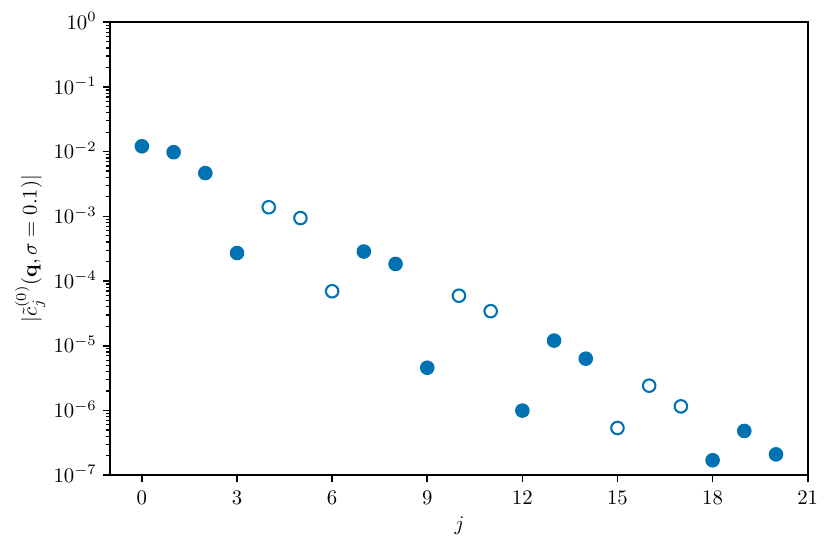}
    	\includegraphics[width=0.6\textwidth]{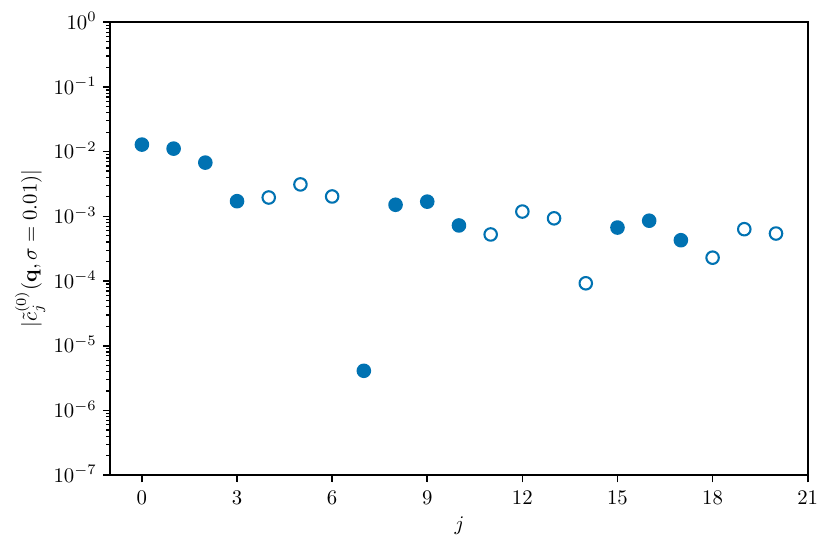}
  	\caption{Coefficients $|\tilde{c}_j^{(0)}(\bm{q},\sigma)~|$ for two choices of the smearing $\sigma = 0.1$ (top) and $\sigma = 0.01$ (bottom) in the kernel function calculated for $\bm{q}^2 = \SI{0.22}{\giga\electronvolt^2}$. Positive and negative coefficients are plotted by filled and empty dots, respectively.}
  	\label{fig:ChebyshevCoeffsLogScale}
\end{figure}

This error bound depends only on the coefficients $\tilde{c}_j(\bm{q};\sigma)$, which are analytically known once the kernel function is given through \eqref{equ:ChebyshevCoefficientsError}. Some examples of the coefficients are shown in Fig.~\ref{fig:ChebyshevCoeffsLogScale}. In general, they are suppressed exponentially for large $j$'s. The exponential fall-off is slower when the kernel function contains high-frequency components, {\it i.e.} smaller $\sigma$'s. The error bound given by the sum of $|\tilde{c}_j(\bm{q};\sigma)|$ converges very slowly for the case of $\sigma=0.01$ given in the plot (bottom).

For larger momenta, the kernel function becomes more singular compared to those plotted in Fig.~\ref{fig:KernelApproximation}, since the upper limit of the $\omega$-integral decreases as $\omega_{\text{max}}=m_{D_s}-|\bm{q}|$. As a consequence, the convergence of the coefficients becomes slower. 



\subsection{Estimating truncation error and the role of the ground state}
\label{sec:SystematicChebApprox}
As discussed in Section~\ref{sec:PracticalChebyshev} and in \cite{Barone:2023tbl, Kellermann:2022mms}, the Chebyshev matrix elements of large order $j$ become nearly identical with the distribution of priors, due to the exponentially growing statistical noise of the lattice data for large time separation. 
In the previous subsection, we discussed the rigorous mathematical bound. Here, we describe how we estimate the truncation error in practice.

Given that the extremum values $\pm 1$ of the Chebyshev polynomials $T_j(x)$ are distributed nearly evenly over $x$ and that there is no correlation with the physical spectrum, it is highly unlikely that the extremum values of the Chebyshev matrix elements appear for many different $j$'s at the same time, and an assumption of uniform distribution of the unknown higher-order matrix elements seems plausible.  Therefore, instead of the discrete $\mathbb{Z}_2$ distribution, we assume a uniform distribution for the unknown Chebyshev matrix elements. The error estimate is then given by
\begin{align}
    \sigma_{\bar{K}, U} = \sqrt{\frac{1}{3} \sum_{k=N_{\text{Cut}}+1}^{N} |\tilde{c}_k(\bm{q};\sigma)|^2} \label{equ:ErrorEstimator}
\end{align}
from the standard deviation of the uniform distribution. 

Although this estimate should allow for a proper estimation of the truncation error for most cases, there are still some critical situations. Namely, when the ground state, which gives the dominant contribution, is very close to the kinematical threshold, the result of the Chebyshev approximation strongly depends on the choice of $\sigma$. Even in this case, the error bound discussed in the previous section covers the correct result for a given $\sigma$, but the resulting error becomes unreasonably large in the limit of $\sigma\to 0$, given that the ground-state energy is precisely known and we can actually obtain the true result without recourse to the Chebyshev approximation. This happens for nearly maximal recoil momenta, where no phase space is left for excited states to enter.

This observation motivates us to treat the ground state separately in the inclusive analysis. In Sec.~\ref{sec:Amplitudes} we describe the fits to extract the ground-state contribution from the four-point functions. We subtract that part from the four-point function and apply the inclusive analysis in the same way as discussed above. The ground-state contribution is estimated without using the Chebyshev approximation of the kernel, and no associated error is expected.

\begin{figure}[tb]
    \centering
    \begin{subfigure}{0.49\textwidth}
        \centering
        \includegraphics[width=\textwidth]{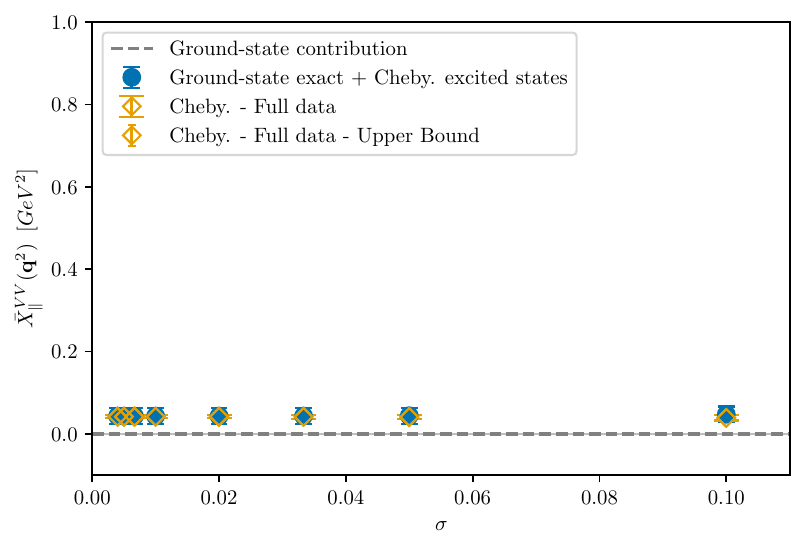}
    \end{subfigure}
    \begin{subfigure}{0.49\textwidth}
    \centering
        \includegraphics[width=\textwidth]{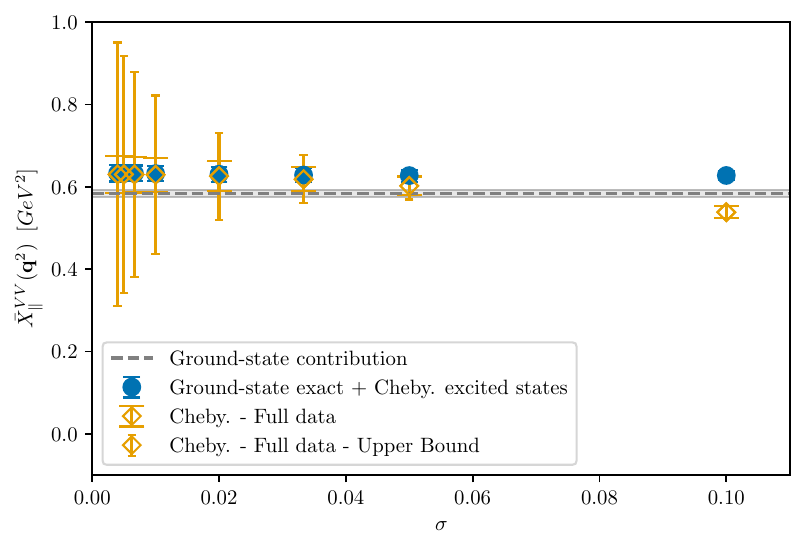}
    \end{subfigure}

    \begin{subfigure}{0.49\textwidth}
        \includegraphics[width=\textwidth]{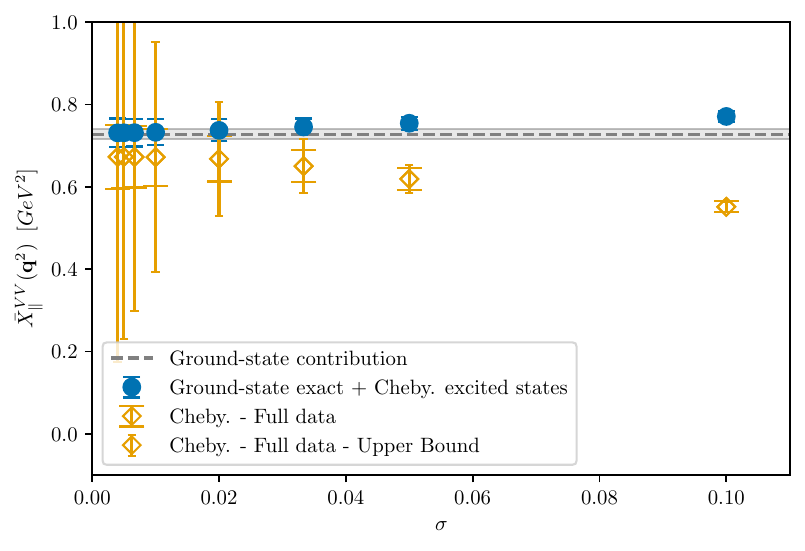}
    \end{subfigure}
    \begin{subfigure}{0.49\textwidth}
        \includegraphics[width=\textwidth]{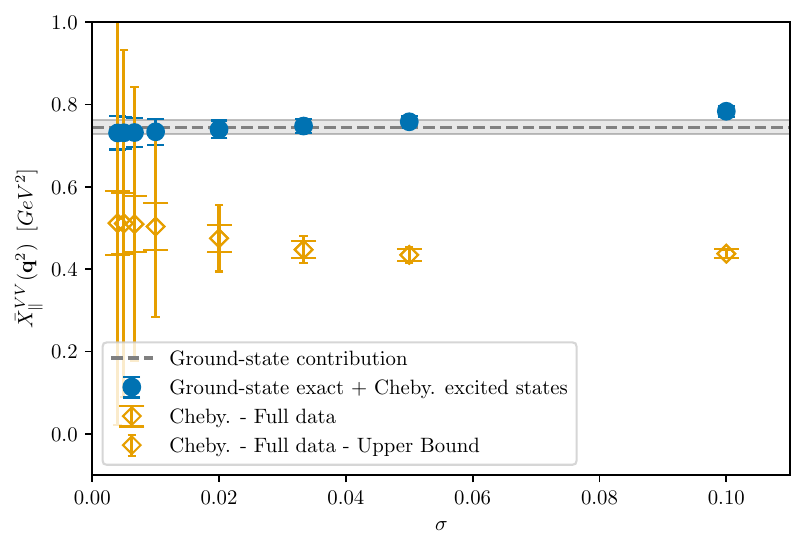}
    \end{subfigure}
    \caption{Comparison of the estimated error of $\bar{X}^{VV}_{\parallel}(\bm{q}^2)$ for $\bm{q} = (0,0,0)$ (top left), $(0,0,1)$ (top right), $(0,1,1)$ (bottom left) and $(1,1,1)$ (bottom right) as a function of $\sigma = 1/N$, with $N$ the polynomial order of the Chebyshev approximation. The gray band represents the expected ground-state contribution. The orange diamonds are determined by applying the Chebyshev approximation on the full data set, while the blue circles treat the ground state exactly and the Chebyshev approximation is only applied on the excited states. For the full data set we compare the error bars obtained assuming the uniform distribution of the Chebyshev matrix elements in \eqref{equ:ErrorEstimator} (inner errors) and the the mathematical upper bound discussed in Sec.~\ref{sec:SysErrApproximation} (outer errors).}
  	\label{fig:NExpansion}
\end{figure}

Fig.~\ref{fig:NExpansion} shows the results for $\bar{X}^{VV}_{\parallel}(\bm{q}^2)$ as a function of $\sigma=1/N$ and the error bars represent the error accumulated up to order $N$. As a reference value, we include the ground-state contribution extracted from a fit of the data by a gray band. We compare the results obtained by applying the Chebyshev approximation on the full data and by subtracting the ground state from the analysis and adding it back at the end, applying the Chebyshev approximation only on the excited states. For the full data set, we show the error estimates using \eqref{equ:ErrorEstimator}, as well as the rigorous upper bound. 

Since the Chebyshev coefficients decrease only slowly for the kernel with a smaller smearing width $\sigma$, the mathematical upper bound grows rapidly if accumulated up to $N=1/\sigma$. It reflects the property of the Chebyshev polynomials that the resolution of the function is given by $1/N$, {\it i.e.} the terms of $O(N)$ play an important role in the approximation of the kernel, as demonstrated in Fig.~\ref{fig:ChebyshevCoeffsLogScale}. On the other hand, with the uniform distribution the unknown Chebyshev matrix elements could be positive or negative and cancel in the sum over $j$, making the error estimate nearly independent of the highest accumulated order. 

As Fig.~\ref{fig:NExpansion} shows, the central values are slightly above the ground-state estimate for $\bm{q} = (0,0,0)$ and $(0,0,1)$, suggesting some excited-state contributions. For higher momenta $\bm{q} = (0,1,1)$ and $(1,1,1)$ the central value for the full data is below the ground-state estimate and, in the case of $\bm{q} = (1,1,1)$, is not covered by the errors from the uniform distribution. This reflects the critical situation described above. 

The results of removing the ground state from the analysis appear to be stable as a function of $\sigma=1/N$, confirming that most of the inclusive rate is saturated by the ground state and the excited-state contribution is insignificant. 


We note that the shift of the central value in Fig.~\ref{fig:NExpansion} as a function of $\sigma$ is due to the dependence of the kernel function and thus the Chebyshev coefficients on $\sigma$. An explicit example is shown in Fig.~\ref{fig:ChebyshevCoeffsLogScale}.

In the following analysis, we employ the error estimate based on the uniform distribution of the unknown Chebyshev matrix elements with $N=250$, where the estimated error is well saturated. Real data are included only up to $N=10$.


\begin{figure}[tb]
    \centering
    \includegraphics[width=0.6\linewidth]{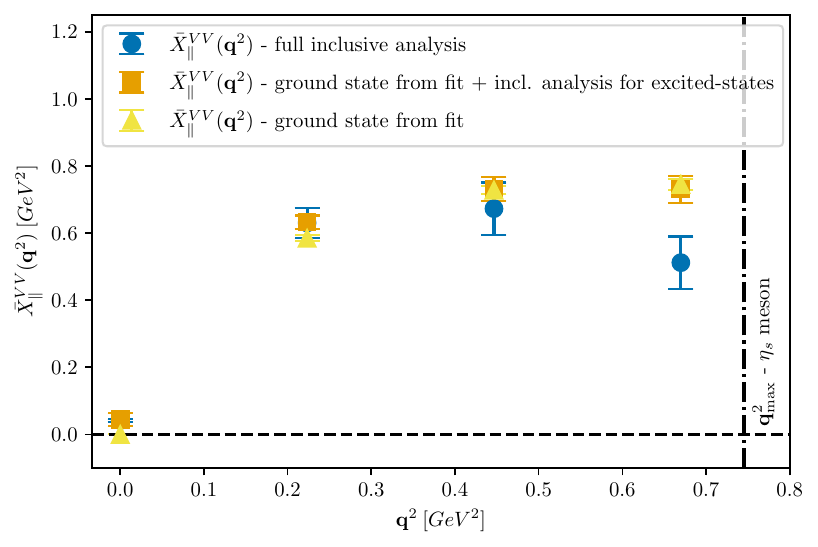}
    \caption{$\bar{X}^{VV}_{\parallel}(\bm{q}^2)$ for all values of $\bm{q}^2$. Application of the inclusive analysis for the full data set (circles) and only for the ground-state subtracted data set (squares) are shown, as well as the ground-state contribution (triangles).
    }
    \label{fig:XVVParCompSystematics}
\end{figure}

In Fig.~\ref{fig:XVVParCompSystematics} we compare the results of the two approaches, {\it i.e.} applying the Chebyshev approximation on the full data set and only on the excited-state contributions, for $\bar{X}_\parallel^{VV}(\bm{q}^2)$ in the $\sigma\to 0$ limit. The expected ground-state contribution extracted from the inclusive data set is also shown for comparison.

The inclusive analysis with the full data goes lower than the ground-state estimate for higher momenta, $\bm{q}=(0,1,1)$ and $(1,1,1)$ as mentioned above, and the problem is circumvented when the Chebyshev approximation is applied only for the excited states. We find a good agreement with the expected ground-state contribution for large $\bm{q}^2$'s. 

Using the fits from Sec.~\ref{sec:GroundStateEstimate} we find that the ground states make up roughly $\SI{90}{\percent}$ of the contribution to $\bar{X}^{VV}_{\parallel}(\bm{q}^2)$ for smaller momenta, {\it e.g.} for $\bm{q} = (0,0,1)$ the ground-state contribution is $\SI{0.5847(57)}{\giga\electronvolt^2}$ while the excited states contribute with $\SI{0.049(17)}{\giga\electronvolt^2}$. For higher momenta, the ground state makes up nearly $\SI{100}{\percent}$ of the contribution, {\it e.g.} for $\bm{q} = (1,1,1)$ the ground and excited state contributions are given by $\SI{0.745(11)}{\giga\electronvolt^2}$ and $\SI{-0.013(38)}{\giga\electronvolt^2}$, respectively.

\section{Finite-volume effects}
\label{sec:SystematicErrors}


\begin{figure}[tb]
  \centering
  \begin{subfigure}{0.49\textwidth}
    \centering
    \includegraphics[width=\textwidth]{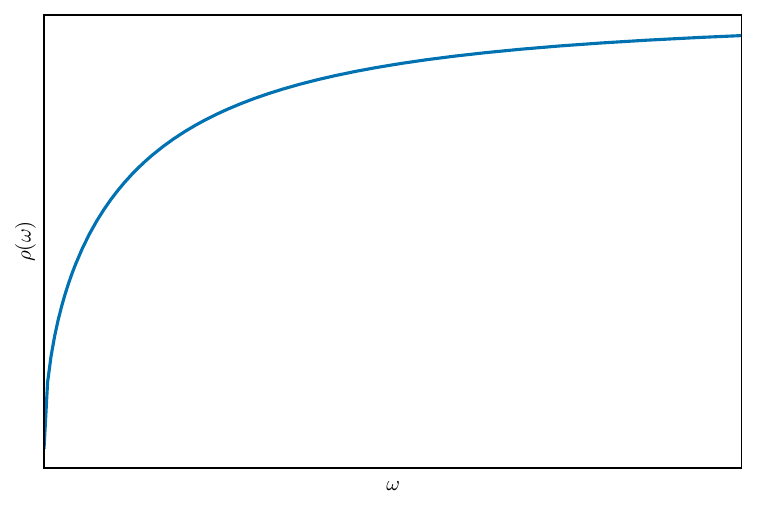}
    \caption{$\rho(\omega)$}
  \end{subfigure}
  \begin{subfigure}{0.49\textwidth}
    \centering
    \includegraphics[width=\textwidth]{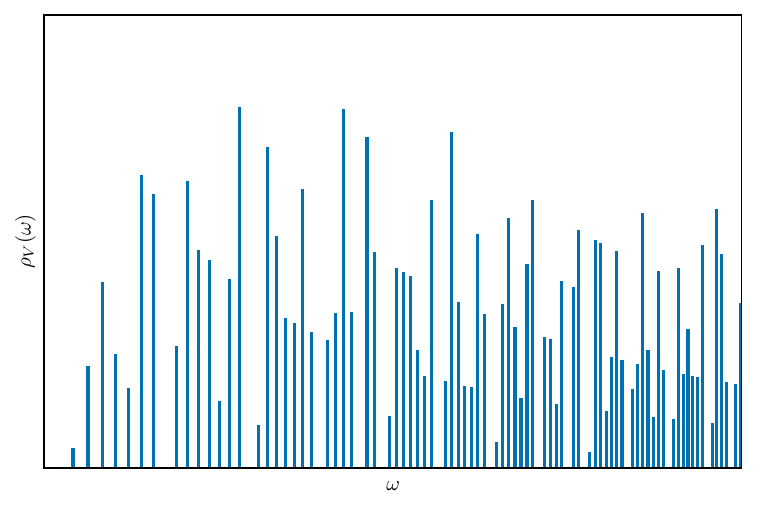}
    \caption{$\rho_V(\omega)$}
  \end{subfigure}
  \caption{Sketch of the infinite-volume spectral density $\rho(\omega)$ (left) and the finite-volume $\rho_V(\omega)$ for a specific volume $V$ (right). The height of $\rho_V(\omega)$ corresponds to the multiplicity of the states for each energy $\omega$.}
  \label{fig:FiniteVolSPectralFunction}
\end{figure}

Since multi-hadron final states are expected to contribute significantly in the inclusive calculation, finite-volume effects, which are suppressed only as a power of the volume $V=L^3$, may be a major concern. (For example, the energy spectrum of two-body states receives a correction of $\mathcal{O}(1/L^3)$ \cite{Luscher:1986pf}.) Two-particle states in a finite volume have a discrete spectrum made of a sum of $\delta$-functions, as depicted in Fig.~\ref{fig:FiniteVolSPectralFunction}, which approaches the continuum spectrum in the infinite-volume limit. The convolution integral with the kernel function that involves the Heaviside function, see \eqref{equ:EnergyIntegralXBar}, therefore becomes discontinuous. In order to uniquely approach the infinite-volume limit of the inclusive decay rate, we must consider an ordered limit
\begin{align}
    \lim_{\sigma\to 0} \lim_{V\to\infty} \bar{X}_{\sigma}(\bm{q}^2) \, .
    \label{equ:OrderedLimits}
\end{align}
The discussion in the previous section is only about the extrapolation to $\sigma\to 0$ combined with $N\to\infty$, but rigorously speaking it should be performed only after the infinite-volume limit is taken.


In practice, the data to take the infinite-volume limit would be numerically expensive and not always available. Instead, we attempt to estimate and correct the potential systematic errors for not taking the $V\to\infty$ limit. We develop a model of the two-body final states and trace the systematic effects in the inclusive analysis. We discuss our model in Sec.~\ref{sec:SysErrFV} and describe the analysis of the lattice data in Sec.~\ref{sec:SystematicFinVol}.

\subsection{A model of two-body states}
\label{sec:SysErrFV}

Among various multi-hadron states, our model treats the two-body final states, specifically $K\bar{K}$ states, which are expected to give the leading contribution. The interaction between $K$ and $\bar{K}$ is ignored, which we assume to be a good approximation for estimating finite-volume effects. 

We consider the imaginary part of the vacuum polarization function involving the $K\bar{K}$ loop
\begin{align}
  \text{Im } \mathcal{M} = \pi^2 \int\! \frac{d^3\bm{k}}{(2\pi)^3} \int\! dk_0\, \delta\left((q+k)^2 - m_K^2\right) \theta(q_0+k_0) \delta(k^2 - m_K^2) \theta(-k_0) \, .
\end{align}
Assuming $q=(\omega, \bm{q})$, the energy and momentum of the final hadronic state, we perform the $k_0$ integration to obtain the finite-volume spectral function
\begin{align}
  \rho_V(\omega,\bm{q}) = \frac{\pi}{V} \sum_{\bm{k}} \frac{\bm{k}\cdot(\bm{q}+\bm{k})}{4\sqrt{\bm{k}^2 + m_K^2}\sqrt{(\bm{q} + \bm{k})^2 + m_K^2}}  \delta\left(\omega - \sqrt{\bm{k}^2 + m_K^2} - \sqrt{(\bm{q} + \bm{k})^2 + m_K^2}\right) \, .
  \label{equ:SpectralDenVec}
\end{align}
The integral over $\bm{k}$ is replaced by a sum $(1/V) \sum_{\bm{k}}$, where $\bm{k} = (2\pi/L) \, \bm{l}$ with $\bm{l} = (l_1,l_2,l_3)$ a vector of integers in $-L/2 < l_i \leq L/2$.
In the rest frame, {\it i.e.} $\bm{q} = \bm{0}$, the infinite-volume spectrum is given by
\begin{align}
  \rho(\omega,\bm{0}) = \frac{1}{64\pi} \omega^2 \left(1 - \frac{4m_K^2}{\omega^2}\right)^{3/2} \, .
  \label{equ:InfVolSpecVector}
\end{align}
This expression represents the production of non-interacting $K\bar{K}$ states from the vacuum through a vector current. The spatial extent of the initial $D_s$ meson is neglected; we incorporate it later through a form factor. 

For this study, we only consider the case of producing the two-body states that could couple to a vector particle. It gives the dominant contribution, as discussed in the previous section. Other channels such as those involving $P$-wave states are sub-dominant, and their finite-volume effects are neglected.

Here we introduce the quantity
\begin{align}
  \bar{X}_{\sigma}^{(l)}(\omega_{\text{th}}) = \int_0^\infty d\omega \, \rho_{V}(\omega,\bm{0}) \, K_\sigma^{(l)}(\omega;\omega_{\text{th}}) \, ,
  \label{equ:AuxilaryFunction}
\end{align}
to study the volume dependence of the inclusive rate. It corresponds to $\bar{X}^{(l)}$ in \eqref{equ:XDecomposition}, and the subscript $\sigma$ denotes the smearing applied on the kernel. The quantity $\bar{X}_\sigma^{(l)}(\omega_{\text{th}})$ is defined as a function of an arbitrary upper limit of the integral $\omega_{\text{th}}$ implemented in the kernel function $K_\sigma^{(l)}(\omega;\omega_{\text{th}})$. We use the threshold $\omega_{\text{th}}$ as a control parameter, although the physical threshold is fixed at $\omega_{\text{th}}^{\text{phys}} = m_{D_s} - \sqrt{\bm{q}^2}$, to investigate the validity of our model. The modified kernel in \eqref{equ:AuxilaryFunction} is defined as
\begin{align}
  K_{\sigma}^{(l)}(\omega;\omega_{\text{th}}) = e^{2\omega t_0} \left(m_{D_s} - \frac{\omega_{\text{th}}^{\text{phys}}}{\omega_{\text{th}}}\omega\right)^l \theta_{\sigma}(\omega_{\text{th}} - \omega) \, .
  \label{equ:VariationKernel}
\end{align}
For simplicity and visualization purposes, we omit the dependence on $\bm{q}^2$, and ignore the factor $\sqrt{\bm{q}^2}^{2-l}$, which is a constant in the $\omega$-integral.

\begin{figure}[tbp]
  \centering
    \centering
    \includegraphics[width=0.6\textwidth]{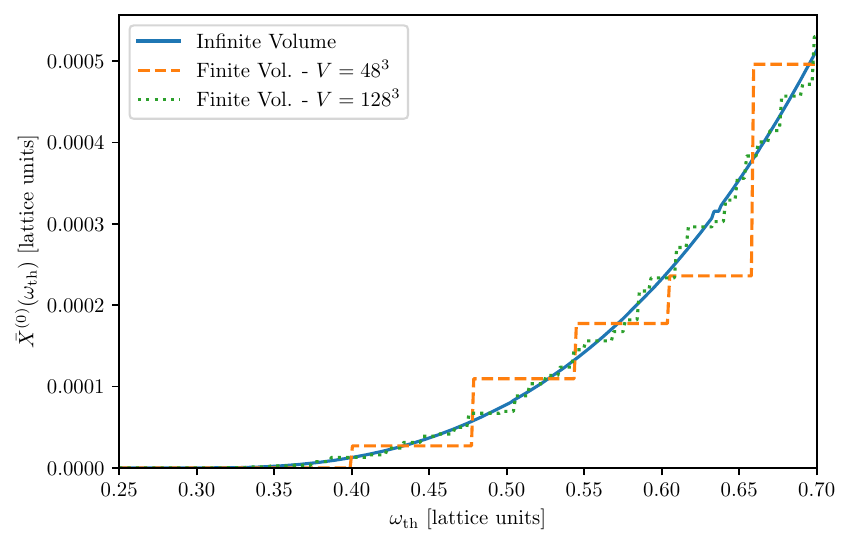}
    \centering
    \includegraphics[width=0.6\textwidth]{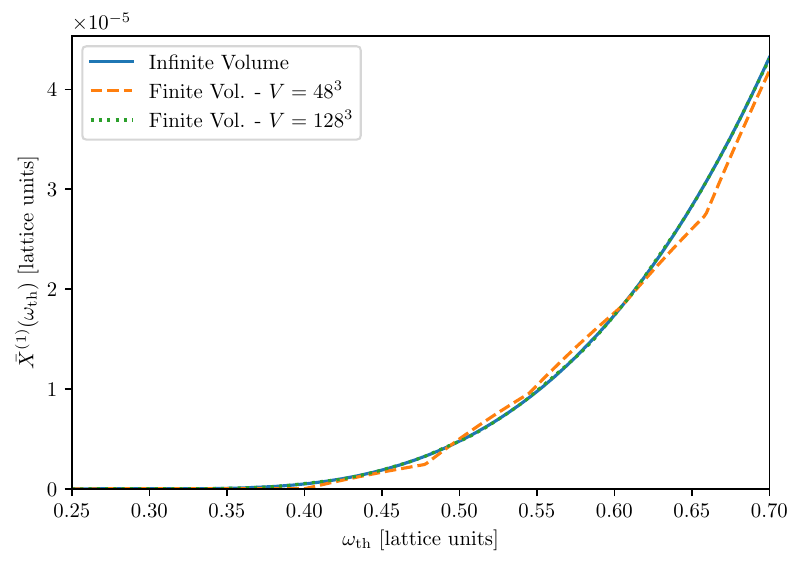}
    \centering
    \includegraphics[width=0.6\textwidth]{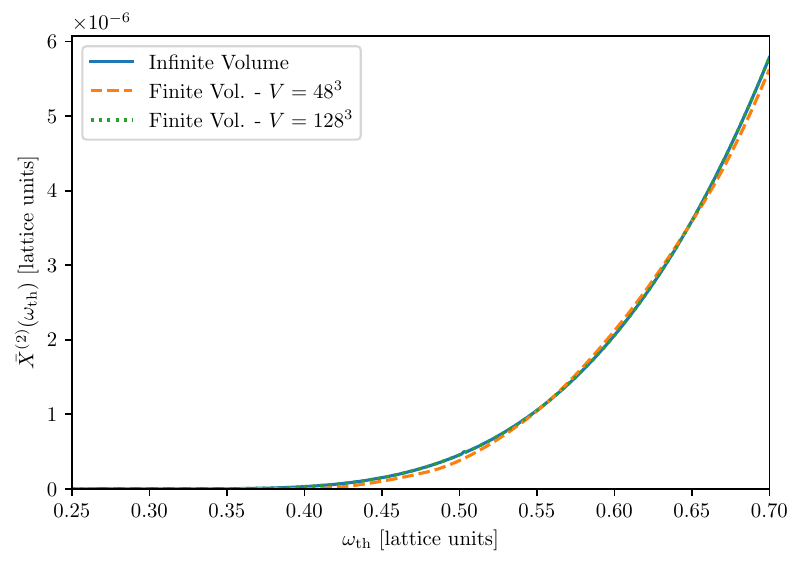}
  \caption{Convolution of the model spectral function with the modified kernel \eqref{equ:VariationKernel} as a function of $\omega_{\text{th}}$. The infinite-volume limit (solid) is shown together with the volume $V=48^3$ (dashed) and $V =256^3$ (dotted). The plots are for $l=0$ (top), $1$ (middle) and $2$ (bottom).}
  \label{fig:VolumeReconstruction}
\end{figure}

In Fig.~\ref{fig:VolumeReconstruction} we show the $K\bar{K}$ contributions to $\bar{X}_{\sigma}^{(l)}(\omega_{\text{th}})$ as obtained from the above model for two choices of finite volume $V = 48^3$ and $128^3$, in addition to the infinite-volume limit. The same lattice spacing as used in the simulations is assumed. We show the results for $l=0$, 1, 2 for the spectral density with the unsmeared kernel, {\it i.e.} $\bar{X}^{(l)}(\omega_{\text{th}})$.

The $\bar{X}^{(l)}(\omega_{\text{th}})$ with $l=0$ shows a strong volume dependence on $\omega_{\text{th}}$, due to the sharp drop in the kernel function; see Fig.~\ref{fig:KernelApproximation}. By increasing the volume from $48^3$ to $128^3$, the step size becomes smaller, and the infinite-volume limit is nearly reproduced. For $l=1$ and 2, the kernel function vanishes linearly or quadratically, respectively, toward the threshold, resulting in a mild volume dependence of $\bar{X}^{(l)}(\omega_{\text{th}})$ and a good reproduction of the infinite-volume limit already at $V=48^3$.

\subsection{Systematic error from finite-volume}
\label{sec:SystematicFinVol}
We now combine this non-interacting model for two-body states with the lattice data. We assume that, other than the contribution of the lowest-lying ground state, the density in the finite volume is proportional to \eqref{equ:SpectralDenVec},
and decompose the spectral density into a ground state and a series of two-body states: 
\begin{align}
C(\bm{q}, t) = A_0(\bm{q}) e^{-E_0(\bm{q})t} + \mathcal{S}(\bm{q}) \sum_{\bm{k}} A_{\bm{k}}(\bm{q}) e^{-E_{\bm{k}}(\bm{q}) t} |F(E_{\bm{k}}(\bm{q}))|^2 \, ,
\label{equ:ModelFitFunction}
\end{align}
where the first term represents the ground state. The remaining terms correspond to the $D_s\to K\bar{K}\ell\bar\nu$ process and are written in terms of a known energy $E_{\bm{k}}(\bm{q})=\sqrt{\bm{k}^2 + m_K^2} + \sqrt{(\bm{q} + \bm{k})^2 + m_K^2}$ and amplitude $A_{\bm{k}}(\bm{q})=\pi/V\biggl[ \bm{k}\cdot (\bm{q} + \bm{k})/\left[4\sqrt{\bm{k}^2 + m_K^2}\sqrt{(\bm{q} + \bm{k})^2+ m_K^2}\right]\biggl]$ from the model \eqref{equ:SpectralDenVec}. The overall factor $\mathcal{S}(\bm{q})$ is a fit parameter that is assumed to be volume independent. We also include a ``form factor'' $F(E)$ in \eqref{equ:ModelFitFunction}, motivated by the idea of vector-meson dominance, $F(E)=1/(E^2-m_\phi^2)$,
which enhances the states close to the vector-meson resonance. The overall normalization is irrelevant as it is absorbed by $\mathcal{S}(\bm{q})$ in \eqref{equ:ModelFitFunction}. We would also expect modifications that take the initial $D_s$ meson into account, but for the purpose of estimating the finite-volume effects, this model should suffice.

We apply this model for the insertions of spatial components of the axial-vector current with vanishing recoil momentum, {\it i.e.} $\bar{X}^{AA}(\bm{q}^2 = \bm{0}^2)$. Since there is no distinction between $\parallel$ and $\perp$ at zero recoil, we averaged over all three directions of the axial-vector current. The decompositions \eqref{equ:C_A0A0}--\eqref{equ:C_APerpAPerp_heavy_meson_FFs_after} indicate that these channels are saturated by the vector meson. With finite momenta, the states with different quantum numbers, such as $P$-wave states, also contribute to the correlator. Later, we discuss how the finite-volume effect is estimated for these cases. 

In the analysis, the ground-state energy $E_0(\bm{0})$ and its amplitude $A_0(\bm{0})$ are determined by a fit to the lattice data using a single exponential function, and then included in the fit to the model \eqref{equ:ModelFitFunction} through a Gaussian prior. The excited-state energies $E_{\bm{k}}(\bm{0})$ and their amplitudes $A_{\bm{k}}(\bm{0})$ are fixed through the model. Thus, we extract $E_0(\bm{0})$, $A_0(\bm{0})$ and $\mathcal{S}(\bm{0})$ from the fit.

\begin{figure}[tbp]
	\centering
    	\includegraphics[width=0.6\textwidth]{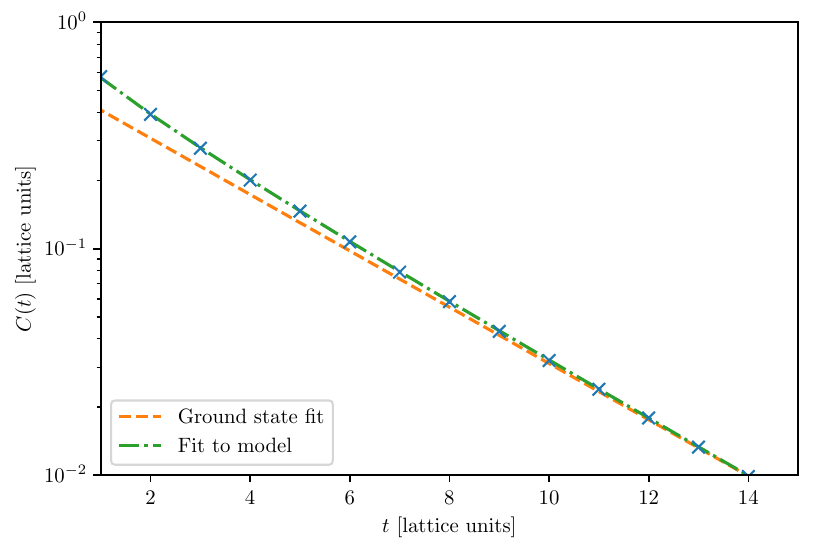}
  	\caption{Four-point correlation function $C_{ii}(\bm{0},t)$ with $A^\dagger_i(t)$ and $ A_i(0)$ inserted. Three spatial directions $i$ are averaged. The fit result with the model \eqref{equ:ModelFitFunction} is shown (dot-dashed). The fit range is given by $t_{\text{min}} = 5$ and $t_{\text{max}} = 12$. The ground-state only fit is given by a dashed line. The statistical error of the lattice data (crosses) is invisible at this scale.}
  	\label{fig:FitToCorrelatorModel}
\end{figure}

Fig.~\ref{fig:FitToCorrelatorModel} shows the four-point correlator \eqref{equ:LatticeCorrelatorRatio}, together with a fit using the model \eqref{equ:ModelFitFunction}. In large time separations, the data are consistent with a ground-state dominance, while the excited-state contribution is clearly seen at short distances and is well described by the model.

\begin{figure}[tbp]
	\centering
    	\includegraphics[width=0.6\textwidth]{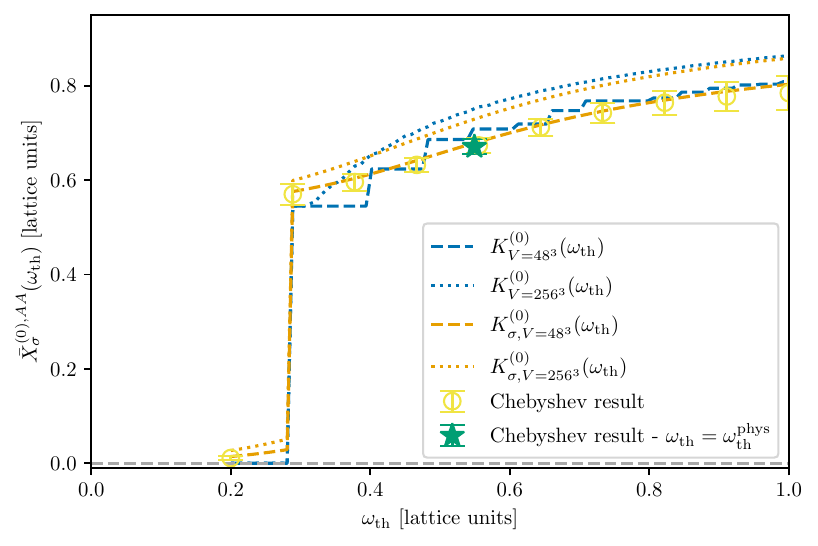}
    	\includegraphics[width=0.6\textwidth]{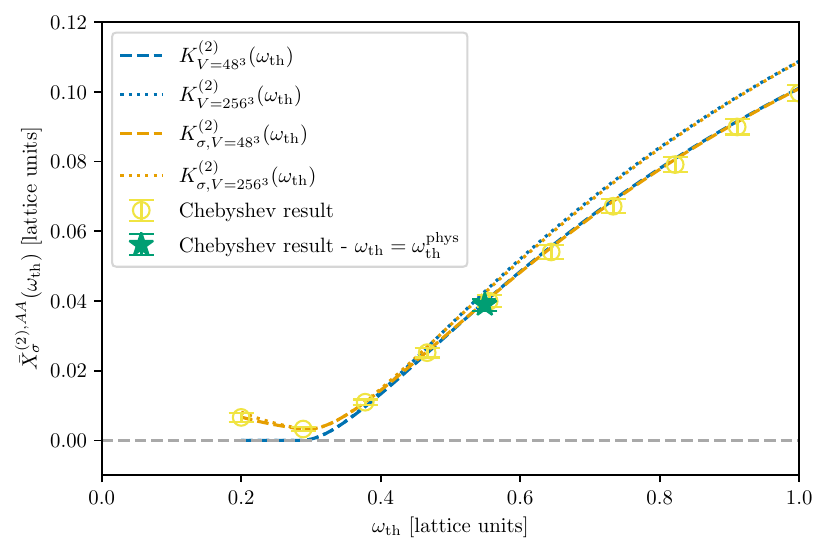}
  	\caption{$\bar{X}_{\sigma}^{(l),AA}(\omega_{\text{th}})$ as a function of $\omega_{\text{th}}$ for $l=0$ (top panel) and 2 (bottom). The estimate using the model \eqref{equ:SpectralDenVec} is shown by dashed line ($48^3$) and dotted line ($256^3$). The results with (orange) and without (blue) smearing are shown. The smearing width is $\sigma=0.1$. The inclusive analysis of the lattice data is given by circles; the physical value of $\omega_{\text{th}}$ is represented by a star.}
  	\label{fig:SpatialComponents}
\end{figure}

To estimate the finite-volume corrections for $\bar{X}_\sigma^{(l)}(\omega_\text{th})$, we consider two volumes, $V=48^3$ and $256^3$. The former corresponds to the lattice data, whereas the latter is used as a proxy for the infinite-volume limit.
The results as a function of $\omega_{\text{th}}$, the upper limit of the integral, are shown in Fig.~\ref{fig:SpatialComponents}. 
Predictions of the model with the smeared (orange) and unsmeared (blue) kernels are plotted. The smearing width is $\sigma=0.1$. The estimate for $\bar{X}_{\sigma}^{(0),AA}(\omega_{\text{th}})$ (top panel, $l=0$) jumps around $\omega_{\text{th}}$ = 0.3 due to the dominant contribution of the ground-state $\phi$ meson. Then, the lines without the smearing (blue) show a stepwise increase as $\omega_{\text{th}}$ crosses the discrete energy levels of the $K\bar{K}$ states (dashed lines). Each step becomes small for larger volumes, and the line for $256^3$ (dotted lines) already looks nearly continuous. With the smeared kernel, the steps are smeared out (orange). Similar observation is found also for $l=2$ (bottom panel).

We find that the infinite-volume limit pushes the estimate upward, {\it i.e.} from dashed ($48^3$) to dotted ($256^3$) lines. This is a consequence of the vector-meson-dominance form factor. Namely, the states close to the $\phi$ meson are highly enhanced; the lowest energy states can show up more closely to the $\phi$ meson mass in larger volumes.\footnote{We note that the $K\bar{K}$ states are always heavier than $\phi$ in our setup, where the light-quark mass is heavier than physical, and the form factor does not diverge even at zero relative momentum. When the $\phi$ meson can decay to $K\bar{K}$, we have to introduce a width to circumvent the divergence.} This enhancement is more prominent for $l=0$ in comparison to $l=2$, because the near-threshold contributions are suppressed by the kernel for $l=2$.

The inclusive analysis of the lattice data performed at different $\omega_{\text{th}}$'s are also shown in Fig.~\ref{fig:SpatialComponents}. We find a good agreement with the model estimate (dashed orange) for the smeared kernel. This is not surprising because the lattice correlator is fitted with the model nearly perfectly, as shown in Fig.~\ref{fig:FitToCorrelatorModel}. We also include the result at $\omega_{\text{th}} = \omega_{\text{th}}^{\text{phys}}$ (stars) in the plot. 

We estimate the size of the finite-volume effect from the difference between $48^3$ and $256^3$ of the model with the smeared kernel at $\omega_{\text{th}} = \omega_{\text{th}}^{\text{phys}}$. The differences are $\num{0.0504(87)}$ and $\num{0.00263(45)}$ in lattice units, or $7.4(1.3)\%$ and $6.6(1.1)\%$, for $\bar{X}_{\sigma}^{(0),AA}(\omega_{\text{th}})$ and $\bar{X}_{\sigma}^{(2),AA}(\omega_{\text{th}})$, respectively.

For nonzero recoil momenta, there exists contamination of the $P$-wave states, which are composed of three-body $K\bar{K}\pi$ states and have different finite-volume effects. Since these states are heavier and their overall contribution to the inclusive rate is subdominant, we expect their finite-volume effects to be much smaller compared to the dominant source from the two-body states. 
And since the excited states only make up a few percent of the correlator, based on the fits to extract the ground state in Sec.~\ref{sec:GroundStateEstimate},
we neglect them in this analysis. The same is applied for the finite-volume effects for vector-current insertions. 

To estimate the finite-volume correction to $\bar{X}^{AA}_{\sigma}(\bm{q}^2)$ for nonzero $\bm{q}^2$, we perform a simultaneous fit of the correlators of $A_0 A_0$, $A_0 A_\parallel$ $A_\parallel A_\parallel$ (or $A_\perp A_\perp$ and $V_\perp V_\perp$\footnote{$V_\perp V_\perp$ is included in the simultaneous fit to better constraint the ground-state $\phi$ meson, see Eqs. \eqref{equ:C_VPerpVPerp_heavy_meson_FFs_after} and \eqref{equ:C_APerpAPerp_heavy_meson_FFs_after}.}) insertions as described in Sec.~\ref{sec:GroundStateEstimate}. The fit functions contain the terms to represent the contribution of the ground-state $\phi$ meson as well as those of the possible $P$-wave states; see Eqs. \eqref{equ:C_A0A0}--\eqref{equ:C_APerpAPerp_heavy_meson_FFs_after}. The amplitude $A_0(\bm{q})$ of the $\phi$ meson pole in \eqref{equ:ModelFitFunction} is thus identified from the fit. We then introduce the model of the two-body final states, assuming that the term of $\mathcal{S}(\bm{q})$ in \eqref{equ:ModelFitFunction} exists with the same fraction $\mathcal{S}(\bm{q})/A_0(\bm{q})$ as in the zero-recoil limit $\bm{q}=\bm{0}$. We note that this two-body contribution is numerically much smaller than the contamination of $P$-wave states and does not impact the simultaneous fit significantly.

\begin{figure}[tbp]
	\centering
    	\includegraphics[width=0.6\textwidth]{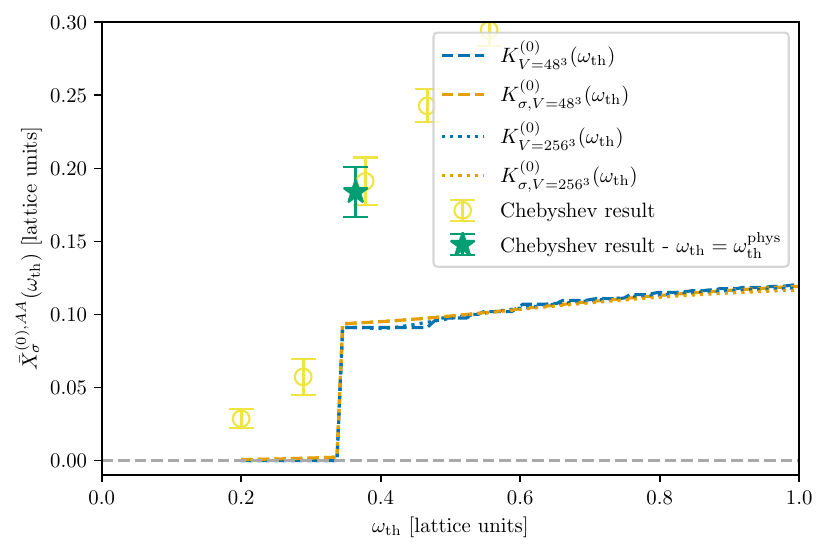}
    	\includegraphics[width=0.6\textwidth]{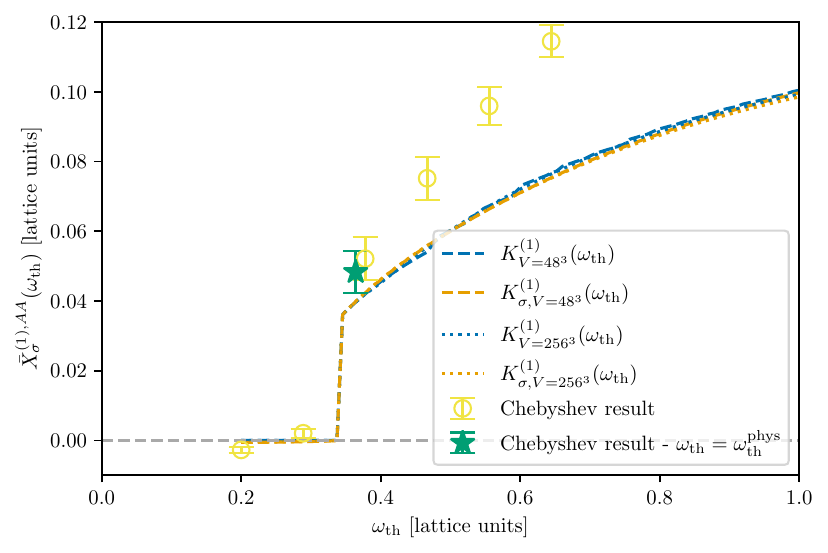}
            \includegraphics[width=0.6\textwidth]{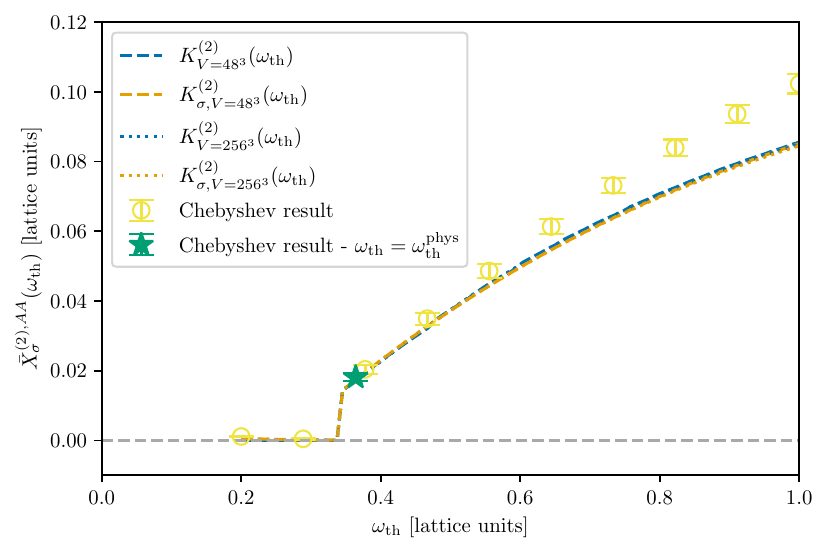}
  	\caption{$\bar{X}_{\parallel,\sigma}^{(l),AA}(\omega_{\text{th}})$ as a function of $\omega_{\text{th}}$ for $l=0$ (top panel) 1 (middle) and 2 (bottom). The momentum inserted is $\bm{q}=(0,1,1)$. Other details are the same as in Fig.~\ref{fig:SpatialComponents}.}
  	\label{fig:Modelq011}
\end{figure}

In Fig.~\ref{fig:Modelq011} we apply the method described above to $\bar{X}^{(l),AA}_{\parallel,\sigma}(\omega_{\text{th}})$ with $\bm{q} = (0,1,1)$. We show all values of $l$ = 0, 1, 2; each correlator contributes to a different $l$, {\it i.e.} $A_0 A_0$ contributes to $l=0$, $A_0 A_\parallel$ to $l=1$ and $A_\parallel A_\parallel$ to $l=2$. Unlike the case of vanishing recoil momentum $\bm{q}=\bm{0}$, Fig.~\ref{fig:SpatialComponents}, we find inconsistencies between the model (dashed lines) and data (circles). This is likely due to the $P$-wave contribution, which is not taken into account in the model. As \eqref{equ:C_A0A0}--\eqref{equ:C_APerpAPerp_heavy_meson_FFs_after} indicate, the contribution of the $\phi$ meson is suppressed by a factor of $\bm{q}^2$ or $|\bm{q}|$ for $A_0 A_0$ and $A_0 A_\parallel$, respectively, while it is the leading contribution for $A_\parallel A_\parallel$. Therefore, we expect poorer agreement between model prediction and data for $l=0,1$, where the $P$-wave contribution is relatively more significant. We therefore assume that the model describes the finite-volume effect properly, and correct the error with an uncertainty of 100\% for the correction.


\section{Determination of the inclusive rate}
\label{sec:Results}

Finally, we present the results for $\bar{X}(\bm{q}^2)$, which is the central object in the analysis of inclusive decays, as it encapsulates the energy integral over hadronic final states. We first provide the values for each component of $\bar{X}(\bm{q}^2)$ without including systematic corrections. We then perform the $V\to\infty$ and $\sigma\to 0$ limits based on the methods discussed above, before constructing the differential decay rate $\sqrt{\bm{q}^2} \bar{X}(\bm{q}^2)$, see \eqref{equ:TotalRateInclusive}, and performing the final $\bm{q}^2$ integration leading to $\Gamma/|V_{cs}|^2$.

The numerical results are given in Tables~\ref{tab:XVVParallel}, \ref{tab:XVVPerp}, \ref{tab:XAAParallel}, \ref{tab:XAAPerp} for $\bar{X}_\parallel^{VV}(\bm{q}^2)$, $\bar{X}_\perp^{VV}(\bm{q}^2)$, $\bar{X}_\parallel^{AA}(\bm{q}^2)$, $\bar{X}_\perp^{AA}(\bm{q}^2)$, respectively. For each channel, we perform a simultaneous fit of the relevant correlators including the ground- and excited-state contributions.
In each cell, corresponding to a value of $\bm{q}^2$ and $l$, the results for $\bar{X}(\bm{q}^2)$ are provided. The energy-integral kernel is treated exactly for the ground state, while the Chebyshev reconstruction is used for the excited states. The Chebyshev approximation is truncated at $N=10$ with the smearing width $\sigma=0.1$. The analysis in \cite{Barone:2023tbl} concluded that the Chebyshev approximation favors a starting point closer to the lowest-lying energy state, and we use the results obtained for $\omega_0 = 0.9\omega_{\text{min}}$.

Next, the model estimate of the finite-volume correction is given for the excited-state contribution (the number in a square bracket). As discussed in the previous section we only present results for the $AA$ contributions. Lastly, the result of the $\sigma\to 0$ extrapolation of the excited-state contribution is given, where the finite-volume correction mentioned above is included. The rightmost column lists the total results summed over $l$, before and after the corrections are applied. 

\begin{table}[tbp]
	\centering
	\begin{tabular}{c | r r r | c }
	\multirow{2}{*}{$\bm{q}^2$} & \multicolumn{3}{c|}{$\bar{X}_{\parallel}^{(l), VV}(\bm{q}^2)$} &  \multirow{2}{*}{$\sum_l \bar{X}_{\parallel}^{(l), VV}(\bm{q}^2)$} \\ \cline{2-4}
	& \multicolumn{1}{c}{$l=0$} & \multicolumn{1}{c}{$l=1$} &   \multicolumn{1}{c|}{$l=2$} &   \\
	\hline
	\multirow{3}{*}{(0,0,0)} & \multirow{3}{*}{\fcolorbox{black}{gray!20}{\shortstack[l]{Ground state \\ Excited states \\ $\sigma\to 0$ extrapolation}}} & & $0.045(19)$ &  \\
	 & & &  $+ 0.0032(35)$ & $0.048(19)$  \\
         & & &  $-0.0011(20)(4)$ & $0.044(19)$  \\
	\hline
        \multirow{3}{*}{(0,0,1)} & $0.1914(30)$ & $0.2863(43)$ & $0.1070(23)$ &   \\
	& $+ 0.0101(18)$ & $+ 0.0179(50)$ & $+ 0.0152(67)$ & $0.628(12)$   \\
        & $+ 0.0070(47)(25)$ & $+0.023(10)(5)$ & $+0.019(11)(5)$ & $0.633(20)$   \\
	\hline
        \multirow{3}{*}{(0,1,1)} & $0.2862(50)$ & $0.3406(63)$ & $0.1013(30)$ &  \\
	& $+ 0.0155(22)$ & $+ 0.0191(58)$ & $+ 0.0085(49)$ & $0.771(13)$ \\
        & $-0.0002(95)(60)$ & $+0.008(20)(12)$ & $-0.005(16)(10)$ & $0.731(35)$ \\
	\hline
        \multirow{3}{*}{(1,1,1)}& $0.3341(69)$ & $0.3296(85)$ &  $0.0813(34)$ &    \\
	& $+ 0.0191(27)$ & $+ 0.0134(54)$ & $+ 0.0062(33)$ & $0.784(14)$   \\
        & $-0.002(12)(9)$ & $-0.007(22)(18)$ & $-0.004(15)(13)$ & $0.731(40)$    \\
	\hline
	\end{tabular}
    \caption{Numerical results for $\bar{X}_{\parallel}^{VV}(\bm{q}^2)$ in $\text{GeV}^2$ for all $\bm{q}^2$ used in Figs. \ref{fig:ContributionsParPerpGSExact} and \ref{fig:ContributionsParPerpErrorwFit}. They are divided into $\bar{X}_{\parallel}^{(l),VV}(\bm{q}^2)$ of $l$ = 0, 1, and 2. In each cell, the ground-state and excited-state contributions are shown in the first and second line, respectively. The excited states are reconstructed using the Chebyshev approximation at finite smearing $\sigma =0.1$. The third line contains the excited-state contribution after the $\sigma \to 0$ extrapolation, where the first error is statistical and the second error is the expected correction from \eqref{equ:ErrorEstimator}. The last column lists the sum over all $l$. Empty cells mean that the corresponding contribution is zero.}
	\label{tab:XVVParallel}
\end{table}
\begin{table}[tbp]
	\centering
	\begin{tabular}{c | r r | c }
	\multirow{2}{*}{$\bm{q}^2$} & \multicolumn{2}{c|}{$\bar{X}_{\perp}^{(l), VV}(\bm{q}^2)$} &  \multirow{2}{*}{$\sum_l \bar{X}_{\perp}^{(l), VV}(\bm{q}^2)$} \\ \cline{2-3}
	& \multicolumn{1}{c}{$l=0$} &  \multicolumn{1}{c|}{$l=2$} &   \\
	\hline
	\multirow{3}{*}{(0,0,0)} & \multirow{3}{*}{\fcolorbox{black}{gray!20}{\shortstack[l]{Ground state \\ Excited states \\ $\sigma\to 0$ extrapolation}}} & $0.089(38)$ &  \\
	 & &  $+ 0.0064(71)$ & $0.096(38)$  \\
      & &  $-0.0022(40)(7)$ & $0.087(38)$  \\
	\hline
	\multirow{3}{*}{(0,0,1)} & $-0.0251(57)$ & $0.081(18)$ &    \\
	& $-0.0434(49)$ & $+ 0.036(15)$ & $0.049(12)$  \\
        & $-0.033(16)(10)$ & $+0.047(27)(11)$ & $0.069(28)$   \\
	\hline
	\multirow{3}{*}{(0,1,1)} & $-0.043(13)$ & $0.056(17)$ & \\
	& $-0.0464(74)$ & $+ 0.0154(80)$ & $-0.0185(68)$ \\
        & $-0.005(20)(13)$ & $-0.007(19)(14)$ & $0.0004(324)$ \\
	\hline
	\multirow{3}{*}{(1,1,1)}& $0(0)$ &  $0(0)$ &    \\
	& $-0.0453(85)$ & $+ 0.0062(40)$ & $-0.0391(69)$  \\
        & $+0.0021(63)(39)$ & $-0.0020(59)(37)$ & $0.0001(98)$   \\
	\hline
	\end{tabular}
    \caption{Same as Tab. \ref{tab:XVVParallel} but for $\bar{X}_{\perp}^{VV}(\bm{q}^2)$. $l=1$ does not contribute and is hence omitted.}
	\label{tab:XVVPerp}
\end{table}
\begin{table}[tbp]
	\centering
	\begin{tabular}{c | r r r | c }
	\multirow{2}{*}{$\bm{q}^2$} & \multicolumn{3}{c|}{$\bar{X}_{\parallel}^{(l), AA}(\bm{q}^2)$} &  \multirow{2}{*}{$\sum_l \bar{X}_{\parallel}^{(l), AA}(\bm{q}^2)$} \\ \cline{2-4}
	& \multicolumn{1}{c}{$l=0$} & \multicolumn{1}{c}{$l=1$} & \multicolumn{1}{c||}{$l=2$} &   \\
	\hline
	\multirow{4}{*}{(0,0,0)} & \multirow{4}{*}{\fcolorbox{black}{gray!20}{\shortstack[l]{Ground state \\ Excited states \\ {[}infinite-volume limit{]} \\ $\sigma\to 0$ extrapolation}}} & & $0.502(27)$ &  \\
         & & &  $+ 0.015(11)$ & $0.518(23)$ \\
	 & & &  $[+ 0.034(34)]$ & \multirow{2}{*}{$0.553(40)$}  \\
	 & & &  $+ 0.0160(110)(7)$ &   \\
	\hline
	\multirow{4}{*}{(0,0,1)} & $0.0210(52)$ & $0.167(24)$ & $0.332(13)$ &    \\
        & $+ 0.0256(39)$ & $+ 0.027(13)$ & $+ 0.0016(56)$ & $0.574(24)$  \\
	& $[+ 0.0003(3)]$ & $[+ 0.002(2)]$ & $[+ 0.0034(34)]$ & \multirow{2}{*}{$0.579(25)$}   \\
	& $+ 0.0214(77)(43)$ & $+ 0.035(19)(4)$ & $-0.0038(96)(39)$ &    \\
	\hline
	\multirow{4}{*}{(0,1,1)} & $0.040(18)$ & $0.190(55)$ & $0.228(26)$ & \\
        & $+ 0.044(10)$ & $+ 0.049(24)$ & $+ 0.015(10)$ & $0.566(60)$  \\
	& $[+ 0.000(0)]$ & $[+ 0.000(0)]$ & $[+ 0.000(0)]$ & \multirow{2}{*}{$0.567(61)$}  \\
	& $0.034(17)(8)$ & $+ 0.060(30)(11)$ & $+ 0.016(17)(7)$ &    \\
	\hline
	\multirow{4}{*}{(1,1,1)} & $0(0)$ & $0(0)$ & $0(0)$ & \\
        & $+ 0.0575(65)$ & $+ 0.060(19)$ & $+ 0.0215(92)$ & $0.139(34)$ \\
	& $[+ 0.0000(0)]$ & $[+ 0.000(0)]$ & $[- 0.0001(1)]$ & \multirow{2}{*}{$0.008(14)$}  \\
	& $+ 0.0030(54)(32)$ & $+ 0.0032(88)(53)$ & $+ 0.0022(52)(30)$ &    \\
	\hline
	\end{tabular}
    \caption{Same as Tab. \ref{tab:XVVParallel} but for $\bar{X}_{\parallel}^{AA}(\bm{q}^2)$. We additionally include the corrections from the infinite-volume limit in the third line of each cell which is added to the final value in addition to the $\sigma\to 0$ extrapolation result.}
	\label{tab:XAAParallel}
\end{table}
\begin{table}[tbp]
	\centering
	\begin{tabular}{c | r r | c }
	\multirow{2}{*}{$\bm{q}^2$} & \multicolumn{2}{c|}{$\bar{X}_{\perp}^{(l), AA}(\bm{q}^2)$} &  \multirow{2}{*}{$\sum_l \bar{X}_{\perp}^{(l), AA}(\bm{q}^2)$} \\ \cline{2-3}
	& \multicolumn{1}{c}{$l=0$} &  \multicolumn{1}{c|}{$l=2$} &   \\
	\hline
	\multirow{4}{*}{(0,0,0)} & \multirow{4}{*}{\fcolorbox{black}{gray!20}{\shortstack[l]{Ground state \\ Excited states \\ {[}infinite-volume limit{]} \\ $\sigma\to 0$ extrapolation}}} & $1.004(53)$ & \\
        & &  $+ 0.031(22)$ & $1.035(46)$  \\
	& &  $[+ 0.069(69)]$ & \multirow{2}{*}{$1.105(80)$} \\
	  & &  $+ 0.032(23)(1)$ &   \\
	\hline
	\multirow{3}{*}{(0,0,1)} & $-0.188(12)$ & $0.604(39)$ &    \\
        & $-0.0384(80)$ & $+ 0.034(14)$ & $0.412(28)$ \\
	& $[-0.0020(20)]$ & $[+ 0.005(5)]$ & \multirow{2}{*}{$0.435(37)$} \\
	& $-0.024(17)(9)$ & $+ 0.039(26)(10)$ & \\
	\hline
	\multirow{3}{*}{(0,1,1)} & $-0.210(19)$ & $0.270(24)$ & \\
        & $-0.0452(93)$ & $+ 0.0215(83)$ & $0.037(10)$ \\
	& $[-0.0001(1)]$ & $[+0.0001(1)]$ & \multirow{2}{*}{$0.063(33)$ } \\
	& $-0.011(20)(13)$ & $+0.013(21)(14)$ & \\
	\hline
	\multirow{3}{*}{(1,1,1)}& $0(0)$ & $0(0)$ &   \\
        & $-0.044(10)$ & $+ 0.0062(50)$ & $-0.0380(85)$ \\
	& $[-0.0005(5)]$ & $[-0.0002(2)]$ & \multirow{2}{*}{$-0.0005(108)$} \\
	& $-0.0039(66)(41)$ & $-0.0017(66)(39)$ & \\
	\hline
	\end{tabular}
    \caption{Same as Tab. \ref{tab:XAAParallel} but for $\bar{X}_{\perp}^{AA}(\bm{q}^2)$.}
	\label{tab:XAAPerp}
\end{table}

\begin{figure}[tbp]
	\centering
	\includegraphics[width=0.8\textwidth]{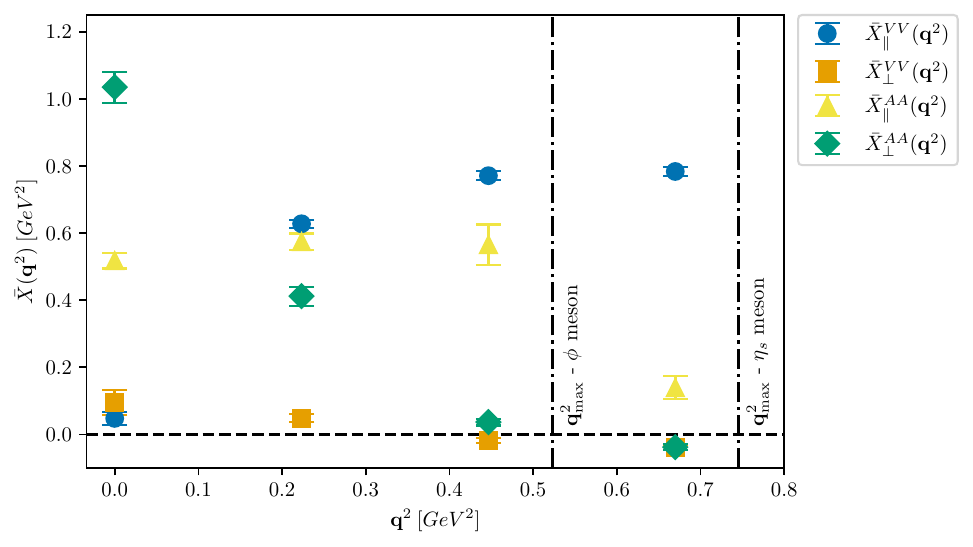}
	\caption{Contributions to the total $\bar{X}(\bm{q}^2)$ decomposed into longitudinal and transverse components of the channels $V_\mu^\dagger V_\mu$ and $A_\mu^\dagger A_\mu$. The results are before the finite-volume and finite $\sigma$ corrections are applied. The black dash-dotted lines represent the kinematical thresholds for the vector and pseudoscalar mesons.}
	\label{fig:ContributionsParPerpGSExact}
\end{figure}

Contributions to $\bar{X}(\bm{q}^2)$, the differential inclusive rate divided by $\sqrt{\bm{q}^2}$, from each channel are shown in Fig.~\ref{fig:ContributionsParPerpGSExact}. Here, the results are those before the finite-volume and $\sigma$ corrections. 
In the plot, we include the kinematical threshold $\bm{q}_{\text{max}, i}^2 = (m_{D_s}^2 - m_{i}^2)^2/(4m_{D_s}^2)$ with $i = \eta_s$ or $\phi$ depending on the lightest final state. With a finite $\sigma$ (or $N$), we find inconsistencies remaining, {\it e.g.} $\bar{X}^{AA}_\parallel(\bm{q}^2)$ above the $\phi$ threshold being nonzero or $\bar{X}^{AA}_\perp(\bm{q}^2)$ being negative.

\begin{figure}[tbp]
	\centering
	\includegraphics[width=0.8\textwidth]{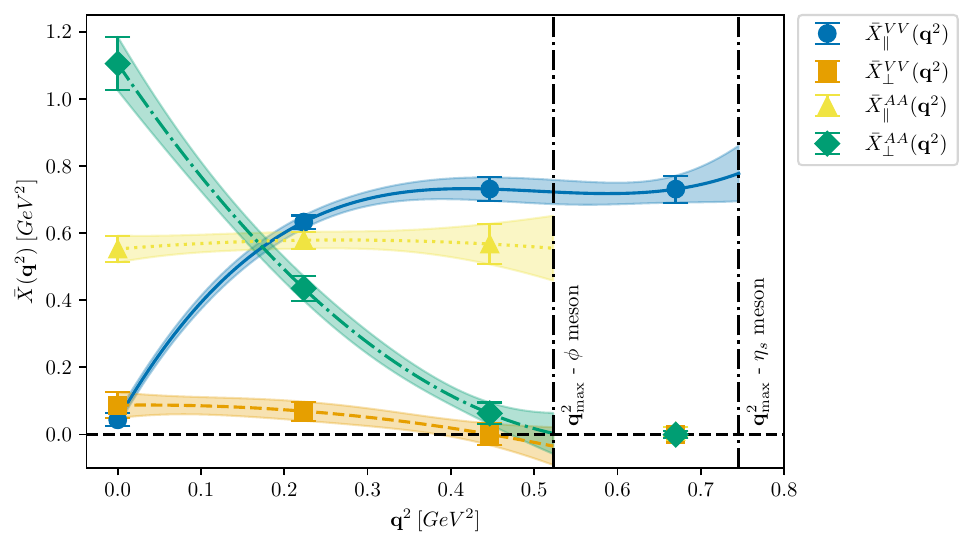}
	\caption{Contributions to the total $\bar{X}(\bm{q}^2)$ decomposed into longitudinal and transverse components of the channels $V_\mu^\dagger V_\mu$ and $A_\mu^\dagger A_\mu$ after performing the infinite-volume limit (for $AA$) and the smearing to zero limit. We also include polynomial fits to each contribution.}
	\label{fig:ContributionsParPerpErrorwFit}
\end{figure}

The result of the $\sigma\to 0$ extrapolation is shown in Fig.~\ref{fig:ContributionsParPerpErrorwFit}. The infinite-volume limit is taken for the axial-vector channel using the model for two-body final states. 
For the vector channel, the finite-volume effects are expected to be negligible and only the $\sigma\to 0$ limit is performed. 

\begin{figure}[tbp]
	\centering
	\includegraphics[width=0.8\textwidth]{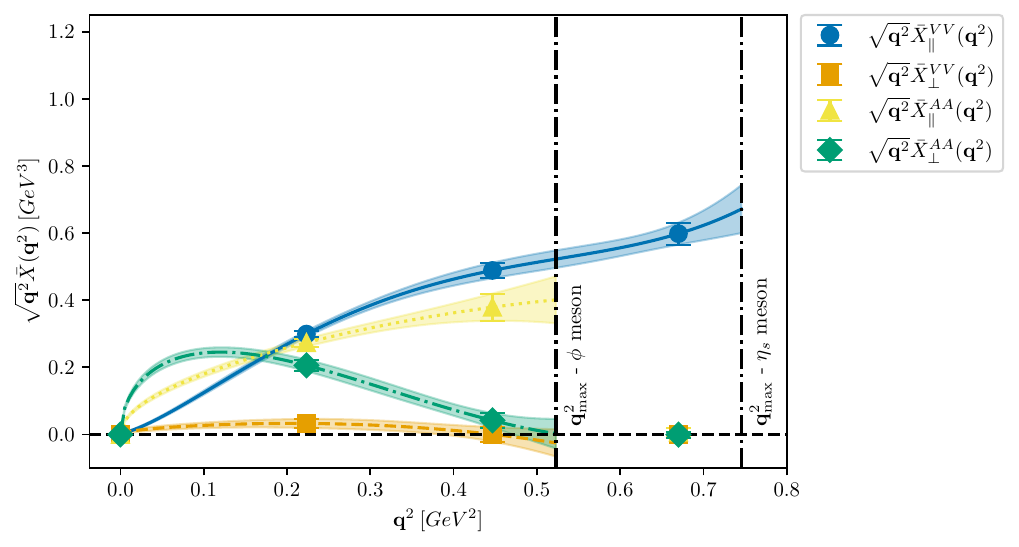}
	\caption{Differential decay rate $\sqrt{\bm{q}^2} \bar{X}(\bm{q}^2)$ constituting the integrand in \eqref{equ:TotalRateInclusive}. We retain the decomposition into longitudinal and transverse components of the channels $V_\mu^\dagger V_\mu$ and $A_\mu^\dagger A_\mu$ after performing the infinite-volume limit and the smearing to zero limit.}
	\label{fig:CombContributionsParPerpErrorwFit}
\end{figure}

We also include a polynomial interpolation of the $\bm{q}^2$ dependence for each channel. The order of the polynomial is given by the number of points available, {\it i.e.} up to $(\bm{q}^2)^3$ for $\bar{X}_\parallel^{VV}(\bm{q}^2)$ or up to $(\bm{q}^2)^2$ for others because the point $\bm{q}=(1,1,1)$ is above the kinematical threshold. To reconstruct the differential decay rate, see \eqref{equ:TotalRateInclusive}, the factor $\sqrt{\bm{q}^2}$ is multiplied and the resulting curves are shown in Fig.~\ref{fig:CombContributionsParPerpErrorwFit}. 

\begin{table}[tbp]
	\centering
	\begin{tabular}{c | l }
	& $\frac{\Gamma}{|V_{cs}|^2} \times 10^{13}$ \\
	\hline
	$\bar{X}^{VV}_{\parallel}(\bm{q}^2)$ & $0.495(16)$ \\
	$\bar{X}^{VV}_{\perp}(\bm{q}^2)$ & $0.0162(87)$ \\
	$\bar{X}^{AA}_{\parallel}(\bm{q}^2)$ & $0.245(16)$ \\
	$\bar{X}^{AA}_{\perp}(\bm{q}^2)$ & $0.130(10)$ \\
    \hline
    total & $0.886(27)$
	\end{tabular}
	\caption{Results of the $\bm{q}^2$ integral for $\Gamma/|V_{cs}|^2 = G_F^2/(24\pi^3)\int_{0}^{\bm{q}^2_{\text{max}}} d\bm{q}^2 \sqrt{\bm{q}^2} \bar{X}(\bm{q}^2)$, in $\text{GeV}$, for individual channels. Statistical and systematic errors are added in quadrature. The renormalization constant $Z_V$ is multiplied for the integrated values.}
	\label{tab:XBarIntegral}
\end{table}

The $\bm{q}^2$-integration for each contribution is performed analytically, and the total decay rate is constructed as the sum of the individual contributions. The integration results for the individual channels are given in Table \ref{tab:XBarIntegral}. The values given in the Table include the renormalization constant. We find that the error for the largest channel, $VV\parallel$, is about 3\% including the uncertainty of infinite volume and $\sigma\to 0$ extrapolations. This uncertainty turns out to be less than that for $\bar{X}_\parallel^{VV}(\bm{q}^2)$ at each value of $\bm{q}^2$, which is about 5\% (see Table~\ref{tab:XVVParallel}), because the statistical correlation among different $\bm{q}^2$ is not strong. The dominant source of systematic uncertainty is the $\sigma\to 0$ extrapolation, which is estimated by assuming the unknown higher-order Chebyshev matrix elements to be within $\pm 1$, and there is no reason why they would have significant correlations among different $\bm{q}^2$.

For the inclusive decay rate, we obtain $\Gamma/|V_{cs}|^2 = \SI{0.886(27)e-13}{\giga\electronvolt}$, where the statistical and systematic errors are added in quadrature. This result is obtained from a single lattice ensemble, and the continuum and chiral extrapolations are yet to be performed. It is also important to mention that the disconnected diagrams are not included in our calculation, and final states including $\eta$ and $\eta'$ may affect the final results. Another potential issue is that the $\phi$ meson is stable for our simulation parameters; the finite-volume effects may therefore behave differently once the $\phi\to K\bar{K}$ decays are possible. Given these caveats, though, a comparison with the experimental data from the BESIII collaboration \cite{BESIII:2021duu}, $\Gamma=\SI{0.827(22)e-13}{\giga\electronvolt}$, indicates that the result for $|V_{cs}|$ determined from this inclusive decay rate is in the expected region. We expect that both experimental and lattice uncertainties will improve in the near future.

\section{Conclusions and Outlook}
\label{sec:ConclusionOutlook}

We performed a lattice QCD analysis of the inclusive semileptonic decay of the $D_s$ meson. We build on our previous work \cite{Barone:2023tbl} using the Chebyshev-polynomial approximation for the energy integral; in this work we focus on systematic errors, especially the finite-volume effects as well as the error associated with the finite polynomial approximation, as well as the smearing of the kernel. We have introduced a model of two-body states and combined it with the lattice data to study the infinite-volume extrapolation. We found that our model predicts minor corrections from finite-volume effects.

As for the approximation error due to the smearing and truncation, we utilized the properties of the Chebyshev polynomials to combine the two limits and estimate their effects. In particular, the uncertainty due to the truncation of the polynomial is under control by using the mathematical constraint on the corresponding matrix elements as well as the analytically known coefficients. Larger uncertainties are expected for large $\bm{q}^2$ due to the limited phase space, and we resolve the issue by extracting the ground-state contribution from the correlator and applying the Chebyshev approximation only for the excited-state contributions. The ground state, which becomes dominant in this kinematical regime, is treated exactly under the energy integral. 

The work presented here constitutes a step towards a fully controlled lattice prediction for the inclusive decay rate, although some issues still remain unresolved, such as the discretization effects and the chiral extrapolation. In the future, we are planning to include additional ensembles to perform both extrapolations. The work on the $B_{(s)}$ decay, along the direction of \cite{Barone:2023tbl}, is also underway.
Furthermore, the analysis can be extended to other observables, such as lepton-energy and $q^2$ moments, which can then be used to compare with experiments or with the OPE-based calculations, see {\it e.g.} \cite{Gambino:2022dvu}.  Another direction may be to use the four-point correlation functions to study the processes involving $P$-wave final states, as attempted in \cite{Hu:2025hpn}.

\acknowledgments
The numerical calculations of the JLQCD collaboration were performed on SX-Aurora TSUBASA at the High Energy Accelerator Research Organization (KEK) under its Particle, Nuclear and Astrophysics Simulation Program, as well as on Fugaku through the HPCI System Research Project (Project ID: hp220056). The works of S.H. and T.K. are supported in part by JSPS KAKENHI Grant Numbers 22H00138 and 23K20846, respectively, and by the Post-K and Fugaku supercomputer project through the Joint Institute for Computational Fundamental Science (JICFuS). T.K. is also supported by the U.S.-Japan Science and Technology Cooperation Program in High Energy Physics (Project ID: 2024-40-2). R.K. is supported in part by JSPS KAKENHI Grant Number 22K21347.


\appendix

\section{Chebyshev Polynomials}
\label{sec:AppChebyshev}

This section focuses on the introduction and discussion of important properties of the standard Chebyshev polynomials relevant to this work. Furthermore, we put a special emphasis on the generalization for the shifted Chebyshev polynomials which play an integral role in the results presented in this work. Nonetheless, we are not able to give a full review on this topic and refer to \cite{RevModPhys.78.275} for more details.

\subsection{Standard Polynomials}
\label{sec:AppStandardChebyshev}

First, let us start by giving the definition of the first kind of Chebyshev polynomials
\begin{align}
	T_k: [-1,1] \rightarrow [-1,1] \, , \quad T_k(x) = \cos \left(k\cos^{-1}(x)\right) \, , \quad k\in \mathbb{N} \, .
\end{align}
By definition, they are orthogonal with respect to the scalar product
\begin{align}
	\int_{-1}^{1} T_r(x)T_s(x)\Omega(x) dx = \frac{\pi}{2} \delta_{rs} (1+\delta_{r0}) \, ,
	\label{equ:OrthogonalityRelation}
\end{align}
where $\Omega(x) = 1/\sqrt{1-x^2}$ defines a weight function. In terms of $x^k$, the polynomial expansion is defined as
\begin{align}
	T_n(x) = \sum_{k=0}^{n} t_{k}^{(n)} x^k \, ,
	\label{equ:ChebyshevExpansion}
\end{align}
where the definition of the coefficients $t_{k}^{(n)}$ is given by
\begin{equation}
	\begin{alignedat}{2}
		t_{0}^{(n)} &= (-1)^{n/2} &&\quad \text{if $n$ even} \, , \\
		t_{k}^{(n)} &= 0 &&\quad \text{if $n-k$ odd} \, , \\
		t_{k}^{(n)} &= (-1)^{(n-k)/2} 2^{k-1} \frac{n}{\frac{n+k}{2}} \binom{\frac{n+k}{2}}{\frac{n-k}{2}} &&\quad \text{if $k\neq 0$ and $n-k$ even} \, .
	\end{alignedat}
	\label{equ:tCoefficientsChebyshev}
\end{equation}

Another useful property involves the reverse formula, i.e. the representation of $x^{k}$ in terms of the standard Chebyshev polynomials
\begin{align}
	\begin{split}
		p_n(x) \equiv x^{n} &= 2^{1-n} \left[ \frac{1}{2} \binom{n}{\frac{n}{2}} T_0(x) + \sum_{\substack{k=1 \\ n-k \: \text{even}}}^{n} \binom{n}{\frac{n-k}{2}} T_{k}(x)\right] \\
		&= \sideset{}{'}\sum_{\substack{k=0 \\ n-k \: \text{even}}}^{n} \binom{n}{\frac{n-k}{2}} T_{k}(x) \, ,
	\end{split}
	\label{equ:InverseFormula}
\end{align}
where in the second line we absorb the first term into the sum and highlight the fact that it has to be halved by the prime, unless it is skipped.

\subsubsection{Approximation using Chebyshev expansion}
\label{sec:AppApproximationChebyshev}

For a function $f: [-1,1] \rightarrow \mathbb{R}$, the Chebyshev polynomials provide a near-\textit{minmax} approximation for any given order $N$ of the approximation.
A convenient feature for the functions considered in this work is the fact that the Chebyshev approximation is guaranteed to converge for the $n \rightarrow \infty$ limit. The actual application of the Chebyshev polynomial approximation is given through
\begin{align}
	f(x) \simeq \frac{1}{2} c_0 T_0(x) + \sum_{j=1}^{N} c_j T_j(x) \, , \quad c_j = \frac{2}{\pi} \int_{-1}^{1} dx f(x) T_j(x)\Omega(x) \, ,
	\label{equ:AppChebyshevCoeffs}
\end{align}
where $T_0(x) = 1$ by definition. Furthermore, as can be seen in their definition, the coefficients are given by the projection of the target function $f$ onto the basis of Chebyshev polynomials.

\subsection{Shifted Chebyshev polynomials}
\label{sec:AppShiftedChebyshev}

In this work, instead of the standard Chebyshev polynomial approach discussed previously, we consider generic functions $f(\omega)$ defined in an interval $[a,b]$, which we approximate through Chebyshev polynomials in $x = \exp(-\omega)$. This requires us to define shifted polynomials $\tilde{T}_j(x)$ with $x \in [\exp(-a),\exp(-b)]$, so that their domain and the one of the target function matches. The standard polynomials and their shifted variant are related through
\begin{align}
	\tilde{T}_j(x) = T_j(h(x)) \, .
\end{align}
Here, $h : [\exp(-a),\exp(-b)] \rightarrow [-1,1]$ is an invertible function which maps the domain $[\exp(-a),\exp(-b)]$ onto the domain where the standard Chebyshev polynomials are defined
\begin{align}
	h(x) = Ax + B \, .
	\label{equ:MappingFunction}
\end{align} 
We now impose the conditions $h(\exp(-a))=-1$ and $h(\exp(-b)) = 1$ to obtain expression for the coefficients $A$ and $B$
\begin{align}
	A = - \frac{2}{e^{-a} - e^{-b}} \, , \quad B = \frac{e^{-a} + e^{-b}}{e^{-a} - e^{-b}} \, .
	\label{equ:GeneralCoefficients}
\end{align}

Furthermore, the orthogonality relation in \eqref{equ:OrthogonalityRelation} is now given by
\begin{align}
	\int_a^b dx \tilde{T}_r(x) \tilde{T}_s(x) \Omega_h(x) = \int_a^b dx T_r(h(x)) T_s(h(x)) \Omega_h(x) \, ,
	\label{equ:NewOrthogonality}
\end{align}
where the map dependent weight for the shifted $\tilde{T}_k(x)$ is given by $\Omega_h(x)$. The original integral \eqref{equ:OrthogonalityRelation} can be recovered by setting $x = h^{-1}(y)$ and $dx = 1/h'(h^{-1}(y)) dy$, so that the integral \eqref{equ:NewOrthogonality} becomes
\begin{align}
	\int_{h(a)}^{h(b)} dy T_r(y) T_s(y) \frac{\Omega_h(h^{-1}(y))}{h'(h^{-1}(y))} \, .
\end{align}
To continue, we write the weight as
\begin{align}
	\Omega_h(x) = \Omega(h(x))|h'(x)| \, ,
\end{align}
so that we finally arrive at
\begin{align}
	\int_a^b dx \tilde{T}_r(x) \tilde{T}_s(x) \Omega_h(x) = \int_{-1}^{1} dy T_r(y) T_s(y) \Omega(y) \, .
\end{align}

The next step is to generalize the polynomial expression and their properties. The generalized polynomial expression of \eqref{equ:ChebyshevExpansion} can be written as
\begin{align}
	\tilde{T}_n (x) = \sum_{j=0}^{n} t_j^{(n)} h(x)^j = \sum_{j=0}^{n} t_j^{(n)} (Ax + B)^j = \sum_{j=0}^{n} t_j^{(n)} \sum_{k=0}^{j} \binom{j}{k} A^k B^{j-k} x^{k}
\end{align}

We can furthermore isolate the coefficients of $\exp(-kx)$ by expanding this sum explicitly and re-summing
\begin{align}
	\begin{split}
		\tilde{T}_n(x) &= A^n x^n \left[\binom{n}{n} t_n^{(n)}\right] + A^{n-1} x^{n-1}\left[ \binom{n-1}{n-1} t_{n-1}^{(n)} + \binom{n}{n-1} t_n^{(n)} B^1 \right] \\
		&+ A^{n-2} x^{n-2} \left[\binom{n-2}{n-2} t_{n-2}^{(n)} + \binom{n-1}{n-2} t_{n-1}^{(n)} B + \binom{n}{n-2} t_{n}^{(n)} B^2\right] + \cdots \\
		&+ A^1 x^1 \left[ \binom{1}{1} t_1^{(n)} + \binom{2}{1} t_2^{(n)} B + \binom{3}{1} t_3^{(n)} B^2 + \cdots + \binom{n}{1} t_n^{(n)} B^{n-1}\right] \\
		&+ \left[\binom{0}{0} t_0^{(n)} + \binom{1}{0} t_1^{(n)} B + \cdots + \binom{n}{0} t_n^{(n)} B^n \right] \\
		&= \sum_{k=0}^{n} x^k A^k \sum_{j=k}^{n} \binom{j}{k} t_j^{(n)} B^{j-k} \\
		&= \sum_{k=0}^{n} \tilde{t}_k^{(n)} x^k \, ,
	\end{split}
	\label{equ:ExpandedSumChebyshev}
\end{align}
where in the last step we introduce the short-hand notation for the coefficients $\tilde{t}_k^{(n)}$ given by
\begin{align}
	\tilde{t}_k^{(n)} = A^k \sum_{j=k}^{n} \binom{j}{k} t_{j}^{(n)} B^{j-k} = \left(\frac{A}{B}\right)^k \sum_{j=k}^{n} \binom{j}{k} t_j^{(n)} B^j \, .
	\label{equ:CoefficientExpandedSum}
\end{align}
In a similar way, we can also generalize the inverse formula defined in \eqref{equ:InverseFormula}
\begin{align}
	\tilde{p}_n(x) \equiv h(x)^n = 2^{1-2n} \sideset{}{'}\sum_{\substack{j=0 \\ n-j \: \text{even}}}^{n} \binom{n}{\frac{n-j}{2}} \tilde{T}_j(x) \, , \quad x \in [e^{-a},e^{-b}] \, ,
	\label{equ:AppInverseFormulaGeneral}
\end{align}
where, once again, the prime denotes that the term at $j=0$ has to be halved.

We may also define an iterative expression through which we can generally determine expressions for $x^n$. For this, we assume $\tilde{p}_0 (x) = 1$ and
\begin{align}
	\tilde{p}_n (x) = \left(Ax + B\right)^n = \sum_{k=0}^{n} \binom{n}{k} A^{k} B^{n-k} x^k \, ,
\end{align}
so that our expression for $x^n$ reads
\begin{align}
	x^n = \frac{1}{A^n} \left[ \tilde{p}_n(x) - \sum_{k=0}^{n-1} \binom{n}{k} A^{k} B^{n-k} x^k \right] \, .
\end{align}

As a final step, we rewrite everything in terms of the shifted Chebyshev polynomials by collecting all the numerical coefficients
\begin{align}
	x^n = \sum_{j=0}^{n} \tilde{a}_j^{(n)} \tilde{T}_j(x) \, ,
	\label{equ:InverseRepresentation}
\end{align}
where we introduce a set of coefficients $\{\tilde{a}_0^{(n)}, \tilde{a}_1^{(n)}, \cdots, \tilde{a}_n^{(n)} \}$, which can be straightforwardly determined numerically for each value of $n$. We will give some additional information on these coefficients in Sec. \ref{sec:AppMatrixRelations}

\subsubsection{Chebyshev expansion with exponential map}
\label{sec:AppApproxChebyMap}

Following the previous section, we now have all necessary tools to formulate the polynomial approximation of a generic function in $x=\exp(-\omega)$, similar to Sec. \ref{sec:AppApproximationChebyshev}. To stay true to the application in this work, we will only consider the case of $f : [\omega_0, \infty) \rightarrow \mathbb{R}$, for which the approximation reads
\begin{align}
	f(\omega)  \simeq \frac{1}{2} \tilde{c}_0 \tilde{T}_0(x) + \sum_{k=1}^{N} \tilde{c}_k \tilde{T}_k(x) \, , \quad \tilde{c}_k = \frac{2}{\pi} \int_{\omega_0}^{\infty} dx f(\omega) \tilde{T}_k(x) \Omega_h(x) \, .
\end{align}
In a more explicit way, the coefficients can be rewritten as
\begin{align}
	\tilde{c}_k = \frac{2}{\pi} \int_{\omega_0}^{\infty} d\theta f(h^{-1}(\cos\theta)) (\cos k\theta) = \frac{2}{\pi} \int_{\omega_0}^{\infty} d\theta f\left(-\text{ln}\left(\frac{\cos\theta - B}{A}\right)\right) (\cos k\theta) \label{equ:ChebyshevCoeffDefinition} \, ,
\end{align}
where in the second step we set $y = h(x)$, so that through inversion
\begin{align}
	x = h^{-1}(y) = -\log\left(\frac{y - B}{A}\right) \, .
\end{align}

For this specific choice of the limits $[\omega_0, \infty)$ the coefficients defined in \eqref{equ:GeneralCoefficients} read
\begin{align}
	A = -2e^{\omega_0} \, , \quad B = 1
	\label{equ:CoefficientsDomain}
\end{align}

\subsubsection{Matrix relations}
\label{sec:AppMatrixRelations}

As previously discussed, in this work we only consider the case in which the domain of the target function is given by $[\omega_0, \infty)$. For this case we have given the definition of the coefficients $A$ and $B$ in \eqref{equ:CoefficientsDomain}. We will now discuss some useful properties that arise by setting $\omega_0 \neq 0$ and then apply those to the case relevant for this work, although a generalization is trivial.

Towards this end, let us start by expressing \eqref{equ:ExpandedSumChebyshev} in matrix notation
\begin{align}
	\begin{pmatrix}
		\tilde{T}_0(x) \\
		\tilde{T}_1(x) \\
		\vdots \\
		\vdots \\
		\tilde{T}_n(x)
	\end{pmatrix} 
	=
	\begin{pmatrix}
		\tilde{t}_0^{(0)} & 0 & \cdots & \cdots & 0 \\
		\tilde{t}_0^{(1)} & \tilde{t}_1^{(1)} & 0 & \cdots & 0 \\
		\vdots & \vdots & \ddots & \ddots & \vdots \\
		\vdots & \vdots & & \ddots & 0 \\
		\tilde{t}_0^{(n)} & \tilde{t}_1^{(n)} & \cdots & \cdots & 0
	\end{pmatrix}
	\begin{pmatrix}
		1 \\
		x \\
		\vdots \\
		\vdots \\
		x^n
	\end{pmatrix}
	\, .
\end{align}
Additionally, the same representation can be written for \eqref{equ:InverseRepresentation}
\begin{align}
	\begin{pmatrix}
		1 \\
		x \\
		\vdots \\
		\vdots \\
		x^n
	\end{pmatrix}
	=
	\begin{pmatrix}
		\tilde{a}_0^{(0)} & 0 & \cdots & \cdots & 0 \\
		\tilde{a}_0^{(1)} & \tilde{a}_1^{(1)} & 0 & \cdots & 0 \\
		\vdots & \vdots & \ddots & \ddots & \vdots \\
		\vdots & \vdots & & \ddots & 0 \\
		\tilde{a}_0^{(n)} & \tilde{a}_1^{(n)} & \cdots & \cdots & 0
	\end{pmatrix}
	\begin{pmatrix}
		\tilde{T}_0(x) \\
		\tilde{T}_1(x) \\
		\vdots \\
		\vdots \\
		\tilde{T}_n(x)
	\end{pmatrix} 
	\, .
\end{align}

By comparing these two $(n+1) \times (n+1)$ matrices $T_M$ with $(T_M)_{ij} = \tilde{t}_j^{(i)}$ and $A_M$ with $(A_M)_{ij} = \tilde{a}_j^{(i)}$, it becomes obvious that they are related through
\begin{align}
	T_M = A_M^{-1} \, ,
\end{align}
i.e. the two matrices are the inverse of each other. We can also obtain an additional decomposition of the matrix $T_M$ by looking at \eqref{equ:ExpandedSumChebyshev} and \eqref{equ:CoefficientExpandedSum}
\begin{align}
	T_M = A P t \, ,
\end{align}
where we introduce the diagonal matrix $A_{kk} = A^k$, the lower triangular Pascal matrix $P_{jk} = \binom{j}{k}$ and the matrix $t$ which follows from the definition given in \eqref{equ:tCoefficientsChebyshev}. This decomposition allows to easily visualize the effect of changing the lower bound of the domain $\omega_0$, with $x=\exp(-\omega_0)$ and $A$ as in \eqref{equ:CoefficientsDomain}. By setting $\left.A_{kk}\right|_{\omega_0\neq 0} = e^{\omega_0 k} \left.A_{kk}\right|_{\omega_0 = 0}$, it follows
\begin{align}
	\left.(T_M)_{nk}\right|_{\omega_0\neq 0} = \left.\tilde{t}_k^{(n)}\right|_{\omega_0\neq 0} = e^{\omega_0 n} \left.\tilde{t}_k^{(n)}\right|_{\omega_0 = 0} \quad , \quad \left.(A_M)_{nk}\right|_{\omega_0\neq 0} = \left.\tilde{a}_k^{(n)}\right|_{\omega_0\neq 0} = e^{-\omega_0 n} \left.\tilde{a}_k^{(n)}\right|_{\omega_0 = 0} \, .
	\label{equ:ChebyshevX0Effect}
\end{align}



\begin{thebibliography}{10}

\bibitem{FlavourLatticeAveragingGroupFLAG:2021npn}
{\scshape Flavour Lattice Averaging Group (FLAG)} collaboration, \emph{{FLAG
  Review 2021}},
  \href{https://doi.org/10.1140/epjc/s10052-022-10536-1}{\emph{Eur. Phys. J. C}
  {\bfseries 82} (2022) 869}
  [\href{https://arxiv.org/abs/2111.09849}{{\ttfamily 2111.09849}}].

\bibitem{Kaneko:2023kxx}
T.~Kaneko, \emph{{Heavy flavor physics from lattice QCD}},
  \href{https://doi.org/10.22323/1.430.0238}{\emph{PoS} {\bfseries LATTICE2022}
  (2023) 238} [\href{https://arxiv.org/abs/2304.01618}{{\ttfamily
  2304.01618}}].

\bibitem{Fael:2024rys}
M.~Fael, M.~Prim and K.K.~Vos, \emph{{Inclusive $B\rightarrow X_c \ell
  {\bar{\nu }}_\ell$ and $B\rightarrow X_u \ell {\bar{\nu }}_\ell$ decays:
  current status and future prospects}},
  \href{https://doi.org/10.1140/epjs/s11734-024-01090-w}{\emph{Eur. Phys. J.
  ST} {\bfseries 233} (2024) 325}.

\bibitem{Finauri:2023kte}
G.~Finauri and P.~Gambino, \emph{{The q$^{2}$ moments in inclusive semileptonic
  B decays}}, \href{https://doi.org/10.1007/JHEP02(2024)206}{\emph{JHEP}
  {\bfseries 02} (2024) 206}
  [\href{https://arxiv.org/abs/2310.20324}{{\ttfamily 2310.20324}}].

\bibitem{Hashimoto:2017wqo}
S.~Hashimoto, \emph{{Inclusive semi-leptonic B meson decay structure functions
  from lattice QCD}}, \href{https://doi.org/10.1093/ptep/ptx052}{\emph{PTEP}
  {\bfseries 2017} (2017) 053B03}
  [\href{https://arxiv.org/abs/1703.01881}{{\ttfamily 1703.01881}}].

\bibitem{Hansen:2017mnd}
M.T.~Hansen, H.B.~Meyer and D.~Robaina, \emph{{From deep inelastic scattering
  to heavy-flavor semileptonic decays: Total rates into multihadron final
  states from lattice QCD}},
  \href{https://doi.org/10.1103/PhysRevD.96.094513}{\emph{Phys. Rev. D}
  {\bfseries 96} (2017) 094513}
  [\href{https://arxiv.org/abs/1704.08993}{{\ttfamily 1704.08993}}].

\bibitem{Gambino:2020crt}
P.~Gambino and S.~Hashimoto, \emph{{Inclusive Semileptonic Decays from Lattice
  QCD}}, \href{https://doi.org/10.1103/PhysRevLett.125.032001}{\emph{Phys. Rev.
  Lett.} {\bfseries 125} (2020) 032001}
  [\href{https://arxiv.org/abs/2005.13730}{{\ttfamily 2005.13730}}].

\bibitem{Gambino:2022dvu}
P.~Gambino, S.~Hashimoto, S.~M\"achler, M.~Panero, F.~Sanfilippo, S.~Simula
  et~al., \emph{{Lattice QCD study of inclusive semileptonic decays of heavy
  mesons}}, \href{https://doi.org/10.1007/JHEP07(2022)083}{\emph{JHEP}
  {\bfseries 07} (2022) 083}
  [\href{https://arxiv.org/abs/2203.11762}{{\ttfamily 2203.11762}}].

\bibitem{Barone:2023tbl}
A.~Barone, S.~Hashimoto, A.~J\"uttner, T.~Kaneko and R.~Kellermann,
  \emph{{Approaches to inclusive semileptonic B$_{(s)}$-meson decays from
  Lattice QCD}}, \href{https://doi.org/10.1007/JHEP07(2023)145}{\emph{JHEP}
  {\bfseries 07} (2023) 145}
  [\href{https://arxiv.org/abs/2305.14092}{{\ttfamily 2305.14092}}].

\bibitem{Evangelista:2023fmt}
{\scshape Extended Twisted Mass} collaboration, \emph{{Inclusive hadronic decay
  rate of the \ensuremath{\tau} lepton from lattice QCD}},
  \href{https://doi.org/10.1103/PhysRevD.108.074513}{\emph{Phys. Rev. D}
  {\bfseries 108} (2023) 074513}
  [\href{https://arxiv.org/abs/2308.03125}{{\ttfamily 2308.03125}}].

\bibitem{ExtendedTwistedMass:2024myu}
{\scshape Extended Twisted Mass} collaboration, \emph{{Inclusive Hadronic Decay
  Rate of the \ensuremath{\tau} Lepton from Lattice QCD: The
  u\textasciimacron{}s Flavor Channel and the Cabibbo Angle}},
  \href{https://doi.org/10.1103/PhysRevLett.132.261901}{\emph{Phys. Rev. Lett.}
  {\bfseries 132} (2024) 261901}
  [\href{https://arxiv.org/abs/2403.05404}{{\ttfamily 2403.05404}}].

\bibitem{Fukaya:2020wpp}
H.~Fukaya, S.~Hashimoto, T.~Kaneko and H.~Ohki, \emph{{Towards fully
  nonperturbative computations of inelastic $\ell N$ scattering cross sections
  from lattice QCD}},
  \href{https://doi.org/10.1103/PhysRevD.102.114516}{\emph{Phys. Rev. D}
  {\bfseries 102} (2020) 114516}
  [\href{https://arxiv.org/abs/2010.01253}{{\ttfamily 2010.01253}}].

\bibitem{Barone:2022gkn}
A.~Barone, A.~J\"uttner, S.~Hashimoto, T.~Kaneko and R.~Kellermann,
  \emph{{Inclusive semi-leptonic $B_{(s)}$ mesons decay at the physical $b$
  quark mass}}, \href{https://doi.org/10.22323/1.430.0403}{\emph{PoS}
  {\bfseries LATTICE2022} (2023) 403}
  [\href{https://arxiv.org/abs/2211.15623}{{\ttfamily 2211.15623}}].

\bibitem{Kellermann:2022mms}
R.~Kellermann, A.~Barone, S.~Hashimoto, A.~J\"uttner and T.~Kaneko,
  \emph{{Inclusive semi-leptonic decays of charmed mesons with M\"obius domain
  wall fermions}}, \href{https://doi.org/10.22323/1.430.0414}{\emph{PoS}
  {\bfseries LATTICE2022} (2023) 414}
  [\href{https://arxiv.org/abs/2211.16830}{{\ttfamily 2211.16830}}].

\bibitem{Barone:2023iat}
A.~Barone, S.~Hashimoto, A.~J\"uttner, T.~Kaneko and R.~Kellermann,
  \emph{{Chebyshev and Backus-Gilbert reconstruction for inclusive semileptonic
  $B_{(s)}$-meson decays from Lattice QCD}},
  \href{https://doi.org/10.22323/1.453.0236}{\emph{PoS} {\bfseries LATTICE2023}
  (2024) 236} [\href{https://arxiv.org/abs/2312.17401}{{\ttfamily
  2312.17401}}].

\bibitem{Kellermann:2023yec}
R.~Kellermann, A.~Barone, S.~Hashimoto, A.~J\"uttner\ensuremath{\mathit{c}} and
  T.~Kaneko\ensuremath{\mathit{a}}, \emph{{Studies on finite-volume effects in
  the inclusive semileptonic decays of charmed mesons}},
  \href{https://doi.org/10.22323/1.453.0272}{\emph{PoS} {\bfseries LATTICE2023}
  (2024) 272} [\href{https://arxiv.org/abs/2312.16442}{{\ttfamily
  2312.16442}}].

\bibitem{Kellermann:2024jqg}
R.~Kellermann, A.~Barone, A.~Elgaziari, S.~Hashimoto, Z.~Hu, A.~J\"uttnerc
  et~al., \emph{{Systematic effects in the lattice calculation of inclusive
  semileptonic decays}},  in \emph{{41st International Symposium on Lattice
  Field Theory}}, 11, 2024 [\href{https://arxiv.org/abs/2411.18058}{{\ttfamily
  2411.18058}}].

\bibitem{Barata:1990rn}
J.C.A.~Barata and K.~Fredenhagen, \emph{{Particle scattering in Euclidean
  lattice field theories}},
  \href{https://doi.org/10.1007/BF02102039}{\emph{Commun. Math. Phys.}
  {\bfseries 138} (1991) 507}.

\bibitem{Bailas:2020qmv}
G.~Bailas, S.~Hashimoto and T.~Ishikawa, \emph{{Reconstruction of smeared
  spectral function from Euclidean correlation functions}},
  \href{https://doi.org/10.1093/ptep/ptaa044}{\emph{PTEP} {\bfseries 2020}
  (2020) 043B07} [\href{https://arxiv.org/abs/2001.11779}{{\ttfamily
  2001.11779}}].

\bibitem{Hansen:2019idp}
M.~Hansen, A.~Lupo and N.~Tantalo, \emph{{Extraction of spectral densities from
  lattice correlators}},
  \href{https://doi.org/10.1103/PhysRevD.99.094508}{\emph{Phys. Rev. D}
  {\bfseries 99} (2019) 094508}
  [\href{https://arxiv.org/abs/1903.06476}{{\ttfamily 1903.06476}}].

\bibitem{Luscher:1985dn}
M.~Luscher, \emph{{Volume Dependence of the Energy Spectrum in Massive Quantum
  Field Theories. 1. Stable Particle States}},
  \href{https://doi.org/10.1007/BF01211589}{\emph{Commun. Math. Phys.}
  {\bfseries 104} (1986) 177}.

\bibitem{Luscher:1986pf}
M.~Luscher, \emph{{Volume Dependence of the Energy Spectrum in Massive Quantum
  Field Theories. 2. Scattering States}},
  \href{https://doi.org/10.1007/BF01211097}{\emph{Commun. Math. Phys.}
  {\bfseries 105} (1986) 153}.

\bibitem{Shamir:1993zy}
Y.~Shamir, \emph{{Chiral fermions from lattice boundaries}},
  \href{https://doi.org/10.1016/0550-3213(93)90162-I}{\emph{Nucl. Phys. B}
  {\bfseries 406} (1993) 90}
  [\href{https://arxiv.org/abs/hep-lat/9303005}{{\ttfamily hep-lat/9303005}}].

\bibitem{Furman:1994ky}
V.~Furman and Y.~Shamir, \emph{{Axial symmetries in lattice QCD with Kaplan
  fermions}}, \href{https://doi.org/10.1016/0550-3213(95)00031-M}{\emph{Nucl.
  Phys. B} {\bfseries 439} (1995) 54}
  [\href{https://arxiv.org/abs/hep-lat/9405004}{{\ttfamily hep-lat/9405004}}].

\bibitem{Brower:2012vk}
R.C.~Brower, H.~Neff and K.~Orginos, \emph{{The M\"obius domain wall fermion
  algorithm}}, \href{https://doi.org/10.1016/j.cpc.2017.01.024}{\emph{Comput.
  Phys. Commun.} {\bfseries 220} (2017) 1}
  [\href{https://arxiv.org/abs/1206.5214}{{\ttfamily 1206.5214}}].

\bibitem{Colquhoun:2022atw}
{\scshape JLQCD} collaboration, \emph{{Form factors of
  B\textrightarrow{}\ensuremath{\pi}\ensuremath{\ell}\ensuremath{\nu} and a
  determination of |Vub| with M\"obius domain-wall fermions}},
  \href{https://doi.org/10.1103/PhysRevD.106.054502}{\emph{Phys. Rev. D}
  {\bfseries 106} (2022) 054502}
  [\href{https://arxiv.org/abs/2203.04938}{{\ttfamily 2203.04938}}].

\bibitem{Aoki:2023qpa}
{\scshape JLQCD} collaboration,
  \emph{{B\textrightarrow{}D*\ensuremath{\ell}\ensuremath{\nu}\ensuremath{\ell}
  semileptonic form factors from lattice QCD with M\"obius domain-wall
  quarks}}, \href{https://doi.org/10.1103/PhysRevD.109.074503}{\emph{Phys. Rev.
  D} {\bfseries 109} (2024) 074503}
  [\href{https://arxiv.org/abs/2306.05657}{{\ttfamily 2306.05657}}].

\bibitem{BMW:2012hcm}
{\scshape BMW} collaboration, \emph{{High-precision scale setting in lattice
  QCD}}, \href{https://doi.org/10.1007/JHEP09(2012)010}{\emph{JHEP} {\bfseries
  09} (2012) 010} [\href{https://arxiv.org/abs/1203.4469}{{\ttfamily
  1203.4469}}].

\bibitem{Morningstar:2003gk}
C.~Morningstar and M.J.~Peardon, \emph{{Analytic smearing of SU(3) link
  variables in lattice QCD}},
  \href{https://doi.org/10.1103/PhysRevD.69.054501}{\emph{Phys. Rev. D}
  {\bfseries 69} (2004) 054501}
  [\href{https://arxiv.org/abs/hep-lat/0311018}{{\ttfamily hep-lat/0311018}}].

\bibitem{Tomii:2016xiv}
{\scshape JLQCD} collaboration, \emph{{Renormalization of domain-wall bilinear
  operators with short-distance current correlators}},
  \href{https://doi.org/10.1103/PhysRevD.94.054504}{\emph{Phys. Rev. D}
  {\bfseries 94} (2016) 054504}
  [\href{https://arxiv.org/abs/1604.08702}{{\ttfamily 1604.08702}}].

\bibitem{Boyle:2015vda}
{\scshape UKQCD} collaboration, \emph{{Conserved currents for Mobius Domain
  Wall Fermions}}, \href{https://doi.org/10.22323/1.214.0087}{\emph{PoS}
  {\bfseries LATTICE2014} (2015) 087}.

\bibitem{Nakayama:2016atf}
K.~Nakayama, B.~Fahy and S.~Hashimoto, \emph{{Short-distance charmonium
  correlator on the lattice with M\"obius domain-wall fermion and a
  determination of charm quark mass}},
  \href{https://doi.org/10.1103/PhysRevD.94.054507}{\emph{Phys. Rev. D}
  {\bfseries 94} (2016) 054507}
  [\href{https://arxiv.org/abs/1606.01002}{{\ttfamily 1606.01002}}].

\bibitem{BoyleGrid}
P.~Boyle, A.~Yamaguchi, G.~Cossu and A.~Portelli, ``Grid: Data parallel c++
  mathematical object library.''

\bibitem{Boyle:2015tjk}
P.~Boyle, A.~Yamaguchi, G.~Cossu and A.~Portelli, \emph{{Grid: A next
  generation data parallel C++ QCD library}},
  \href{https://arxiv.org/abs/1512.03487}{{\ttfamily 1512.03487}}.

\bibitem{Yamaguchi:2022feu}
A.~Yamaguchi, P.~Boyle, G.~Cossu, G.~Filaci, C.~Lehner and A.~Portelli,
  \emph{{Grid: OneCode and FourAPIs}},
  \href{https://doi.org/10.22323/1.396.0035}{\emph{PoS} {\bfseries LATTICE2021}
  (2022) 035} [\href{https://arxiv.org/abs/2203.06777}{{\ttfamily
  2203.06777}}].

\bibitem{PortelliHadrons}
A.~Portelli, R.~Abott, N.~Asmussen, A.~Barone, P.~Boyle and F.E.~et~al.,
  \emph{aportelli/hadrons: Hadrons v1.3},  2022.
\newblock 10.5281/zenodo.4063666.

\bibitem{LepageLSQ}
P.~Lepage and C.~Gohlke, \emph{gplepage/lsqfit: lsqfit version 13.2.4},  2024.
\newblock 10.5281/zenodo.592174.

\bibitem{Lepage:2001ym}
{\scshape HPQCD} collaboration, \emph{{Constrained curve fitting}},
  \href{https://doi.org/10.1016/S0920-5632(01)01638-3}{\emph{Nucl. Phys. B
  Proc. Suppl.} {\bfseries 106} (2002) 12}
  [\href{https://arxiv.org/abs/hep-lat/0110175}{{\ttfamily hep-lat/0110175}}].

\bibitem{LepageCORR}
P.~Lepage, \emph{gplepage/corrfitter: corrfitter version 8.2},  2021.
\newblock 10.5281/zenodo.592232.

\bibitem{LepageGVAR}
P.~Lepage, C.~Gohlke and D.~Hackett, \emph{gplepage/gvar: gvar version 13.1.1},
   2024.
\newblock 10.5281/zenodo.592129.

\bibitem{Leibovich:1997em}
A.K.~Leibovich, Z.~Ligeti, I.W.~Stewart and M.B.~Wise, \emph{{Semileptonic B
  decays to excited charmed mesons}},
  \href{https://doi.org/10.1103/PhysRevD.57.308}{\emph{Phys. Rev. D} {\bfseries
  57} (1998) 308} [\href{https://arxiv.org/abs/hep-ph/9705467}{{\ttfamily
  hep-ph/9705467}}].

\bibitem{Fahy:2016qji}
B.~Fahy, G.~Cossu and S.~Hashimoto, \emph{{Approaching the Bottom Using Fine
  Lattices With Domain-Wall Fermions}},
  \href{https://doi.org/10.22323/1.256.0118}{\emph{PoS} {\bfseries LATTICE2016}
  (2016) 118} [\href{https://arxiv.org/abs/1702.02303}{{\ttfamily
  1702.02303}}].

\bibitem{ParticleDataGroup:2024cfk}
{\scshape Particle Data Group} collaboration, \emph{{Review of particle
  physics}}, \href{https://doi.org/10.1103/PhysRevD.110.030001}{\emph{Phys.
  Rev. D} {\bfseries 110} (2024) 030001}.

\bibitem{Kaneko:2017xgg}
{\scshape JLQCD} collaboration, \emph{{D meson semileptonic form factors in
  $N_f$ = 3 QCD with M\"obius domain-wall quarks}},
  \href{https://doi.org/10.1051/epjconf/201817513007}{\emph{EPJ Web Conf.}
  {\bfseries 175} (2018) 13007}
  [\href{https://arxiv.org/abs/1711.11235}{{\ttfamily 1711.11235}}].

\bibitem{Backus:1968}
G.~Backus and F.~Gilbert, \emph{The resolving power of gross earth data},
  \href{https://doi.org/10.1111/j.1365-246X.1968.tb00216.x}{\emph{Geophysical
  Journal of the Royal Astronomical Society} {\bfseries 16} (1968) 169}.

\bibitem{BESIII:2021duu}
{\scshape BESIII} collaboration, \emph{{Measurement of the absolute branching
  fraction of inclusive semielectronic $D_s^+$ decays}},
  \href{https://doi.org/10.1103/PhysRevD.104.012003}{\emph{Phys. Rev. D}
  {\bfseries 104} (2021) 012003}
  [\href{https://arxiv.org/abs/2104.07311}{{\ttfamily 2104.07311}}].

\bibitem{Hu:2025hpn}
Z.~Hu, A.~Barone, A.~Elgaziari, S.~Hashimoto, A.~J\"uttner, T.~Kaneko et~al.,
  \emph{{Study on the $P$-wave form factors contributing to $ B_s $ to $D_s$
  inclusive semileptonic decays from lattice simulations}},  in \emph{{41st
  International Symposium on Lattice Field Theory}}, 1, 2025
  [\href{https://arxiv.org/abs/2501.19284}{{\ttfamily 2501.19284}}].

\bibitem{RevModPhys.78.275}
A.~Wei\ss{}e, G.~Wellein, A.~Alvermann and H.~Fehske, \emph{The kernel
  polynomial method},
  \href{https://doi.org/10.1103/RevModPhys.78.275}{\emph{Rev. Mod. Phys.}
  {\bfseries 78} (2006) 275}.

\end{thebibliography}

\providecommand{\href}[2]{#2}\begingroup\raggedright\endgroup






\end{document}